\documentclass[12pt]{article}
\usepackage{arxiv}

\usepackage[utf8]{inputenc} %
\usepackage[T1]{fontenc}    %
\usepackage{booktabs}       %
\usepackage{amsfonts}       %
\usepackage{nicefrac}       %
\usepackage{microtype}      %

\usepackage[numbers,sort&compress]{natbib}
\usepackage{doi}
\usepackage{algorithm,algpseudocode}
\usepackage{amsmath}
\usepackage{amssymb}
\usepackage{mathtools}
\usepackage{amsthm}
\usepackage{amsfonts,bbm,mathrsfs}
\usepackage{longtable,multirow,threeparttable,tabularx,booktabs,caption}
\usepackage[font=small,skip=4pt]{caption}
\usepackage{subcaption}

\usepackage[dvipsnames]{xcolor}
\usepackage{graphicx}   %
\usepackage{hyperref}   %
\hypersetup{
    colorlinks,
    linkcolor={black},
    citecolor={green!50!black},
    urlcolor={blue!50!black}
}
\usepackage[capitalize,nameinlink]{cleveref}
\DeclareMathOperator{\diag}{diag}

\usepackage{tocloft}
\usepackage{ifthen,filecontents}

\title{Multi-fidelity Bayesian Optimization: A Review}

\author{{\hspace{1mm}Bach Do}\\
	University of Houston\\
	Houston, TX 77004 \\
	\texttt{bdo3@uh.edu} \\
	\And
 {\hspace{1mm}Ruda Zhang}\\ %
	University of Houston\\
	Houston, TX 77004 \\
	\texttt{rudaz@uh.edu} \\
}

\date{}

\renewcommand{\headeright}{ }
\renewcommand{\shorttitle}{MF BO: A Review}

\begin{document}
\maketitle

\begin{abstract}
Resided at the intersection of multi-fidelity optimization (MFO) and Bayesian optimization (BO), MF BO has found a niche in solving expensive engineering design optimization problems, thanks to its advantages in incorporating physical and mathematical understandings of the problems, saving resources, addressing exploitation-exploration trade-off, considering uncertainty, and processing parallel computing.
The increasing number of works dedicated to MF BO suggests the need for a comprehensive review of this advanced optimization technique.   
In this paper, we survey recent developments of two essential ingredients of MF BO: Gaussian process (GP) based MF surrogates and acquisition functions.
We first categorize the existing MF modeling methods and MFO strategies to locate MF BO in a large family of surrogate-based optimization and MFO algorithms.
We then exploit the common properties shared between the methods from each ingredient of MF BO to describe important GP-based MF surrogate models and review various acquisition functions.
By doing so, we expect to provide a structured understanding of MF BO.
Finally, we attempt to reveal important aspects that require further research for applications of MF BO in solving intricate yet important design optimization problems, including constrained optimization, high-dimensional optimization, optimization under uncertainty, and multi-objective optimization.
\end{abstract}


\clearpage

\tableofcontents

\clearpage
\section{Introduction}\label{Sec1}
Engineering design is an iterative process to develop a product or a system for a specific purpose.
At the beginning of the conventional design process, engineers determine product specifications via a set of performance metrics.
The iterative design process starts with an initial design, followed by comprehensive analyses to evaluate the performance metrics of this design.
These metrics allow engineers to assess whether the current design satisfies the established specifications.
If any specification is deemed invalid or in need of refinement, engineers make necessary changes to the design based on their intuition and experience, and all information gathered from trial designs~\cite{Arora2016}.

Design optimization replaces the conventional design process to accelerate the design cycle and obtain better products~\cite{Kochenderfer2019,Martins2021}.
Since it necessitates a formal mathematical formulation of the optimization problem, design optimization encapsulates the performance metrics in an objective to be achieved, design variables to be changed, and constraints to be met.
Once the optimization problem is formulated, changes in designs are automatically made via an optimization algorithm that is systematic and requires less intervention from engineers.

We distinguish the following two concepts in optimization: mathematical optimization and numerical optimization.
\textit{Mathematical optimization} is the problem of finding a set of points where a real-valued function attains its minimum (or maximum) value.
This function is called the objective function, and its domain can be prescribed by a set of equality and/or inequality constraints, each determined by a separate function.
\textit{Numerical optimization} is the use of numerical algorithms to solve optimization problems.
Common optimization algorithms can be broadly split into two groups, depending on whether they require derivatives of the objective function or not: derivative-based methods such as line search and trust region methods, and derivative-free methods such as population-based and direct search methods.

Consider the following engineering design optimization problem:
\begin{equation}\label{eqn1}
	\begin{aligned}
		\underset{\bf x}{\min} \ \ & f(\bf x)\\
		\textrm{s.t.} \ \ 
		& \bf x \in \mathcal{X}, 
	\end{aligned}
\end{equation} 
where ${\bf x} \in \mathbb{R}^d$ is the vector of $d$ design (input) variables selected in some feasible domain $\mathcal{X}$ and $f({\bf x}):\mathbb{R}^d \mapsto \mathbb{R}$ is the objective (output) function.

Solving problem~(\ref{eqn1}) via numerical optimization requires two critical elements: (1) the mathematical modeling of the problem and (2) an optimization algorithm.
The first aspect arises because objective functions for engineering design are often regarded as black boxes that are only queried at specified points of design variables for their values or derivatives.
Selecting optimization algorithms in the second aspect depends on the number of design variables and how the objective can be evaluated. 

\begin{figure*}[t]
	\centering
	\includegraphics[width=\textwidth]{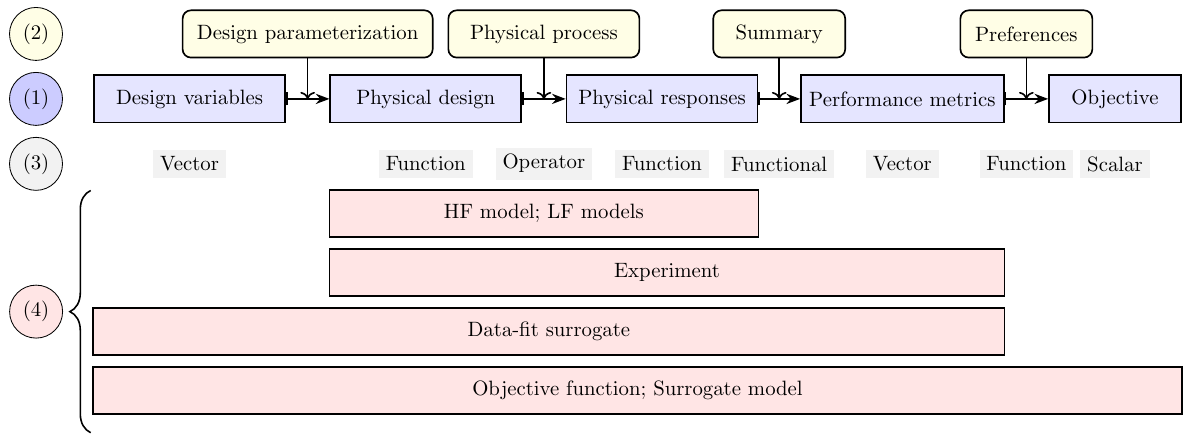}
	\caption{Mathematical modeling for engineering design optimization: (1) fundamental objects, (2) relations, (3) mathematical abstractions, and (4) other concepts in engineering design optimization.}
	\label{Fig-1}
\end{figure*}

The mathematical modeling of problem~(\ref{eqn1}) utilizes a mathematical model derived from our decent understanding of the governing equations of the physical process underlying the relationship between the design objective and design variables; see \cref{Fig-1}.
The expression of this mathematical model is the objective function $f(\bf x)$.
Unfortunately, solving the governing equations of $f(\bf x)$ is often analytically intractable in engineering design.
Thus, \textit{experiments} are considered as the ground truth to measure certain aspects of the physical process.
However, they are subject to uncertainties in design manufacturing, operational control, and measurement errors, which must be considered in experimental data analysis.
Oftentimes the mathematical model is implemented via its computational counterparts, such as finite element (FE) models~\cite{Bathe2006}, which simulate the physical process under hypothetical scenarios.

Depending on their prediction qualities and costs, computational models for the physical process can be classified into high- and low-fidelity models; see \cref{Fig-1}.
A \textit{high-fidelity model} (HF model) is a computational model that predicts the performance metrics with an accuracy level sufficient for a basic understanding of the underlying physical process and evaluation of the objective function.
We denote $f_\text{H}(\bf x)$ as the objective function value at ${\bf x}$ obtained from experiments or an HF model.
We consider both data from experiments and evaluations of an HF model as HF data.
Meanwhile, a \textit{low-fidelity model} (LF model) is a computational model that is designated for applications where a lower level of accuracy is acceptable for evaluating the objective function.
The objective function and data from predictions of an LF model are $f_\text{L}(\bf x)$ and LF data, respectively.
The advantage of the LF model is that it can be much more computationally efficient than its HF counterpart, both in space and time costs.
Examples of LF models include simplified-physics approximations and projection-based reduced order models (ROMs), as described in~\cite{Peherstorfer2018}.

If the objective function is costly to compute, the optimization process may fail to find a good design given a limited budget on resources.
This may be attributed to, for example, the collection of HF data.
In this case, we may consider using a \textit{surrogate model} (i.e., metamodel or model of model), that is, a computationally efficient approximation of the objective function.
In the context of engineering design and throughout this paper, we distinguish surrogate models from LF models.
A surrogate model approximates the objective function, while an LF model approximates an HF model of the physical process.
The data-fit models as defined in~\cite{Peherstorfer2018}, when applied to numerical optimization, usually approximate mappings with low-dimensional input and output such as the objective function; therefore they are surrogates rather than LF models.

\begin{figure*}
	\centering
	\includegraphics[width=\textwidth]{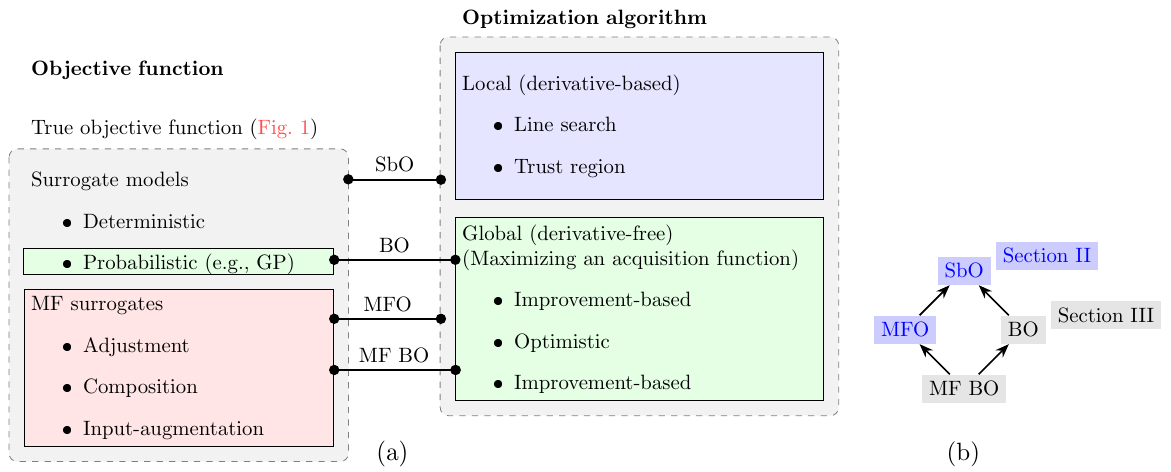}
	\caption{(a) Distinction between SbO, MFO, BO, and MF BO; (b) Hierarchy of these optimization concepts.}
	\label{Fig-2}
\end{figure*}

Once the mathematical modeling of the objective function is settled, we can consider the selection of an optimization algorithm.
Ideally, a good optimization algorithm should possess three properties: robustness, accuracy, and efficiency~\cite{Nocedal2006}.
However, these goals may conflict and trade-offs between them should be addressed.
A prudent approach is to select optimization algorithms based on how the objective function is accessed; see \cref{Fig-2}.

Due to their high accuracy level, HF data are expected to be used in the design optimization process.
However, the collection of HF data discourages direct applications of any optimization algorithms because it requires conducting time-consuming experiments or running HF computational models, thereby reducing efficiency.
For example, one crash simulation reported by Ford Motor Company required 36-160h to complete~\cite{Wang2006}.
In another example, a high-precision structural analysis to capture the elastoplastic dynamic responses of a five-story steel frame under a ground motion of 5s required 23 days~\cite{Ohsaki2009}.

As a successful approach to addressing the efficiency-accuracy trade-off, \textit{surrogate-based optimization} (SbO) approximates the objective function $f(\bf x)$ by a surrogate model~\cite{Queipo2005,Wang2006,Forrester2008,Simpson2008,Forrester2009}.
The surrogate can be a data-fit model constructed by fitting the parameters of an interpolation, regression, or machine learning model, given pairs of design variables and performance metrics or the overall score derived from HF data, thereby favoring the accuracy aspect.
Alternatively, it can be induced from an LF model used as an approximation of the physical process in \cref{Fig-1}, which, however, favors the efficiency aspect.
We can further classify surrogates into deterministic surrogates, such as response surfaces and most neural networks, and probabilistic surrogates, such as Gaussian process (GP)~\cite{Rasmussen2006}.
We can also categorize them into local and global surrogates.
A local surrogate approximates a black-box objective function in a restricted region of the design variable space, e.g., the neighborhood of a design variable vector.
A global surrogate approximates all predictions of that black-box objective function.

To further balance the accuracy-efficiency trade-off in SbO, we can use \textit{multi-fidelity optimization} (MFO) that leverages both LF models and HF data to construct an MF surrogate for the objective function~\cite{Alexandrov1998,Huang2006smo,Forrester2007,Viana2014}.
From this definition, it is clear that MFO is a type of SbO.
The MF surrogate can be constructed, for example, by adjusting an LF model-induced surrogate to HF data via adjustment methods~\cite{Kennedy2000,Han2012,Gratiet2014}.
It can also be constructed by composing an LF model-induced surrogate with an input-input mapping or an output-output mapping~\cite{Bandler1994,Bandler2004,Perdikaris2017}.

\textit{Bayesian optimization} (BO) is a global optimization technique that assumes a prior probabilistic model on the objective function, and combines it with the available data to guide the optimization process.
BO is well suited for small- to moderate-dimensional optimization problems that involve expensive-to-evaluate objective functions that offer no efficient mechanism to estimate their derivatives~\cite{Snoek2012,Shahriari2016,Frazier2018}.
Most of the time, a GP is used as the prior model in BO, due to its tractability and flexibility.
Typically, the GP prior is specified simply by a constant mean function and a covariance function with a closed form and a few hyperparameters.
Given the dataset at each optimization iteration, the posterior model (i.e., the prior model conditioned on the dataset) determines an acquisition function reflecting an optimization policy under uncertainty, which is then numerically maximized to find the next design point.
A derivative-free, global optimization algorithm is usually used for this purpose.
The GP posterior at each iteration serves as a surrogate for the objective function, and therefore BO is a type of SbO.
BO tends to favor accuracy if it constructs the GP posterior based on HF data.
The details of BO and its applications are deferred until \Cref{Sec32}.  

\textit{Multi-fidelity Bayesian optimization} (MF BO) is the intersection of MFO and BO. More specifically, MF BO includes LF models within a prior probabilistic model for the objective function, and combines this prior with HF data to guide the optimization process.

MF BO is applicable whenever (1) the objective function is expensive to evaluate and (2) LF models are available or can be constructed.
In fact, we advocate that MF BO should be applied whenever these two conditions are true because it can use both data and engineering knowledge efficiently.
BO is a well-recognized approach to optimizing expensive objective functions that makes efficient use of the available HF data and therefore reduces the number of HF data points to ensure solution accuracy.
MFO opens up the optimization black box in engineering design, and not only exploits the structure of mathematical modeling details but also uses our knowledge about the underlying physical processes.
Combining BO and MFO further accelerates engineering design optimization processes.

In recent years, there has been a remarkable increase in the number of successful applications of MF BO across diverse domains of engineering design~\cite[see e.g.,][]{Perdikaris2017,Meliani2019,Tran2020cise,Hebbal2021oe,Khatamsaz2021md}.
Additionally, the evolving landscapes of MFO and especially BO, with their continuous advancements in addressing intricate yet important design optimization problems~\cite{Frazier2018,Wang2023}, hold a promise for further enhancing the capabilities of MF BO.
Thus, there is a need for a comprehensive review of MF BO to meet the demand for future development progress and widespread applications of this powerful optimization method.

Our goal in this review is twofold.
First, we comprehensively review the existing techniques from two essential ingredients of MF BO: the GP-based MF surrogates and acquisition functions.
To accomplish this, we exploit the common properties shared between the techniques from each ingredient.
With this approach, we expect to provide a structured understanding of MF BO.
Second, we provide critical research topics on solving intricate yet important design optimization problems, aimed at expanding the horizon of the existing MF BO algorithms.	
    
We organize the rest of this paper as follows.
\Cref{Sec2} surveys two basic elements of MFO: MF modeling methods and MFO strategies.
\Cref{Sec3} provides an overview of MF BO.
\Cref{Sec4} describes important GP-based MF surrogates.
\Cref{Sec5} surveys various acquisition functions of BO and how we can modify them for use of MF BO.
\Cref{Sec6:examples} shows the applications of several MF BO methods for optimizing benchmark functions and standard airfoil shapes.
\Cref{Sec7} reviews the extensions of BO and MFO to address intricate yet important design optimization problems, including constrained optimization, high-dimensional optimization, optimization under uncertainty, and multi-objective optimization.
\Cref{Sec8} summarizes and concludes this paper.

\clearpage
\section{Multi-fidelity optimization}\label{Sec2}

This section starts with a brief overview of SbO (\Cref{Sec21}).
It then surveys two elements of MFO: MF surrogates (\Cref{Sec22}) and MFO strategies (\Cref{Sec23}).

\subsection{Surrogate-based optimization}\label{Sec21}

The development of SbO in engineering design optimization has received a substantial boost since the work \textit{Design and Analysis of Computer Experiments} (DACE) by~\citet{Sacks1989}.
DACE used a cost-effective surrogate model for estimating the output of an expensive computer code.
The best linear unbiased predictor was obtained by minimizing the mean squared error (MSE) of the predictor.
DACE also introduced several classical adaptive design criteria to experimental design processes, such as the integrated MSE, maximum MSE, and expected posterior entropy.

Inspired by DACE, engineering design optimization has used surrogates for objective functions or optimization problems to reduce the computational cost of running expensive computational models.
Based on the size of the input spaces used to construct the surrogates, they can be categorized as local or global surrogates.

The use of local surrogates is a classical approach of SbO.
This approach is classified into local function and local problem approximations~\cite{Barthelemy1993}.
The former guides the optimization process using local surrogates for the objective function evaluated at the current design point.
Taylor series approximations often serve as such local surrogates.
The latter replaces the original optimization problem with a sequence of easier-to-solve subproblems, or attempts to reduce the number of constraints or the number of design variables.

The use of global surrogates, such as response surfaces and most neural networks, to facilitate the optimization process dates back to the 1980s~\cite{Barthelemy1993}.
At that time, there was a limited number of applications of global surrogates in engineering design optimization.
However, the early 1990s witnessed many applications of polynomials-based response surfaces, first introduced by~\citet{Box1951}, to single-discipline and multi-discipline design optimization problems~\cite{Sobieski1997}.
The community also soon recognized a major limitation of these methods that they are only applicable to small-dimensional problems due to the curse of dimensionality.
Consequently, attention began to shift from polynomials-based response surfaces toward alternative global surrogates such as radial basis functions~\cite{Hussain2002}, support vector regression~\cite{Girosi1998}, Kriging~\cite{Cressie1990,Kleijnen2009}, reduced order models~\cite{Antoulas2005}, and ensembles~\cite{Goel2007,LiuH2020}. 

There exist invaluable surveys that provide comprehensive insights into the use of SbO methods in engineering design optimization.
\citet{Queipo2005} discussed the fundamental issues arising in surrogate-based analysis and optimization such as the design of experiments, construction of surrogate models, model selection, and sensitivity analysis.
\citet{Wang2006}, from a practitioner’s perspective, provided an overview of how global surrogate modeling (i.e., metamodeling) can support engineering design optimization. 
\citet{Simpson2008} conducted a detailed review of the development of approximation methods in multi-discipline design optimization from 1980 to 2008.
\citet{Forrester2009} surveyed several state-of-the-art surrogate models and their applications in optimization.
\citet{Viana2014} reviewed the progression of metamodeling techniques in multi-discipline design optimization.

The use of MFO in engineering design was initiated by~\citet{Haftka1991} with a multiplicative MF model (\Cref{Sec22}).
This model broadens the application range of HF responses by adjusting the associated LF responses.
During the optimization process, the multiplicative MF model is imposed \textit{consistency conditions}, which require that the HF response and its derivatives at a given design point must align with those of the corresponding scaled LF response~\cite{Alexandrov2001}.
If the HF derivatives are not available, the consistency can be defined for the response only~\cite{Rodriguez2001}.

Early attempts at MFO relied on local approximations with deterministic MF surrogates valid in the neighborhood of the current point of design variables.
Applications of local MFO methods have been found in supersonic transport design~\cite{Knill1999}, aerodynamic wing design~\cite{Alexandrov2001}, and airfoil design~\cite{Gano2005}.
Since the seminal works by \citet{Huang2006smo} and \citet{Forrester2007}, MFO has shifted its attention to global approximations with probabilistic MF surrogates valid for a large region of the design variable space.
As we will see in \Cref{Sec23}, a majority of recent applications of MFO in engineering design have been from global methods.

\subsection{Multi-fidelity surrogates}\label{Sec22}

There are typically three classes of MF modeling methods:
adjustment, composition, and input augmentation.
\paragraph{Adjustment.}
The class of adjustment MF methods constructs MF surrogates that use adjustment factors to update LF models explaining HF data.
Most adjustment MF surrogates to approximate $f_\text{H}({\bf x})$ take the following form:
\begin{equation}\label{eqn2}
	 f_\text{H}({\bf x}) = w \rho({\bf x}) f_\text{L}({\bf x})\\
	+ (1-w)\left[a + b f_\text{L}({\bf x}) + \delta({\bf x})\right]+ c \delta({\bf x}).
\end{equation}
Here we abuse the notations $f_\text{H}({\bf x})$ and $f_\text{L}(\bf x)$ to represent the surrogates that approximate the objective function predictions of HF and LF models, respectively.
$w$ is a weighting factor, which can be adjusted during the optimization process~\cite{Gano2005}.
$\rho({\bf {x}})$ is an adjustment coefficient function indicating that the adjusted LF response can vary in different regions of the design variable space~\cite{Haftka1991}.
$a$, $b$, and $c$ are three unknown constants.
$\delta({\bf{x}})$ is a discrepancy function~\cite{Kennedy2000}.

Note that the form in \cref{eqn2} does not include cases when multiple LF models are available.
However, we can use it to recover almost all MF surrogates derived from multiple hierarchical LF models listed in a recent review of MF modeling methods~\cite{FernandezGodino2023}.

\begin{figure*}
	\centering
	\includegraphics[width=0.8\textwidth]{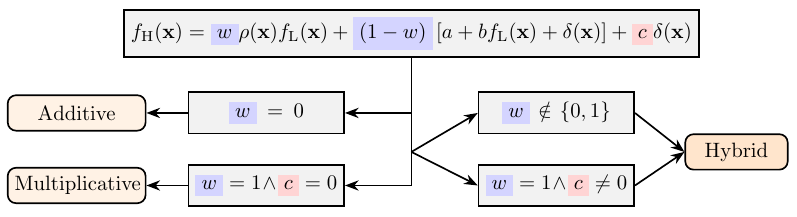}
	\caption{Additive, multiplicative, and hybrid adjustment MF surrogates.}
	\label{Fig-3}
\end{figure*}

Based on the values of $w$ and $c$, we can further classify adjustment MF surrogates into three types as shown in \cref{Fig-3}:
\begin{itemize}
	\item Additive, for $w=0$.
	
	\item Multiplicative, for $w=1$ and $c=0$.
	
	\item Hybrid additive/multiplicative, for either $w \notin \{0,1\}$, or $w=1$ and $c \neq 0$.
\end{itemize}

\Cref{Table1} enumerates important published works for each type of adjustment MF surrogates.
We see that, a substantial portion of these works has relied on additive adjustment MF surrogates. 

\begin{table*}
	\caption{Published works on additive, multiplicative, hybrid adjustment MF surrogates.}
	\label{Table1}
	\centering
	\begin{tabular}{ll}
		\hline \noalign{\smallskip}
		Type & Reference\\
		\hline \noalign{\smallskip}
		Additive & \cite{Lewis2000,Gano2006b,Viana2009,Palar2016,ZhangYiming2018,FernandezGodino2019,Song2019,Kou2019,Meng2020,Viana2014,Durantin2017,Teichert2019,LYan2019,Kennedy2000,Forrester2007,Leary2003,Xiong2008,Kuya2011,Toal2011,Han2012,Keane2012,Goh2013,Zheng2013,deBaar2015,Park2017,ZhangY2018,Rokita2018,Xiao2018,Jiang2019,Serani2019,Zhou2020,Shu2021,Kaps2022,Toal2023,Xu2023,Ribeiro2023,Peng2023,Wiangkham2023}     \\
		\noalign{\smallskip}
		Multiplicative & \cite{Haftka1991,Chang1993,Goldfeld2005,Hino2006,Sun2010}     \\
		\noalign{\smallskip}
		Hybrid & \cite{Gano2005,Gano2006a,Han2013,Tyan2015,Nguyen2015,Hu2016,Absi2016,Rumpfkeil2017,Bryson2017,Rumpfkeil2019,Wang2021,Qian2008,Gratiet2013,Gratiet2014,Parussini2017,Hao2020,Ji2023,Cheng2021}    \\
		\hline \noalign{\smallskip}
	\end{tabular}
\end{table*}

Apart from the form in \cref{eqn2}, adjustment MF surrogates can also be described using weighted average models when there exist a total of $m$ LF models, such that~\cite{Goel2007}
\begin{subequations}\label{eqn3}
	\begin{align}
		f_\text{H}({\bf x}) & = \sum_{i=1}^{m} w_i f_i({\bf x}) + \delta({\bf x}) \label{eqn3-1},\\
		f_\text{H}({\bf x}) & = \sum_{i=1}^{m}\rho_i({\bf x})f_i({\bf x}) + \delta({\bf x}) \label{eqn3-2},
	\end{align}
\end{subequations}
where $f_i({\bf x})$ represents a surrogate for approximating the prediction of LF model $i$, $w_i$ is a weight value corresponding to $f_i({\bf x})$ with $\sum_{i=1}^{m} w_i = 1$, and $\rho_i({\bf x})$ is a weight function corresponding to $f_i({\bf x})$ with $\sum_{i=1}^{m}\rho_i({\bf x})=1$.
We use \cref{eqn3-1} when the weights are fixed in different regions of the design variable space.
We use \cref{eqn3-2} when there is a correlation between $f_\text{H}({\bf x})$ and $f_i({\bf x})$ at any values of the design variables.

\begin{figure*}
	\centering
	\includegraphics[width=0.8\textwidth]{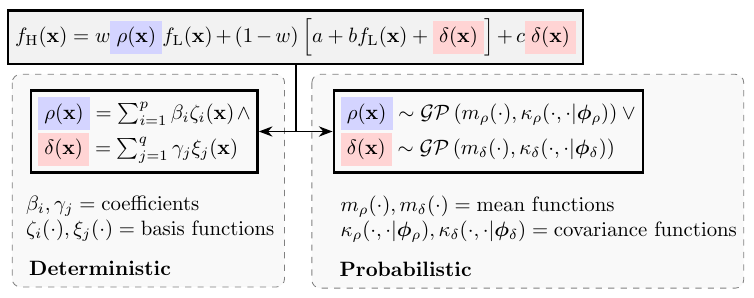}
	\caption{Deterministic versus probabilistic adjustment MF surrogates.}
	\label{Fig-4}
\end{figure*}

Based on how $\rho({\bf {x}})$ and $\delta({\bf{x}})$ are modeled, we can further classify adjustment MF surrogates into deterministic and probabilistic ones; see \cref{Fig-4}.
\Cref{Table2} lists notable published works for each of these classes.

\begin{table*}
	\caption{Published works on deterministic and probabilistic adjustment MF surrogates.}
	\label{Table2}
	\centering
	\begin{tabular}{ll}
		\hline \noalign{\smallskip}
		Type & Reference\\
		\hline \noalign{\smallskip}
		Deterministic & \cite{ZhangYiming2018,FernandezGodino2019,Song2019,Kou2019,Durantin2017,Han2013,Tyan2015,Nguyen2015,Palar2016,LiuB2016,LYan2019,Rumpfkeil2017,Rumpfkeil2019,Wang2021}  \\
		\noalign{\smallskip}
		Probabilistic &    \cite{Kennedy2000,Forrester2007,Xiong2008,Kuya2011,Toal2011,Han2012,Keane2012,Goh2013,Park2017,Serani2019,Qian2008,Gratiet2013,Gratiet2014,Allaire2014,Parussini2017,Zhou2017kbs,Xiao2018,Ji2023,PietrenkoDabrowska2022sr2}\\
		\hline \noalign{\smallskip}
	\end{tabular}
\end{table*}

Deterministic adjustment MF surrogates described in \cref{Fig-4} model both $\rho({\bf {x}})$ and $\delta({\bf{x}})$ as linear combinations of a finite number of basis functions $\zeta_i({\bf {x}})$ and $\xi_j({\bf x})$, respectively.
Popular basis functions include monomials~\cite{ZhangYiming2018,FernandezGodino2019}, neural networks~\cite{Kou2019}, radial basis functions~\cite{Tyan2015,LiuB2016,Durantin2017}, and orthogonal polynomial functions~\cite{Palar2016}. 
The combination coefficients and hyperparameters associated with the basis functions are determined by the least squares or regularized least squares approach. 
In comparison, probabilistic adjustment MF surrogates describe $\rho({\bf {x}})$ or $\delta({\bf{x}})$ as a stochastic process, which is often a GP characterized by a mean function and a covariance function.   
Several GP-based MF surrogates can be found in~\citet{Brevault2020}.
The mathematical expressions of common GP-based MF surrogates are deferred until \Cref{Sec4}.  

It is worth noting that the aforementioned classifications of adjustment MF surrogates do not conflict.
This means, an additive (multiplicative, or hybrid) adjustment MF surrogate can be either deterministic or probabilistic.

\paragraph{Composition.}
The class of composition methods consists of input-input and output-output mapping MF techniques.
MF surrogates from input-input mapping techniques often take the form of $f_\text{H}({\bf x}) \approx f_\text{L}(g({\bf x}))$,
where $g(\cdot)$ maps the input space of the HF model to that of the LF model.
Given an initial set of HF design variable values and the corresponding output values, the set of corresponding LF design variable values is found by a parameter extraction, and a deterministic mapping $g(\cdot)$ can be found iteratively by defining  $g^j(\cdot)$ at iteration $j$ as a linear combination of some predefined and fixed basis functions~\cite{Bandler1994}.
An appropriate mapping $g(\cdot)$ is found when $\| {\bf f}_\text{H}(\cdot) - {\bf f}_\text{L}(\cdot)\| \leq \varepsilon$, where ${\bf f}_\text{L}$ corresponds to $g$ values evaluated at the given HF design variable points, $\|\cdot\|$ the $L_2$ norm, and $\varepsilon$ a small positive constant . 
Meanwhile, MF surrogates from output-output mapping techniques take the form of
$f_\text{H}({\bf x}) \approx h\left(f_\text{L}({\bf x})\right)$,
  where $h(\cdot)$ maps the output space of the LF model to that of the HF model.
The idea is to convert the original many-to-one mapping relationship between the space of HF design variables and the space of HF objective function to a one-to-one mapping relationship between certain regions of the spaces of LF and HF objective functions. 
This is possible because the values of LF objective function $f_\text{L}(\cdot)$ at HF design variable values can be obtained easily at a low computational cost.
Moreover, a composition MF surrogate can be deterministic or probabilistic depending on how we define the mappings $g(\cdot)$ and $h(\cdot)$.
The reader is referred to \Cref{Table3} for a list of published works related to the composition MF modeling methods.

\begin{table*}
	\caption{Published works on composition MF methods.}
	\label{Table3}
	\centering
	\begin{tabular}{ll}
		\hline \noalign{\smallskip}
		Type & Reference\\
		\hline \noalign{\smallskip}
		Input-input mapping & \cite{Bandler1994,Bandler2004,Koziel2006,Robinson2008,TaoSiyu2019} \\
		\noalign{\smallskip}
		Output-output mapping &  \cite{Zheng2014,Zhou2017aei,Perdikaris2017,Jiang2018,Cutajar2019,Hebbal2021smo,Li2023}   \\
		\hline \noalign{\smallskip}
	\end{tabular}
\end{table*}

\paragraph{Input augmentation.}
The class of input-augmentation MF modeling methods describes MF surrogates as functions that depend not only on design variables but also on fidelity variables.
This differs from the class of adjustment MF surrogates in Eqs.~(\ref{eqn2}) and (\ref{eqn3}) which are functions of the design variables only.
The fidelity variables are indeed categorical but can be assumed to be continuous for computational efficiency~\cite{Kandasamy2017}.
If categorical fidelity variables are considered, non-continuous covariance functions are proposed for constructing GP-based MF surrogates. 
\Cref{Sec46} provides different input-augmentation MF modeling methods for both continuous and categorical fidelity variables.
Note that most input-augmentation MF surrogates in the literature are probabilistic and have been predominantly used in BO.

\subsection{Multi-fidelity optimization strategies}\label{Sec23}

MFO strategies take advantage of MF surrogates and/or their derivatives to accelerate the optimization process.
Each MFO strategy is characterized by its way of updating the design points and MF surrogates during the optimization process.
There are three approaches to MFO strategies: derivative-based, model-then-optimize, and model-and-optimize.
Notable published works on the algorithms of each approach are listed in \Cref{Table4}.

\begin{table*}
	\caption{Published works on MFO algorithms.}
	\label{Table4}
	\centering
	\begin{tabular}{ll}
		\hline \noalign{\smallskip}
		Type  & Reference\\
		\hline \noalign{\smallskip}
		Derivative-based  & \cite{Alexandrov1998,Alexandrov2000,Alexandrov2001,Gano2005,Robinson2008,March2011,Elham2015,Bryson2018,De2020,WuN2022,ZhangT2023,PietrenkoDabrowska2022tmtt,PietrenkoDabrowska2022sr1} \\
		\noalign{\smallskip}
		Model-then-optimize   &  \cite{Viana2009,Leusink2015,Singh2017,Yang2018}   \\
		\noalign{\smallskip}
		\vtop{\hbox{\strut Model-and-optimize}\hbox{\strut (MF BO only)}} &  \cite{Sobester2004,Huang2006smo,Forrester2007,Perdikaris2016,Chen2016,Kandasamy2017,Pang2017,Amrit2018,ZhangY2018,Bonfiglio2018a,Bonfiglio2018b,Serani2019,Ghoreishi2019,Bailly2019,Shi2020,Kontogiannis2020,Tran2020jcp,Tran2020cise,Ruan2020,Hao2020,Nachar2020,Fiore2021,WuY2021,He2021,ZhangX2021,Shu2021,Khatamsaz2021aiaa,Sacher2021,WuW2021,Renganathan2021,Kishi2022,Cheng2022,Foumani2023,Huang2023,Grassi2023,Fiore2023,Shintani2023,Lin2023,Winter2023,Ribeiro2023}   \\
		\noalign{\smallskip}
		\hline 
	\end{tabular}
\end{table*}

\paragraph{Derivative-based approach.}
This approach uses a deterministic MF surrogate and its derivatives to inform the step length and search direction. 
The optimization is often via the trust-region framework that solves a trust-region subproblem to find the search direction~\cite{Nocedal2006}.
Specifically, we first restrict the step size (i.e., defining the trust region) for a reliable local model, and then find the search direction within the defined truss region so that it minimizes the local model.
In each optimization iteration, a deterministic MF surrogate serves as the local model, while its derivatives guide the search toward a good solution to the trust-region subproblem~\cite{Alexandrov1998,Alexandrov2001,Robinson2008}.
The MF surrogate is also used for computing a ratio of actual to predicted improvement for checking whether the new design point obtained from solving the trust-region subproblem is accepted or not, and whether the local model agrees with the actual objective function if the new design point is accepted.
This ratio also allows the trust region to adjust its size.
To make the algorithm more robust, the MF surrogate is occasionally
calibrated using consistency conditions ensuring the preservation of both HF response and its derivatives through the use of local MF surrogates~\cite{Alexandrov1998,Alexandrov2001}.
These conditions remain valid under assumptions that the HF model is considered the ground truth, and that the HF model and its derivatives are deterministic.
To further address constrained optimization problems, the MF trust-region framework is equipped with a constraint-handling technique, for example, Lagrange multiplier method~\cite{Robinson2008,March2011}, augmented Lagrangian method~\cite{Alexandrov2001}, and penalty method with either $L_1$ penalty functions~\cite{Alexandrov2001,Gano2005} or quadratic penalty functions~\cite{Elham2015}.

While the derivative-based MFO approach can handle high-dimensional optimization problems, its nature as a derivative-based approach drives it toward several limitations.
\begin{itemize}
	\item First, the calibration of local MF surrogates using HF derivatives may hinder direct applications of the approach to engineering design problems because it is often nontrivial to extract the derivatives of quantities of interest from engineering HF models.
	
	\item Second, the approach demands a high level of expertise in optimization from practicing engineers.
	This is because the performance of local approximations strongly depends on a careful selection of optimization parameters underlying the trust-region framework.
	These parameters, including the threshold values for the ratio of actual to predicted improvement, trust-region scaling factors, and tolerance thresholds, are essential for ensuring not only the solution improvement but also the accuracy of local MF surrogates and solution quality in each iteration.
	
	\item Third, the approach may provide no insight to engineers because its derivative-based nature is generally nontransparent~\cite{Wang2006}.
	
	\item Finally, the approach cannot consider imperfections inherent in an HF model.
\end{itemize}

A possible way to avoid the reliance of an HF model on the local information when solving a trust-region subproblem is to adopt derivative-free trust-region algorithms~\cite{Conn2009,Wild2011}.
These algorithms are designed to determine the search direction for each iteration so that it simply satisfies the conditions of Cauchy decrease and eigenvector decrease of the local model.
Additionally, derivative-free trust-region algorithms do not require fully-linear or fully-quadratic local models in all iterations as their derivative-based counterparts often do, but require fully-linear or fully-quadratic local models during a finite, uniformly bounded, number of iterations to achieve global
convergence to first-order or second-order stationary point, respectively~\cite{Conn2009}.
\citet{March2012smo,March2012aiaa} exploited the concept of derivative-free trust-region algorithms for developing both unconstrained and constrained MFO algorithms.

\paragraph{Model-then-optimize approach.}
As its name implies, this approach separates the construction of an MF surrogate for the costly objective function from the optimization process.
It relies on a strong assumption that the MF surrogate for the objective function possesses sufficient accuracy to ensure both the feasibility and quality of candidate solutions.
However, obtaining such a level of accuracy may be challenging without the support of an optimization algorithm, primarily because not all regions of the design variable space are useful for optimization.
In practice, focusing on learning a high-confidence region of interest can lead to favorable optimal solutions~\cite{ZhangF2023}.
 
The model-then-optimize approach is useful in certain cases.
It allows an examination of how initial sampling designs influence the solution quality.
It also enables the use of population-based algorithms for solving multi-objective optimization problems~\cite{Viana2009,Leusink2015,Singh2017,Yang2018}.

\paragraph{Model-and-optimize approach.}
This approach, also known as sequential model-based optimization approach~\cite{Bossek2020}, iterates between updating global MF surrogates for the costly objective function and using them to propose new design points via maximizing an acquisition function (i.e., infill criterion or figure of merit).
From this definition, MF BO is a type of model-and-optimization approach. 
The model-and-optimize approach, via formulating the acquisition function, minimizes the need for using numerous optimization parameters as the derivative-based approach does.
Factors that affect the performance of model-and-optimize algorithms include the choice of MF surrogates, the selection of acquisition functions, and the construction of initial MF surrogates.
As shown in \Cref{Table4}, this approach has recently attracted a large number of published works on the applications of MFO in design optimization.

\section{Multi-fidelity Bayesian optimization}\label{Sec3}

As illustrated in \cref{Fig-2} and briefly defined in \Cref{Sec1}, BO is a global optimization technique working based on constructing a probabilistic surrogate model 
to represent our belief about an expensive-to-evaluate objective function given its observations, which then defines an optimization policy via an acquisition function that is maximized before a new design point is selected for evaluating the objective function to update our belief.
BO was originated by~\citet{Mockus1975} and popularized by~\citet{Jones1998} and their work on the \textit{Efficient Global Optimization} (EGO) algorithm.
Combining BO with MFO leads to MF BO which further accelerates the optimization process.
 
This section starts with describing the fundamental elements of EGO (\Cref{Sec31}) and BO (\Cref{Sec32}), and how the two concepts connect to each other.
It then elucidates distinctive characteristics that set MF BO apart from BO (\Cref{Sec33}).

\subsection{Efficient global optimization}\label{Sec31}

EGO operates through a sequence of three key steps in each iteration.
Oftentimes it terminates when reaching a pre-specified upper limit on the number of iterations,
which corresponds to a fixed computational budget. 

In the first step, EGO uses a Kriging model~\cite{Sacks1989,Chiles1999}
as a stochastic process to approximate the costly objective function.
The hyperparameters underlying the Kriging model are determined by maximizing
model likelihood given the current training dataset~\cite{Jones1998}.
This enables the derivation of the best linear unbiased predictor and the MSE of the predictor.
Traditionally, in geostatistics, the best linear unbiased predictor of Kriging is found by
minimizing the MSE~\cite{Sacks1989,Chiles1999,Kleijnen2009}. 
However, within the context of EGO, minimizing the MSE is equivalent to maximizing the likelihood
because EGO models the residual term of Kriging as a Gaussian~\cite{Jones1998}.

In the second step, EGO formulates what is known as the expected improvement based on three sources of information: (1) the best design point found in the current training dataset, (2) the best linear unbiased predictor, and (3) the MSE of the predictor.
The concept of expected improvement (\Cref{Sec511}) provides a delicate balance between exploiting the information provided by the predictor and exploring uncertain regions within it.
Specifically, exploitation (i.e., immediate reward) focuses on regions of the design variable space where the objective function values are expected to be small, while exploration (i.e., expected future reward) discovers regions where the predictions of the objective function are highly uncertain.

In the third step, EGO maximizes the expected improvement for a new design point that is added to the current training dataset, leading to the update of both the stochastic model and the best-observed solution in the next iteration.
Notably, maximizing the expected improvement is more straightforward and computationally efficient than computing the derivatives of the costly objective function.
This computational advantage contributes to the overall efficiency of EGO.

In the community of design optimization, a Kriging model used to approximate an objective function is often considered a global surrogate.
However, if we view design points in the current training dataset locally, we can consider a Kriging model as a local surrogate, which knows nothing about the objective function in regions far from the design points (i.e., extrapolation regions).
In this context, EGO is a technique that relies on a local surrogate for global optimization.  

\begin{figure*}
	\centering
	\includegraphics[width=0.9\textwidth]{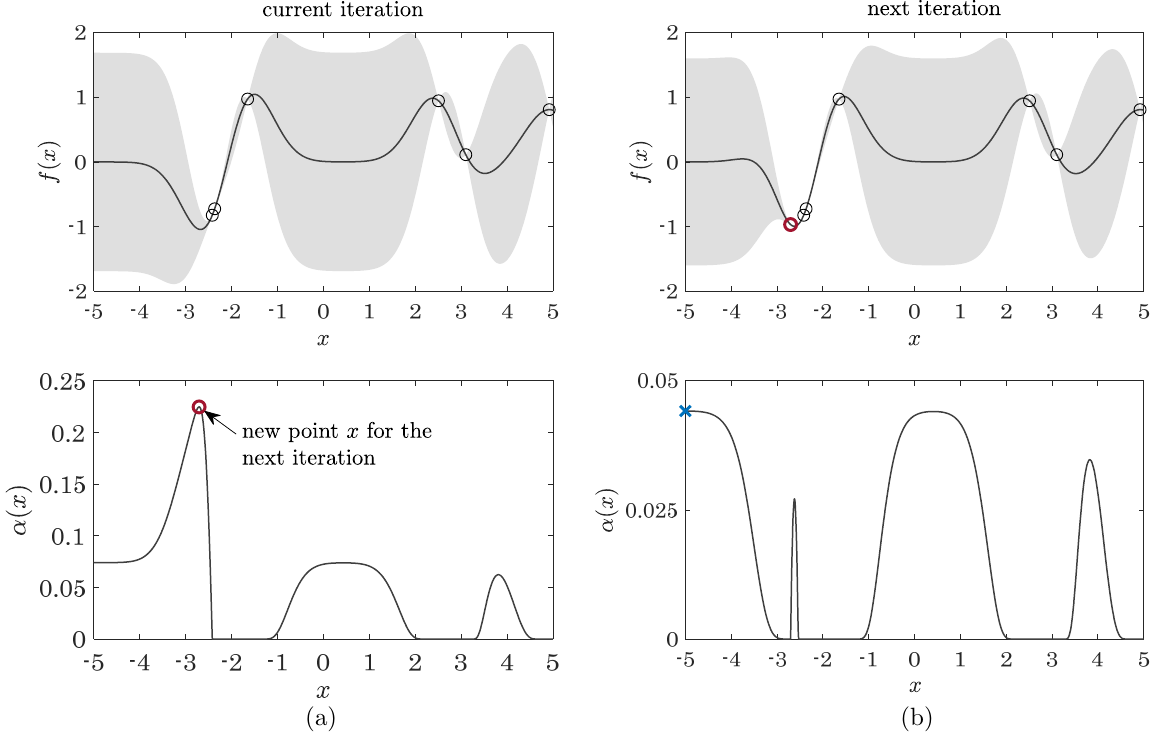}
	\caption{ Illustration of two consecutive iterations (a) and (b) of BO for minimizing a univariate objective function.}
	\label{Fig-5}
\end{figure*}

\subsection{Bayesian optimization}\label{Sec32}

While the fundamental concept behind BO closely resembles that of EGO, the nomenclature of BO is commonly used in the communities of statistics and machine learning~\cite{Bull2011,Snoek2012,Shahriari2016,Frazier2018}.
Oftentimes BO uses a univariate GP (Appendix A) as a surrogate for the costly objective function while an acquisition function guides the optimization process. 
Thus, EGO can be considered a specific instance of BO that adopts the expected improvement as an acquisition function.
Many fundamental similarities shared between Kriging and GP further reinforce the connection between EGO and the broader family of BO algorithms~\cite{Christianson2023}.

\begin{algorithm}[t]
	\caption{Generic BO.}\label{Algo1}
	\begin{algorithmic}[1]
		\State \textbf{Input:} $f_\text{H}(\cdot)$, $\mathcal{X}$, $K$, $N$; \label{Algo1:1}
		\State Generate $N$ samples of ${\bf x}$; 
		
		\For {$i=1:N$} 
		\State $f^i \gets f_\text{H}({\bf x}^i)$; \textcolor{black}{\Comment{Costly step}}
		\EndFor
		
		\State $\mathcal{D}^0 \gets \{{\bf x}^i,f^i\}_{i=1}^N$; \label{Algo1:6}
		
		\State $\{{\bf x}_{\min},f_{\min}\} \gets \min\{f^i,\, i=1,\dots, N\}$;
		
		\For {$k=1:K$} 
		\State Construct $\hat{f}_\text{H}^k({\bf x})$ based on $\mathcal{D}^{k-1}$; \label{Algo1:9}
		\State Formulate $\alpha({\bf x})$; \label{Algo1:10}
		\State ${\bf x}^{k} \gets \underset{{\bf x}}{\mathrm{argmax}} \ \ \alpha({\bf x})$ s.t. ${\bf x} \in \mathcal{X}$; \label{Algo1:11}
		\State $f^{k} \gets f_\text{H}({\bf x}^{k})$;
		\textcolor{black}{\Comment{Costly step}} \label{Algo1:12}
		\State $\mathcal{D}^k\gets\mathcal{D}^{k-1} \cup \{{\bf x}^{k},f^{k}\}$; \label{Algo1:13}
		\State $\{{\bf x}_{\min},f_{\min}\} \gets \min\{f_{\min},f^{k}\}$;
		\EndFor
		
		\State \Return $\{{\bf x}_{\min},f_{\min}\}$.
	\end{algorithmic}
\end{algorithm}

Although we use Kriging and GP interchangeably in this work, as also observed elsewhere~\cite[see e.g.,][]{Forrester2008,Erickson2018}, it is worth noting that there is a subtle difference between the two concepts.
This difference lies in the degrees of human intervention for hyperparameter estimation and treatment of anisotropy~\cite{Christianson2023}.
While Kriging often determines its hyperparameters from a frequentist perspective by minimizing the MSE of the predictor, GP determines its hyperparameters from a Bayesian perspective by maximizing the model likelihood or utilizing a full Bayesian treatment for the mode of hyperparameter vector.
In Kriging, the covariance between two objective function values only depends on the relative point-wise distance of the corresponding design variable vectors, which is called stationary.
If such a relative distance is calculated in a Euclidean space, Kriging's assumption of intrinsic stationarity also implies isotropy, i.e., the covariance is invariant to both translation and rotation.
This property requires a high degree of human intervention to prescale or rotate the coordinate system for ensuring fidelity in spatial modeling.
In comparison, GP requires less human intervention by using different characteristic length scales along each axis or more flexibly parameterized rotations and scales, thus allowing the use of anisotropy covariance functions.
This is motivated by our expectation that a very large value of the length scale should be used for a less influential design variable.

\Cref{Fig-5} illustrates two consecutive iterations of BO for minimizing a univariate objective function $f(x)$.
In \cref{Fig-5}(a), BO fits a GP model for $f(x)$ to a training dataset of six samples (top panel).
Subsequently, it formulates and maximizes an acquisition function $\alpha(x)$ to identify a new design point (bottom panel).
In \cref{Fig-5}(b), BO proceeds by updating the GP model and reformulating $\alpha(x)$ with the design point identified in the previous iteration.

\Cref{Algo1} provides a pseudo-code for solving problem~(\ref{eqn1}) using the generic BO, where the objective function is the prediction $f_\text{H}(\bf x)$ of an HF model.
Parameters $K$ and $N$ in Line 1 represent threshold values for the number of BO iterations and the number of initial training samples, respectively.
$\mathcal{D}^{0}$ in Line 6 is the initial training dataset generated from the HF model.
Note that $\mathcal{D}^{0}$ can be HF data, which contain both predictions of the HF model and measures from experiments.  
$\hat{f}_\text{H}^k({\bf x})$ in Line 9 and $\mathcal{D}^{k}$ in Line 13 are the GP posterior and training dataset associated with iteration $k$ of BO, respectively.
The acquisition function $\alpha({\bf x})$ is formulated in Line 10 based on $\mathcal{D}^{k-1}$ and $\hat{f}_\text{H}^k({\bf x})$, which is detailed in \Cref{Sec5}.
Maximizing $\alpha({\bf x})$ in Line 11 is often straightforward as it does not invoke the HF model.
By doing so, we replace problem~(\ref{eqn1}), which is difficult to solve, with a series of simpler, inexpensive problems of maximizing the acquisition function in Line 11.

At this point, a critical question arises: under what conditions can we ensure that the existence of a unique global solution to a GP-distributed objective function ${f}_\text{H}({\bf x})$?
Addressing this question is crucial because it makes the construction of the GP model $\hat{f}_\text{H}^k({\bf x})$ in Line 9 of \Cref{Algo1} meaningful.
Since the optimization problem is completely defined by the mean and covariance functions of the GP and the design domain $\mathcal{X}$, their properties, intuitively, decide whether a unique global solution is guaranteed.
In fact, we can be certain that a global solution exists and that is unique under the following two mild assumptions: (1) the mean and covariance functions are continuous over a compact design domain $\mathcal{X}$, and (2) there are no two unique points in the domain that can have perfectly correlated function values~\cite{Garnett2023}.
The reader may consult Section~2.7 of \citet{Garnett2023} for detailed discussions on the existence and uniqueness of global solutions to GP-distributed functions.

The unique ability to optimize expensive-to-evaluate objective functions, i.e., making efficient use of the available data for ensuring solution accuracy, has made BO an invaluable tool for various applications in scientific and engineering design.
These applications have been found in machine learning~\cite{Bergstra2011}, aircraft design~\cite{Priem2020}, material design~\cite{Tran2019,Khatamsaz2021md}, experimental design~\cite{Greenhill2020}, material science~\cite{Ueno2016}, structural engineering~\cite{Mathern2021}, transportation~\cite{ShiR2021}, chemical engineering~\cite{Park2018}, electronics engineering~\cite{Torun2018}, environmental engineering~\cite{Manheim2019}, and  physics~\cite{Roussel2021}, to name a few.
Moreover, sophisticated BO algorithms have been developed to
(1) examine the performance of acquisition functions formulated from different optimization policies
when optimizing objective functions in the face of uncertainty and
(2) extend applications of BO to a wide range of design optimization problems.
The first aspect has examined the performance of BO acquisition functions derived from different formulation approaches: improvement-based~\cite{Jones2001}, optimistic~\cite{Srinivas2010}, information-based ~\cite{Hennig2012}, and likelihood-weighted~\cite{Blanchard2021jcp}.
It has also explored the performance of acquisition functions from one-step look-ahead and multi-step look-ahead perspectives (\Cref{Sec51}).
The second aspect has focused on developing advanced BO algorithms to solve intricate yet important design optimization problems such as constrained problems, high-dimensional problems, problems under uncertainty, and multi-objective problems (\Cref{Sec7}).
These problems are difficult to solve because of their own nature and/or limitations of the generic BO.
For the details of recent advances in BO, the reader may refer to comprehensive surveys and tutorials by~\citet{Brochu2010}, \citet{Shahriari2016}, \citet{Frazier2018}, and \citet{Wang2023}.

\subsection{Multi-fidelity Bayesian optimization}\label{Sec33}

As illustrated in \cref{Fig-2}, MF BO is the intersection of MFO and BO, and therefore it inherits the advantages of the two approaches.
There are several reasons that make MF BO powerful, especially when solving engineering optimization problems:
\begin{itemize}
	\item Resource saving. MF BO uses MF surrogates for a costly objective function, thus further reducing the number of HF data points by exploiting the structure of mathematical modeling details and/or using our knowledge about the underlying physical process.
	
	\item Handling exploitation-exploration trade-off. The optimization process can benefit from the attempts to address the exploitation-exploration trade-off by notable acquisition functions of the generic BO.
	
	\item Robustness to noise. Engineering optimization problems often involve noisy objective and constraint functions.
	While LF evaluations are less noisy than their HF counterparts, incorporating evaluations from both LF and HF models into BO can make the optimization process more robust to noise.
	This is justified because LF models are less detailed, therefore exhibit lower variability, whereas HF models capture more intricate details and complexities.
	
	\item Parallelization. It is easy to obtain LF evaluations in parallel, which further accelerates the optimization process.
	
	\item Adaptive optimization of fidelity. MF BO allows for adaptive selections of fidelity levels.
	This means, it can decide when and where to carry out HF evaluations based on the current state of knowledge about the objective function.
	This adaptability makes MF BO well-suited for situations when the cost or availability of HF evaluations varies across different regions of the design variable space.
	
	\item Incorporation of non-GP models. While the generic BO often relies on common GP models that, from a local view of design points in the training dataset, can only construct local surrogates, many LF models provide global information.
	The use of MF models enables BO to expand its horizon beyond the constraint on GP models.
	
\end{itemize}

Compared with the generic BO outlined in~\Cref{Algo1}, MF BO introduces two important modifications.
First, it constructs in Line 9 a GP-based MF surrogate for the objective function using LF models and HF data.
One of the most popular GP-based MF surrogates is the auto-regressive model~\cite{Kennedy2000}; see~\Cref{Sec421}.
Second, MF BO develops in Line 10 an MF acquisition function that is capable of selecting both a new design point and the fidelity level for a computational model to be called in Line 12.
The very first example of such an acquisition function is the so-called augmented expected improvement~\cite{Huang2006smo}; see~\Cref{Sec522}. 

The pioneering works by~\citet{Kennedy2000} and~\citet{Huang2006smo} have initiated three directions of recent research and development efforts of MF BO:
\begin{itemize}
	\item Enhancing MF surrogates; see~\Cref{Sec4}.
	
	\item Innovating new MF acquisition functions based on the existing BO acquisition functions; see~\Cref{Sec5}.
	
	\item Applying MF BO to solving various optimization problems in science and engineering design; see~\Cref{Table4}.	
\end{itemize}

\section{Gaussian process-based multi-fidelity surrogates}\label{Sec4}

Let ${\bf f}({\bf x})=[f_1({\bf x}),\dots,f_T({\bf x})]^\intercal$, $T \geq 2$, denote a vector of $T$ outputs of $T$ computational models with different fidelities that are used for predicting the objective function of an engineering design optimization problem.
We assume that these models share the same input variables so that we do not need input variable mappings to perform information transfers between the fidelities.
If the fidelities are sorted in increasing order, there exist $(T-1)$ LF predictions, and $f_T({\bf x})=f_\text{H}({\bf x})$ is the prediction of the highest-fidelity model.
If there is no obvious fidelity ordering, then there is no hierarchy of the predictions.

Let ${\bf X}_{t}\in \mathbb{R}^{N_t \times d}$, $t=1,\dots,T$, be a set of $N_t$ input samples associated with fidelity $t$ and ${\bf F}_{t}\in \mathbb{R}^{N_t}$ be a set of the corresponding output values.
Let ${\bf X}=[{\bf X}_1;\dots;{\bf X}_T] \in \mathbb{R}^{\sum_{t=1}^{T} N_t \times d}$ and ${\bf F}=[{\bf F};\dots;{\bf F}] \in \mathbb{R}^{\sum_{t=1}^{T} N_t}$ denote sets of input and output data, where ${\bf X}$ and ${\bf F}$ concatenate matrices ${\bf X}_t$ and vectors ${\bf F}_t$, respectively. 
Thus, $\mathcal{D}_t=[{\bf X}_t,{\bf F}_t] \in \mathbb{R}^{N_t \times (d+1)}$ and $\mathcal{D}=[\mathcal{D}_1;\dots;\mathcal{D}_T] \in \mathbb{R}^{\sum_{t=1}^{T} N_t \times (d+1)}$ are training datasets associated with fidelity $t$ and all fidelities, respectively.

In this section, we describe popular GP-based MF surrogates constructed from $\mathcal{D}$ for use of MF BO.
They include multivariate GP via linear model of coregionalization (\Cref{Sec41}), auto-regressive model and its variants (\Cref{Sec42}), graphical MF GP model (\Cref{Sec43}), Bayesian hierarchical model (\Cref{Sec44}), composition of GP models (\Cref{Sec45}), and input-augmentation GP-based MF surrogate models (\Cref{Sec46}).
To obtain a structured understanding of these models, we attempt to exploit common properties shared between them.

\subsection{Linear model of coregionalization}\label{Sec41}

Without ordering the model fidelities, consider the problem of constructing a $T$-variate GP surrogate to approximate the output vector ${\bf f}({\bf x})=\left[f_1({\bf x}),\dots,f_T({\bf x})\right]^\intercal$, which can be described as
\begin{equation}\label{eqn4}
	{\bf f}({\bf x}) = {\bf m}({\bf x}) + {\boldsymbol \delta}({\bf x}),
\end{equation}
where ${\bf m}({\bf x})$: $\mathbb{R}^d \mapsto \mathbb{R}^T$  and ${\boldsymbol \delta}({\bf x})$: $\mathbb{R}^d \mapsto \mathbb{R}^T$ are mean and discrepancy vectors, respectively.

The \textit{multivariate GP via linear model of coregionalization} (LMC)~\cite[see e.g.,][]{Chiles1999,Fricker2013} describes each element of ${\boldsymbol \delta}({\bf x})$ as a linear combination of $T$ independent zero-mean GPs with covariance functions $\kappa_1(\cdot,\cdot|{\boldsymbol \phi}_1),\dots,\kappa_T(\cdot,\cdot|{\boldsymbol \phi}_T)$, where ${\boldsymbol \phi}_t$ contains hyperparameters of $\kappa_t$.
Thus, the discrepancy vector can be written as ${\boldsymbol \delta}({\bf x}) = {\bf R} {\bf z}({\bf x})$, where ${\bf R} \in \mathbb{R}^{T \times T}$ consists of combination coefficients, and ${\bf z}(\cdot) \sim \mathcal{GP}\left({\bf 0},{\boldsymbol \Sigma}(\cdot,\cdot)\right)$ is a $T$-variate GP with ${\boldsymbol \Sigma}(\cdot,\cdot) = \diag\{\kappa_1(\cdot,\cdot|{\boldsymbol \phi}_1),\dots,\kappa_T(\cdot,\cdot|{\boldsymbol \phi}_T)\}$.
The GP prior for ${\bf f}(\cdot)$ in \cref{eqn4} reads
\begin{equation}\label{eqn5}
	{\bf f}(\cdot) \sim \mathcal{GP}\left({\bf m}(\cdot),{\bf S}(\cdot,\cdot)\right).
\end{equation}
Here 
${\bf S}(\cdot,\cdot) = {\bf R} {\boldsymbol \Sigma}(\cdot,\cdot) {\bf R}^\intercal = \sum_{t=1}^T {\bf C}_t \kappa_t(\cdot,\cdot|{\boldsymbol \phi}_t)$ is the so-called inter-group covariance matrix, where ${\bf C}_t = {\bf r}_t {\bf r}_t^\intercal$ is the $t$-th coregionalization matrix and ${\bf r}_t$ the $t$-th column of ${\bf R}$.
To determine ${\bf R}$, ${\boldsymbol \phi}_t$, and the parameters underlying ${\bf m}({\bf x})$, we condition the GP prior in \cref{eqn5} on the training dataset $\mathcal{D}$, which is called the training process, as detailed in Appendix B. 
Note that the inter-group covariance matrix ${\bf S}(\cdot,\cdot) \in \mathbb{R}^{T \times T}$ should not be confused with a significantly larger covariance matrix ${\bf K}({\bf X},{\bf X}) \in \mathbb{R}^{\sum_{t=1}^{T}N_t \times \sum_{t=1}^{T}N_t}$ that is used for the training process and prediction equations, i.e., Eqs.~(\ref{eqnB5}) and (\ref{eqnB6}) of Appendix B.

Once trained, the LMC model can provide predictions for the outputs of all fidelities at unseen input variable vectors.
An advantage of LMC is that it is capable of using nonseparable covariance structures for describing the outputs.
This results in sufficient good predictions and the capability of capturing joint uncertainty about the outputs, especially when they are different physical quantities~\cite{Fricker2013}.
A main disadvantage of the method is the huge computational cost it requires for training, which scales cubically with $\sum_{t=1}^{T}N_t$.

\subsection{Auto-regressive model and variants}\label{Sec42}

\subsubsection{KOH auto-regressive model}\label{Sec421}

If the fidelities are ordered, the surrogate for predicting the output of fidelity $t$ ($t=2,\dots,T$) can be constructed from the surrogate that approximates the output of fidelity $(t-1)$ using the \textit{auto-regressive model} (i.e., KOH model). The KOH model reads~\cite{Kennedy2000}
\begin{subequations}\label{eqn6}
	\begin{align}
		& f_1({\bf x}) = \delta_{1}({\bf x}), \label{eqn6-1}\\
		& f_t({\bf x}) =
		 b_{t-1} f_{t-1}({\bf x}) + \delta_{t}({\bf x}), \ \  t=2,\dots,T \label{eqn6-2},\\
		& \text{cov}\left[f_{t}({\bf x}),f_{t-1}({\bf x}')|f_{t-1}({\bf x})\right]=0 
		\label{eqn6-3},
	\end{align}
\end{subequations}
where $b_{t-1}$ in \cref{eqn6-2} is an unknown correlation coefficient and $\delta_{t}(\cdot) \sim \mathcal{GP}\left(m_{\delta,t}(\cdot),\kappa_{\delta,t}(\cdot,\cdot|{\boldsymbol \phi}_{\delta,t})\right)$ is the discrepancy function independent of $f_{t-1}(\cdot),\dots,f_{1}(\cdot)$.
Here $b_{t-1}$ and $\delta_{t}(\cdot)$ play similar roles as $b$ and $\delta(\cdot)$ in \cref{eqn2}, respectively.
A common choice for the mean function of $\delta_{t}(\cdot)$ is $m_{\delta,t}({\bf x})=\sum_{i=1}^p \beta_i \zeta_i({\bf x})$, where $\beta_i$ and  $\zeta_i({\bf x})$ are combination coefficients and basis functions, respectively.
The covariance function $\kappa_{\delta,t}(\cdot,\cdot|{\boldsymbol \phi}_{\delta,t})$ is often the squared exponential covariance function~\cite{Kennedy2000,Forrester2008}.
Other covariance functions for use of GP-based MF modeling include Matern~\cite{Pang2017} and composite covariance functions~\cite{Palar2023}. 
$\text{cov}\left[f_{t}({\bf x}),f_{t-1}({\bf x}')|f_{t-1}({\bf x})\right]$ in \cref{eqn6-3} represents the covariance of two random variables $f_{t}({\bf x})$ and $f_{t-1}({\bf x}')$ given that $f_{t-1}({\bf x})$ is known. 
The Markov property in \cref{eqn6-3} implies that observing $f_{t-1}({\bf x}')$ provides no information for predicting $f_{t}({\bf x})$ if $f_{t-1}({\bf x})$ is observed.
 
For simplicity, consider two fidelities $f_{1}({\bf x})=f_\text{L}({\bf x})$ and $f_{2}({\bf x})=f_\text{H}({\bf x})$.
In this case, the KOH model reads
\begin{subequations}\label{eqn7}
	\begin{align}
		& f_1({\bf x})  = \delta_1({\bf x}) 
		\label{eqn7-1},\\
		& f_2({\bf x}) = b_{1} f_{1}({\bf x}) + \delta_2({\bf x}) \label{eqn7-2},\\
            & \text{cov}\left[f_{2}({\bf x}),f_{1}({\bf x}')|f_{1}({\bf x})\right]=0 \label{eqn7-3}.
	\end{align}
\end{subequations}
Let $z_1({\cdot}) \sim \mathcal{GP}\left(0,\kappa_{\delta,1}(\cdot,\cdot|{\boldsymbol \phi}_{\delta,1})\right)$ and $z_2({\cdot}) \sim \mathcal{GP}\left(0,\kappa_{\delta,2}(\cdot,\cdot|{\boldsymbol \phi}_{\delta,2})\right)$, where the covariance functions $\kappa_{\delta,1}$ and $\kappa_{\delta,2}$ are parameterized by the hyperparameter vectors ${\boldsymbol \phi}_{\delta,1}$ and ${\boldsymbol \phi}_{\delta,2}$, respectively.
\cref{eqn7} can be rewritten as
\begin{subequations}\label{eqn8}
	\begin{align}
		& f_1({\bf x})  = m_{\delta,1}({\bf x}) + z_1({\bf x}) 
		\label{eqn8-1},\\
		& f_2({\bf x})  = b_1  m_{\delta,1}({\bf x}) + m_{\delta,2}({\bf x}) + b_1 z_1({\bf x}) + z_2({\bf x})
		\label{eqn8-2},\\
         & \text{cov}\left[f_{2}({\bf x}),f_{1}({\bf x}')|f_{1}({\bf x})\right]=0 \label{eqn8-3}.
	\end{align}
\end{subequations}
This is equivalent to
\begin{equation}\label{eqn9}
	{\bf f}({\bf x}) =  {\bf m}({\bf x}) + {\bf R} {\bf z}({\bf x}),
\end{equation}
where
\begin{subequations}\label{eqn10}
	\begin{align}
		{\bf f}({\bf x}) &= \left[f_1({\bf x}),f_2({\bf x})\right]^\intercal 
		\label{eqn10-1},\\
		{\bf m}({\bf x}) &= {\bf R} \boldsymbol{\mu}({\bf x}), \label{eqn10-2}\\
		{\bf R} &= \left[\begin{matrix}
			1 & 0 \\
			b_1 & 1
		\end{matrix}\right] = {\bf I}_2 + b_1 {\bf e}_2 {\bf e}^\intercal_1, \label{eqn10-3}\\
		\boldsymbol{\mu}({\bf x}) & = [m_{\delta,1}({\bf x}),m_{\delta,2}({\bf x})]^\intercal,
		\label{eqn10-4}\\
		{\bf z}({\bf x}) & = \left[z_1({\bf x}),z_2({\bf x})\right]^\intercal, \label{eqn10-5}\\
		{\bf z}(\cdot) & \sim \mathcal{GP}\left({\bf 0},{\boldsymbol \Sigma}(\cdot,\cdot)\right),
		\label{eqn10-6}\\
		{\boldsymbol \Sigma}(\cdot,\cdot) & = \left[\begin{matrix}
			\kappa_{\delta,1}(\cdot,\cdot|{\boldsymbol \phi}_{\delta,1}) & 0 \\
			0 & \kappa_{\delta,2}(\cdot,\cdot|{\boldsymbol \phi}_{\delta,2})
		\end{matrix}\right] \label{eqn10-7}.
	\end{align}
\end{subequations}
Here ${\bf I}_2$ denotes the $2$-by-$2$ identity matrix, ${\bf e}_1 = [1,0]^\intercal$, and ${\bf e}_2 = [0,1]^\intercal$. 

Thus, the GP prior for ${\bf f}({\bf x})$ in \cref{eqn9} can be written using the form in \cref{eqn5} with the inter-group covariance matrix
\begin{equation}\label{eqn11}
	\begin{aligned}
		{\bf S}(\cdot,\cdot)  = {\bf R} {\boldsymbol \Sigma}(\cdot,\cdot) {\bf R}^\intercal
		 &= \left[\begin{matrix}
		 	1 & 0 \\
		 	b_1 & 1
		 \end{matrix}\right] 
		 \left[\begin{matrix}
		 	\kappa_{\delta,1}(\cdot,\cdot|{\boldsymbol \phi}_{\delta,1}) & 0 \\
		 	0 & \kappa_{\delta,2}(\cdot,\cdot|{\boldsymbol \phi}_{\delta,2})
		 \end{matrix}\right]
		 \left[\begin{matrix}
		 	1 & 0 \\
		 	b_1 & 1
		 \end{matrix}\right]^\intercal \\
	 & = \left[\begin{matrix}
	 	\kappa_{\delta,1}(\cdot,\cdot|{\boldsymbol \phi}_{\delta,1}) & b_1 \kappa_{\delta,1}(\cdot,\cdot|{\boldsymbol \phi}_{\delta,1}) \\
	 	b_1 \kappa_{\delta,1}(\cdot,\cdot|{\boldsymbol \phi}_{\delta,1}) & b_1^2 \kappa_{\delta,1}(\cdot,\cdot|{\boldsymbol \phi}_{\delta,1})+\kappa_{\delta,2}(\cdot,\cdot|{\boldsymbol \phi}_{\delta,2})
	 \end{matrix}\right], 
	\end{aligned}
\end{equation}
which is the inter-group covariance matrix for the coKriging model derived by~\citet{Forrester2007}.

In the general case when there exist $T$ model fidelities ($T \geq 2$), the terms of \cref{eqn9} are
\begin{subequations}\label{eqn12}
	\begin{align}
		{\bf f}({\bf x}) & = \left[f_1({\bf x}),\dots,f_T({\bf x})\right]^\intercal, \label{eqn12-1}\\
		{\bf m}({\bf x}) & = {\bf R} \boldsymbol{\mu}({\bf x}), \label{eqn12-2}\\
		{\bf R} & = {\bf R}_T\dots {\bf R}_2, \, {\bf R}_t = {\bf I}_T + b_{t-1} {\bf e}_t {\bf e}^\intercal_{t-1}, \, t=2,\dots,T, \label{eqn12-3}\\
 		\boldsymbol{\mu}({\bf x}) & = [m_{\delta,1}({\bf x}),\dots,m_{\delta,T}({\bf x})]^\intercal \label{eqn12-4},
	\end{align}
\end{subequations}
where ${\bf I}_T$ denotes the $T$-by-$T$ identity matrix, ${\bf e}_t$ denotes the $T$-dimensional standard unit vector with a 1 in the $t$th coordinate, and the coefficient matrix ${\bf R}$ is a $T$-by-$T$ low triangular matrix~\cite{Garland2020}.

\subsubsection{Hierarchical Kriging model}\label{Sec422}

\citet{Han2012} proposed a hierarchical Kriging model by making three modifications to the KOH model in \cref{eqn6}.
First, they modeled the surrogate that approximates the lowest-fidelity output $f_1({\bf x})$ in \cref{eqn6-1} as a sum of an unknown constant and a stationary random process.
Second, they replaced $f_{t-1}({\bf x})$ in \cref{eqn6-2} with the best linear unbiased predictor of Kriging, denoted by $\mu_{\text{f},t-1}({\bf x})$~\cite{Sacks1989}.
Finally, they used $z_t({\cdot}) \sim \mathcal{GP}\left(0,\kappa_{\text{z},t}(\cdot,\cdot|{\boldsymbol \phi}_{\text{z},t})\right)$ instead of $\delta_{t}(\cdot) \sim \mathcal{GP}\left(m_{\delta,t}(\cdot),\kappa_{\delta,t}(\cdot,\cdot|{\boldsymbol \phi}_{\delta,t})\right)$, $t=1,\dots,T$.
The \textit{hierarchical Kriging model} reads
\begin{subequations}\label{eqn13}
	\begin{align}
		f_1({\bf x}) & = a + z_1({\bf x}) 
		\label{eqn13-1},\\
		f_t({\bf x}) & =
		b_{t-1} \mu_{\text{f},t-1}({\bf x}) + z_{t}({\bf x}), \ \  t=2,\dots,T
		\label{eqn13-2},
	\end{align}
\end{subequations}
where the unknown constant $a$ in \cref{eqn13-1} is given in \cref{eqn2}.
The GP prior for the hierarchical Kriging model can be derived from the model in \cref{eqn9} for ${\bf R}={\bf I}_T$ and $\boldsymbol{\mu}({\bf x}) = [a,b_1 \mu_{\text{f},1}({\bf x}),\cdots,b_{T-1} \mu_{\text{f},T-1}({\bf x})]^\intercal$.

\subsubsection{Recursive model}\label{Sec423}

\citet{Gratiet2014} introduced the recursive model by making two modifications to the KOH model.
First, they adopted the hybrid additive/multiplicative approach by replacing the regression coefficient $b_{t-1}$ in \cref{eqn6-2} with an adjustment coefficient function $\rho_{t-1}({\bf x})$, which is similar to $\rho({\bf x})$ in \cref{eqn2}.
Second, rather than using the GP prior $f_{t-1}({\bf x})$ in \cref{eqn6-2}, they used the GP posterior $\hat{f}_{t-1}({\bf x})$ constructed from the data $\mathcal{D}_{t-1}$ of fidelity $(t-1)$.
Accordingly, the \textit{recursive model} reads
\begin{subequations}\label{eqn14}
	\begin{align}
		& f_1({\bf x}) = \delta_1({\bf x}), \label{eqn14-1}\\
		& f_t({\bf x}) =
		\rho_{t-1}({\bf x}) \hat{f}_{t-1}({\bf x}) + \delta_{t}({\bf x}), \ \ t=2,\dots,T \label{eqn14-2},\\
		&\text{cov}\left[f_{t}({\bf x}),\hat{f}_{t-1}({\bf x}')|\hat{f}_{t-1}({\bf x})\right]=0 
		\label{eqn14-3}.
	\end{align}
\end{subequations}
Here $\rho_{t-1}({\bf x}) = {\boldsymbol \zeta}_{t-1}^\intercal({\bf x}) {\boldsymbol \beta}_{t-1}$, as shown in \cref{Fig-4}, is the linear combination of a finite number of basis functions, and
\begin{subequations}\label{eqn15}
	\begin{align}
		\delta_{t}(\cdot) & \sim \mathcal{GP}\left(m_{\delta,t}(\cdot),\kappa_{\delta,t}(\cdot,\cdot|{\boldsymbol \phi}_{\delta,t})\right)
		\label{eqn15-1},\\
		\hat{f}_{t-1}(\cdot) & \sim \mathcal{GP}\left(m_{\text{f},t-1}(\cdot),\kappa_{\text{f},t-1}(\cdot,\cdot)\right) 
		\label{eqn15-2},
	\end{align}
\end{subequations}
where $m_{\text{f},t-1}(\cdot)$ and $\kappa_{\text{f},t-1}(\cdot,\cdot)$ are posterior mean and posterior covariance functions, respectively.

An advantage of the recursive model is that the computational cost it requires for training the GP model that approximates the highest-fidelity output $f_T({\bf x})$ is much less than that the KOH model requires, while the predictive efficiency is preserved.
Specifically, the computational costs for training the recursive and KOH models are $\mathcal{O}\left(T \times \max\{N_t^3,t=1,\dots,T\}\right)$ and $\mathcal{O}(\sum_{t=1}^T N_t)^3$ respectively, where $N_t$ is the sample size of data $\mathcal{D}_t$~\cite{Gratiet2014}. 

\subsection{Graphical multi-fidelity Gaussian process}\label{Sec43}

\begin{figure}
	\centering
	\includegraphics[width=0.5\textwidth]{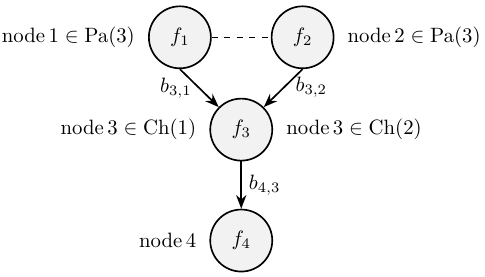}
	\caption{An example of a directed acyclic graph for MF modeling (bottom-up descends fidelity order).}
	\label{Fig-6}
\end{figure}

In an attempt to generalize the KOH model, \citet{Ji2023} developed the \textit{graphical multi-fidelity Gaussian process} (GMGP) for cases when the hierarchy of LF models remains unclear.
The GMGP approach involves constructing a directed acyclic graph (DAG) where each node represents a computational model, and each directed edge connecting two nodes represents the hierarchy of their fidelities.
If the graph is in a topological ordering such that every directed edge $(t',t)$ from $t'$ to $t$, then $t'<t$ and the prediction of node $t'$ is made before making the prediction of node $t$.  
\Cref{Fig-6} shows an example of a topological-ordering DAG of four nodes.

Consider node $t$ and let $t' \in \text{Pa}(t)$ be a parent node of $t$.
Thus, $t \in \text{Ch}(t')$ is a child node of $t'$.
Let $V_\text{s}$ denote a set of all source nodes that have no lower-fidelity node, and $\bar{V}_\text{s}$ denote a set of non-source nodes.
The DAG in \Cref{Fig-6}, for example, has $V_\text{s}=\{1,2\}$ and $\bar{V}_\text{s}=\{3,4\}$.

GMGP models the surrogate for the output at node $t$ as a weighted sum of the surrogates predicting the outputs of its parent nodes, plus a discrepancy term $\delta_t(\cdot)\sim \mathcal{GP}\left(m_{\delta,t}(\cdot),\kappa_{\delta,t}(\cdot,\cdot|{\boldsymbol \phi}_{\delta,t})\right)$.
This can be generalized using the following form for the predictions at a total of $T$ nodes:
\begin{subequations}\label{eqn16}
	\begin{align}
		&{\bf f}({\bf x}) =  {\bf R} {\boldsymbol \mu}({\bf x}) + {\bf R} {\bf z}({\bf x}) \label{eqn16-1},\\
		&\text{cov}\left[f_{t}({\bf x}),f_{t'}({\bf x}')|f_{t'}({\bf x})\right]=0 \label{eqn16-2}.
\end{align}	
\end{subequations}
Here ${\bf f}({\bf x}) \in \mathbb{R}^T$ and ${\boldsymbol \mu}({\bf x}) \in \mathbb{R}^T$ are described in Eqs.~(\ref{eqn12-1}) and (\ref{eqn12-4}), respectively.
 ${\bf z}(\cdot) \sim \mathcal{GP}\left({\bf 0},{\boldsymbol \Sigma}(\cdot,\cdot)\right)$ is a $T$-variate GP with ${\boldsymbol \Sigma}(\cdot,\cdot) = \diag\{\kappa_{\delta,1}(\cdot,\cdot|{\boldsymbol \phi}_{\delta,1}),\dots,\kappa_{\delta,T}(\cdot,\cdot|{\boldsymbol \phi}_{\delta,T})\}$.
The condition in \cref{eqn16-2} holds for $t' \in  \text{Pa}(t)$ or for $t,t' \in  V_\text{s}$, with $t \neq t'$.
The coefficient matrix ${\bf R}$ in \cref{eqn16-1} is defined as
\begin{equation}\label{eqn17}
		{\bf R} = \prod_{t=T-1}^1 \prod_{j \in \text{Ch}(t)} {\bf R}_{jt}, \ \ {\bf R}_{jt} = {\bf I}_T + b_{j,t} {\bf e}_j {\bf e}^\intercal_{t},
\end{equation}
where the correlation coefficient $b_{j,t}$ relates nodes $j$ and $t$ as indicated in~\cref{Fig-6}.
We see that, while the GMGP model is a type of the LMC model described in~\Cref{Sec41}, it is more general than the KOH model in \cref{eqn12} as the coefficient matrix ${\bf R}$ in \cref{eqn17} becomes that in \cref{eqn12-3} if the fidelities of all nodes are ordered.
		
Inspired by~\citet{Gratiet2014}, \citet{Ji2023} also provided the following recursive formulation for the GMGP:
\begin{subequations}\label{eqn18}
	\begin{align}
		f_t({\bf x}) = \sum_{t' \in \text{Pa}(t)}
		b_{t,t'} \hat{f}_{t'}({\bf x}) + \delta_{t}({\bf x}), \ \  t \in \bar{V}_\text{s}
		\label{eqn18-1},\\
		\text{cov}\left[f_{t}({\bf x}),\hat{f}_{t'}({\bf x}')|\hat{f}_{t'}({\bf x})\right]=0, \ \ t' \in  \text{Pa}(t)
		\label{eqn18-2},
	\end{align}
\end{subequations}
where $\hat{f}_{t'}({\bf x})$ is the GP posterior for predicting the output of node $t'$, which is conditioned on the data associated with node $t'$ and its ancestor nodes.

While the computational cost for training the GMGP in \cref{eqn16} is $\mathcal{O}(\sum_{t=1}^T N_t)^3$, where $N_t$ is the sample size of data associated with node $t$, that for training the model in \cref{eqn18} is $\mathcal{O}\left(T \times \max\{N_t^3,t=1,\dots,T\}\right)$.
Inference in GMGP and its applications to several numerical experiments and to emulation of heavy-ion collisions are detailed in~\citet{Ji2023}.

\subsection{Bayesian hierarchical model}\label{Sec44}

Given two computational models with different fidelity for predicting an objective function, and the corresponding predictions $f_1({\bf x})= f_\text{L}({\bf x})$ and $f_2({\bf x})= f_\text{H}({\bf x})$,
\citet{Qian2008} proposed a Bayesian hierarchical model for approximating the prediction of $f_\text{2}({\bf x})$.
Using the KOH model, the \textit{Bayesian hierarchical model} adds a measurement error to the hybrid additive/multiplicative form in \cref{eqn2} for $w=1$ and $c \neq 0$, such that
\begin{subequations}\label{eqn19}
	\begin{align}
		f_1({\bf x}) & = \delta_1({\bf x}) \label{eqn19-1},\\
		f_\text{2}({\bf x}) & = \rho({\bf x})f_\text{1}({\bf x}) + \delta_2({\bf x})+\varepsilon_2({\bf x}) \label{eqn19-2}.
	\end{align}
\end{subequations}
Here the adjustment coefficient function $\rho({\bf x})$ is a GP characterized by a scalar mean $m_\rho$, variance parameter $\sigma^2_\rho$, and correlation function $k_\rho(\cdot,\cdot|{\boldsymbol \phi}_\rho)$.
The discrepancy function $\delta_i({\bf x})$ $(i=1,2)$ is a GP characterized by a mean function $m_{\delta,i}({\bf x})$, variance parameter $\sigma^2_{\delta,i}$, and correlation function $k_{\delta,i}(\cdot,\cdot|{\boldsymbol \phi}_{\delta,i})$.
The measurement error $\varepsilon_\text{2}(\cdot) \sim \mathcal{N}(0,\sigma^2_{\varepsilon,2})$ is a Gaussian with zero mean and variance $\sigma^2_{\varepsilon,2}$.
Thus, hyperparameters underlying the MF surrogate predicting $f_\text{2}({\bf x})$ can be encapsulated in a mean parameter vector ${\boldsymbol \phi}_1$, a variance parameter vector ${\boldsymbol \phi}_2$, and
a correlation parameter vector ${\boldsymbol \phi}_3$~\cite{Qian2008}. 
The parameter of each vector is then assigned to a pre-specified prior distribution to enable a full Bayesian treatment.

It is computationally expensive to predict $f_2$ at an unseen input vector using the full Bayesian treatment because it requires the posterior samples of ${\boldsymbol \phi}_1$, ${\boldsymbol \phi}_2$, and ${\boldsymbol \phi}_3$, while sampling the posterior of ${\boldsymbol \phi}_3$ is nontrivial as the form of its conditional distribution is irregular.
To address this, \citet{Qian2008} used the posterior samples of ${\boldsymbol \phi}_1$ and ${\boldsymbol \phi}_2$ for the prediction while fixing ${\boldsymbol \phi}_3$ at its mode value.
Unfortunately, finding the mode of ${\boldsymbol \phi}_3$ is still elaborate because it involves integration to estimate the associated unnormalized posterior, which in turn requires the use of sample average approximation for an approximate solution~\cite{Verweij2003}.
Although it is one of the important GP-based MF surrogates, the Bayesian hierarchical model may not be a good choice for use of MF BO due to its computational complexity.

\begin{figure}
	\centering
	\includegraphics[width=0.425\textwidth]{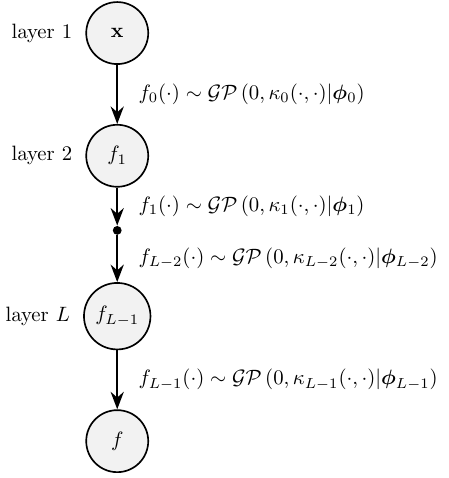}
	\caption{General architecture of DGP.}
	\label{Fig-7}
\end{figure}

\subsection{Composition of Gaussian processes}\label{Sec45}

\subsubsection{Deep Gaussian processes}\label{Sec451}
\textit{A deep Gaussian process} (DGP) describes the mapping between the input variables and the output using a composition of GPs~\cite{Damianou2013}.
\Cref{Fig-7} shows a general architecture of DGP with $L$ hidden layers in which a GP relates two consecutive layers.
The architecture in~\cref{Fig-7} is mathematically described as follows:
\begin{equation}\label{eqn20}
	f({\bf x}) = f_{L-1}\left(\dots f_1(f_0({\bf x}))\right)+\varepsilon_{L-1},
\end{equation}
where $\varepsilon_{L-1}$ is often additive zero-mean Gaussian noise corresponding to layer $L$, and
\begin{equation}\label{eqn21}
	f_{l-1}(\cdot) \sim \mathcal{GP}\left(0,\kappa_{l-1}(\cdot,\cdot|{\boldsymbol \phi}_{l-1})\right), \ \ l=1,\dots,L,
\end{equation}
where $\kappa_{l-1}(\cdot,\cdot|{\boldsymbol \phi}_{l-1})$ denotes the covariance function parameterized by hyperparameters ${\boldsymbol \phi}_{l-1}$.

Unlike the standard GP, a DGP is capable of handling structured data that encapsulates hierarchical features, enabling it to accurately assess the similarity between pairs of data points properly.
Unfortunately, learning the hyperparameters underlying DGPs is rather complicated. 

\subsubsection{Nonlinear auto-regressive model}\label{Sec452}

Based on the KOH model and an output-output mapping technique, \citet{Perdikaris2017} proposed the \textit{nonlinear auto-regressive model} for learning complex nonlinear correlations between models of variable fidelity.
Let $h_{t}(\cdot):\mathbb{R}^{d+1} \mapsto \mathbb{R}$ be a mapping that relates the vector of input and output of fidelity $(t-1)$ to the output of fidelity $t$.
Let ${\bf z}_{t-1}({\bf x})=\left[{\bf x}^\intercal,\hat{f}_{t-1}({\bf x})\right]^\intercal$ be the input vector of the surrogate predicting the output of fidelity $t$, where $\hat{f}_{t-1}({\bf x})$ is the GP posterior corresponding to fidelity $(t-1)$.
The nonlinear auto-regressive model reads~\cite{Perdikaris2017}
\begin{subequations}\label{eqn22}
	\begin{align}
		&f_1({\bf x})=\delta_1({\bf x}) \label{eqn22-1},\\
		&f_{t}({\bf x})=h_{t}\left({{\bf z}_{t-1}({\bf x})}\right), \ \  t=2,\dots,T \label{eqn22-2},\\
		&\text{cov}\left[f_{t}({\bf x}),{\bf z}_{t-1}({\bf x}')|{\bf z}_{t-1}({\bf x})\right]=0 
		\label{eqn22-3},
	\end{align}
\end{subequations}
where $\delta_1(\cdot) \sim \mathcal{GP}\left(0, \kappa_{\delta,1}(\cdot,\cdot|{\boldsymbol \phi}_{\delta,1})\right)$ and the condition in \cref{eqn22-3} is to describe the Markov property in \cref{eqn6-3}.

In a probabilistic framework, the mapping $h_{t}(\cdot)$ in \cref{eqn22-2} is modeled as a GP with zero mean and covariance function $\kappa_{\text{h},t}(\cdot,\cdot|{\boldsymbol \phi}_{\text{h},t})$, such that
\begin{equation}\label{eqn23}
	h_{t}(\cdot) \sim \mathcal{GP}\left(0, \kappa_{\text{h},t}(\cdot,\cdot|{\boldsymbol \phi}_{\text{h},t})\right).
\end{equation} 
Here the covariance function $\kappa_{\text{h},t}(\cdot,\cdot|{\boldsymbol \phi}_{\text{h},t})$ treats the input and output variables separately, as
\begin{equation}\label{eqn24}
		\kappa_{\text{h},t}(({\bf x},f),({\bf x'},f')|{\boldsymbol \phi}_{\text{h},t})  
		 =\kappa_{\text{x},t}({\bf x},{\bf x}'|{\boldsymbol \phi}_{\text{x},t})\kappa_{\text{f},t}(f,f'|{\boldsymbol \phi}_{\text{f},t}) + \kappa_{\delta,t}({\bf x},{\bf x}'|{\boldsymbol \phi}_{\delta,t}),
\end{equation}
where ${\boldsymbol \phi}_{\text{h},t}=[{\boldsymbol \phi}_{\text{x},t}^\intercal,{\boldsymbol \phi}_{\text{f},t}^\intercal,{\boldsymbol \phi}_{\delta,t}^\intercal]^\intercal$, and $\kappa_{\text{x},t}$, $\kappa_{\text{f},t}$, $\kappa_{\delta,t}$ are covariance functions corresponding to ${\bf x}$, $f({\bf x})$, and $\delta({\bf x})$ of fidelity $t$, respectively.

The GP surrogate for the model in \cref{eqn22-2} is a DGP when $\delta_1(\bf x)$ and $h_{t}(\bf x)$ are GPs.
However, the number of hyperparameters for this DGP is much smaller than that for the full DPG described in~\Cref{Sec451}~\cite{Perdikaris2017}. 
  
For $t>2$, we cannot analytically marginalize the likelihood function described in \cref{eqn25}.
As a result, the statistical estimates of $f_{t}$ at an unseen input vector can be obtained via the Monte-Carlo integration.

\subsubsection{Deep multi-fidelity Gaussian process}\label{Sec453}

Using the DGP modeling framework, \citet{Cutajar2019} proposed a \textit{multi-fidelity deep Gaussian process} (MF DGP), with each layer of the DGP corresponding to a fidelity level. Specifically, layers $2,\dots,L$ of the DGP in~\cref{Fig-7}
correspond to the surrogate models for fidelities $1,\dots,T-1$, respectively.
This yields the following MF DGP marginal likelihood for the surrogate that approximates the highest-fidelity output:
\begin{equation}\label{eqn25}
	p(f_{T}|{\bf x}) = \int p(f_{T}|f_{T-1})\dots p(f_{1}|{\bf x}) \text{d}f_{1} \dots \text{d}f_{T-1},
\end{equation}
where $p(\cdot)$ denotes the probability density function (PDF).

Since $p(f_{T}|{\bf x})$ is computationally intractable, \citet{Cutajar2019} relied on an approximate inference via what is called doubly stochastic variational inference method, which is a variational inference technique based on sparse GPs~\cite{Salimbeni2017}.

For this variational inference, let  ${\bf Z}_{t-1} \in \mathbb{R}^{N_t \times {(d+t-1)}}$, $t=1,\dots,T$, denote the set of $N_t$ samples of inducing input variables corresponding to the surrogate associated with fidelity $t$, and ${\bf U}_{t} \in \mathbb{R}^{N_t}$ denote the corresponding function values, where $p({\bf U}_{t}) = \mathcal{N}({\boldsymbol \mu}_{\text{u},t},{\boldsymbol \Sigma}_{\text{u},t})$.
Here the inducing input variables and the corresponding function values serve as the representative for the observed data.
The selection and optimization of inducing input variables can be found in~\citet{Titsias2009}.
Note that we have ${\bf Z}_{t-1} \in \mathbb{R}^{N_t \times {(d+t-1)}}$ as the input vector of surrogate for fidelity $t$ consists of the input vectors and the outputs of surrogates for lower fidelities.

In the doubly stochastic variational inference method, the posterior distribution of $\{{\bf U}_t\}_{t=1}^T$ is factorized between
layers. Therefore, the joint posterior $p({\bf F}, {\bf U})$ is simply the product of the joint posteriors at $T$ fidelities~\cite{Salimbeni2017}.
Since we are interested in large datasets with non-Gaussian likelihoods, we wish to find a variational joint posterior at each fidelity.
By further maximizing the lower bound on the marginal likelihood, we obtain an approximate variational joint posterior at fidelity $t$, which reads~\cite{Salimbeni2017}
\begin{equation}\label{eqn26}
	q({\bf F}_t,{\bf U}_t) = p({\bf F}_t|{\bf U}_t) p({\bf U}_t).
\end{equation}

Since the joint GP prior $p({\bf F}_t, {\bf U}_t)$ and the prior $p({\bf U}_t)$ are Gaussians, the conditional $p({\bf F}_t|{\bf U}_t)$ is also a Gaussian.
Thus, both terms of the variational joint posterior in \cref{eqn26} are Gaussians.
This enables marginalizing ${\bf U}_t$ from each fidelity level to obtain an analytical form of the variational marginal $q({\bf F}_t)$, as
\begin{equation}\label{eqn27}
	q({\bf F}_t|{\boldsymbol \mu}_{\text{u},t},{\boldsymbol \Sigma}_{\text{u},t}) = \mathcal{N}({\boldsymbol \mu}_{\text{f},t},{\boldsymbol \Sigma}_{\text{f},t}),
\end{equation}
which is fully coupled within and between layers.

As a result, the marginal associated with the highest fidelity depends only
on the marginals associated with the other fidelities, such that
\begin{equation}\label{eqn28}
	q(f_{T}) = \int \prod_{t=1}^{T} q({\bf F}_t|{\boldsymbol \mu}_{\text{f},t},{\boldsymbol \Sigma}_{\text{f},t}) \text{d}f_{1} \dots \text{d}f_{T-1},
\end{equation}
which has no analytical form, but is computationally tractable via a sampling method.

Some applications of DGP and MF DGP to engineering design optimization can be found in~\citet{Hebbal2021oe,Hebbal2021smo}.
Nevertheless, it is worth noting that the use of DGPs or MF DGPs as surrogates for performing BO may compromise the beauty of BO because of the intricate nature of inference in DGPs, let alone non-analytical predictions from posterior DGPs.

\subsection{Input-augmentation multi-fidelity Gaussian processes}\label{Sec46}

The GP-based MF surrogates we have seen so far are built in the space of input variables, or an augmented space of input and output variables.
An alternative approach, as briefly described in \Cref{Sec22}, constructs GP-based MF surrogates as functions of both input and fidelity variables.
We call this approach input-augmentation GP-based MF modeling, and which has been used in BO predominantly. 
Once an input-augmentation MF surrogate has been trained, the highest-fidelity output at any set of input variables can be predicted by setting the fidelity-level variables at their highest values.

Depending on how the fidelity variables are described, we classify the input-augmentation methods into two groups.
The first group relies on continuous approximations of the fidelity variables (\Cref{Sec461}), while the second group treats them as categorical variables (\Cref{Sec462}).

\subsubsection{Continuous approximations}\label{Sec461}

\citet{Kandasamy2017} proposed the method of \textit{continuous approximations} when solving a problem of finding hyperparameters ${\bf x}$ that maximizes a validation accuracy $f({\bf x})$.
The validation accuracy depends not only on ${\bf x}$ but also on the number of data points $t_1$ and the number of optimization iterations $t_2$.
A pair of $t_1$ and $t_2$, therefore, defines a fidelity level.
Let ${\bf t}=[t_1,t_2]^\intercal$ denote the vector of fidelity variables, where $t_1 \in [1,N_{\max}]$, $t_2 \in [1,I_{\max}]$, and $N_{\max}$ and $I_{\max}$ are threshold values of the number of data points and the number of optimization iterations, respectively.
If we use a function $g({\bf t},{\bf x})$ to define the validation accuracy on in a larger space, then $f({\bf x})$, under continuous approximations, is a slice of $g({\bf t},{\bf x})$ at ${\bf t}$, such that
\begin{equation}\label{eqn29}
	f({\bf x}) = g({\bf t},{\bf x}).
\end{equation}

To construct a GP-based MF surrogate, \citet{Kandasamy2017} further assigned a GP prior to $g({\bf t},{\bf x})$, as
\begin{equation}\label{eqn30}
	g(\cdot) \sim \mathcal{GP}\left(0,\kappa_\text{g}\left(\cdot,\cdot|{\boldsymbol \phi}_\text{g}\right)\right),
\end{equation}
where $\kappa_\text{g}\left(\cdot,\cdot|{\boldsymbol \phi}_\text{g}\right)$ is defined using the following point-wise product:
\begin{equation}\label{eqn31}
	\kappa_\text{g}\left(({\bf t},{\bf x}),({\bf t}',{\bf x}')|{\boldsymbol \phi}_\text{g}\right)=\kappa_\text{t}\left({\bf t},{\bf t}'|{\boldsymbol \phi}_\text{t}\right) \kappa_\text{x}\left({\bf x},{\bf x}'|{\boldsymbol \phi}_\text{x}\right),
\end{equation}
with ${\boldsymbol \phi}_\text{g} = [{\boldsymbol \phi}_\text{t}^\intercal,{\boldsymbol \phi}_\text{x}^\intercal]^\intercal$ representing the hyperparameter vector of $\kappa_\text{g}$. 

\subsubsection{Use of non-continuous covariance functions}\label{Sec462}

The construction of GP-based MF surrogates for categorical fidelity variables using the input-augmentation approach requires non-continuous covariance functions $\kappa_\text{t}\left({\bf t},{\bf t}'|{\boldsymbol \phi}_\text{t}\right)$.
Let ${\bf t}=[t_1,\dots,t_{n_\text{t}}]^\intercal$ denotes the vector of categorical fidelity variables, where each element $t_i$ ($i=1,\dots,n_\text{t}$) has $l_i$ categories, i.e., $t_i \in \{t_{i,1},\dots,t_{i,l_i}\}$.

\citet{Zhou2011} proposed the \textit{hypersphere decomposition} method that defines the non-continuous covariance function $\kappa_\text{t}\left({\bf t},{\bf t}'|{\boldsymbol \phi}_\text{t}\right)$ as a product of the univariate covariance functions associated with individual elements of ${\bf t}$.
The univariate covariance function for each element is formulated by mapping each of its $l_i$ categories onto a point on the surface of $l_i$-dimensional unit hypersphere.
Accordingly, the non-continuous covariance reads
\begin{subequations}\label{eqn32}
	\begin{align}
		\kappa_\text{t}\left({\bf t},{\bf t}'|{\boldsymbol \phi}_\text{t}\right) & = \prod_{i=1}^{n_t}\kappa(t_i,t_i'|{\boldsymbol \phi}_{\text{t},i})
		\label{eqn32-1},\\
		\kappa(t_i,t_i'|{\boldsymbol \phi}_{\text{t},i})& =\sigma_{\text{t},i}^2\varphi(t_i)^\intercal \varphi(t_i')
		\label{eqn32-2},
	\end{align}
\end{subequations}
where $\varphi(t_i)=[z_{i,0},\dots,z_{i,l_i}]^\intercal$ defines the hypersphere mapping, and $\sigma_{\text{t},i}$ represents the cross-correlation between categories of $t_i$ and $t_i'$.
The determination of $[z_{i,0},\dots,z_{i,l_i}]^\intercal$ can be found in~\citet{Zhou2011} and~\citet{Pelamatti2021}.

\citet{Roustant2020} used a so-called \textit{compound symmetry} covariance function for $\kappa(t_i,t_i'|{\boldsymbol \phi}_{\text{t},i})$ in \cref{eqn32-2} by assuming a common correlation for all categories.
Accordingly, $\kappa(t_i,t_i'|{\boldsymbol \phi}_{t,i})$ reads the following parsimonious form: 
\begin{equation}\label{eqn33}
	\kappa(t_i,t_i'|{\boldsymbol \phi}_{\text{t},i}) = 
	\begin{cases}
		\sigma_{\text{t},i}^2, & \text{for } t_i = t_i', \\
		\theta \sigma_{\text{t},i}^2, & \text{for } t_i \neq t_i',
	\end{cases}
\end{equation}
where $ 0 < \theta < 1$.

Alternatively, \citet{Oune2021} mapped the vector ${\bf t}$ of fidelity variables onto a \textit{latent space} of continuous variables ${\bf z}({\bf t})$ using a mapping matrix ${\bf A}$.
This mapping reads
\begin{equation}\label{eqn34}
	{\bf z}({\bf t}) = {\boldsymbol \zeta}({\bf t}) {\bf A},
\end{equation}
where ${\boldsymbol \zeta}({\bf t})$ is the prior vector representation of ${\bf t}$ which is defined using either the random initialization or the one-hot encoding technique, and the elements of ${\bf A}$ are found via maximum likelihood estimation.
Once ${\bf z}({\bf t})$ has been established, the non-continuous covariance function $\kappa_\text{t}\left({\bf t},{\bf t}'|{\boldsymbol \phi}_t\right)$ is defined as
\begin{equation}\label{eqn35}
	\kappa_\text{t}\left({\bf t},{\bf t}'|{\boldsymbol \phi}_\text{t}\right)=\kappa_\text{z}\left({\bf z}({\bf t}),{\bf z}({\bf t}')|{\boldsymbol \phi}_\text{z}\right),
\end{equation}
where $\kappa_\text{z}\left({\bf z}({\bf t}),{\bf z}({\bf t}')|{\boldsymbol \phi}_\text{z}\right)$ is a standard covariance function defined in the continuous space of ${\bf z}$.

To further select an appropriate non-continuous covariance function for a specific problem, \citet{Pelamatti2021} tested and compared the modeling performances of the hypersphere decomposition, compound symmetry, and latent mapping covariance functions.
The main characteristics of these covariance functions were also compared to point out their advantages and disadvantages.

Instead of relying on a point-wise product to define $\kappa_\text{g}$ as in \cref{eqn31}, \citet{Poloczek2017} used a sum of covariance functions to define $\kappa_\text{g}$ when maximizing an objective function $f({\bf x}) = g(t,{\bf x}) - \delta_t({\bf x})$.
Here $t \in \mathcal{T} = \{t_1,\dots,t_T\}$ is a categorical variable indicating the noisy information source, and therefore can be seen as a fidelity variable.
$g(t,{\bf x})$ is the noisy value of $f({\bf x})$ when observing $t$ at ${\bf x}$, thus $g(t_T,{\bf x}) = f({\bf x})$ is the noise-free observation.
$\delta_t({\bf x})$ is the discrepancy function defined as a GP independent of $f$ such that $\delta_t(\cdot) \sim \mathcal{GP}(m_{\delta,t}(\cdot),\kappa_{\delta,t}\left(\cdot,\cdot|{\boldsymbol \phi}_{\delta,t}\right))$.
The objective function is $f(\cdot) \sim \mathcal{GP}(m_\text{f}(\cdot),\kappa_\text{x}\left(\cdot,\cdot|{\boldsymbol \phi}_\text{x}\right))$.
Thus, $\kappa_\text{g}\left((t,{\bf x}),(t',{\bf x}')|{\boldsymbol \phi}_\text{g}\right)$ can be derived from $g(t,{\bf x}) = f({\bf x}) + \delta_t({\bf x})$, as
\begin{equation}\label{eqn36}
	\kappa_\text{g}\left((t,{\bf x}),(t',{\bf x}')|{\boldsymbol \phi}_\text{g}\right) = \text{cov}[g(t,{\bf x}),g(t',{\bf x}')]
		=\kappa_\text{x}\left({\bf x},{\bf x}'|{\boldsymbol \phi}_\text{x}\right) +  \mathbbm{1}_{t,t'} \kappa_{\delta,t}\left({\bf x},{\bf x}'|{\boldsymbol \phi}_{\delta,t}\right),
\end{equation}
where $\mathbbm{1}_{t,t'}$ with $t,t' \in \mathcal{T}$ denotes the Kronecker delta, and $\kappa_\text{x}$ and $\kappa_{\delta,t}$ are standard parameterized covariance functions.

\section{Acquisition functions}\label{Sec5}

An acquisition function formulated in each iteration of BO maps one of our preferences for the next design point to each point in the design variable space. 
Our preferences, under incomplete knowledge about the objective function, often include an improvement in the objective function value, a gain in information on the true minimizer, and a gain in information on the true minimum.
Maximizing the acquisition function, therefore, guides BO toward a new design point at which our preference achieves the highest score.
In this section, we first review notable acquisition functions of the generic BO (\Cref{Sec51}).
We then describe several ways of modifying these acquisition functions for use of MF BO (\Cref{Sec52}), followed by a brief discussion on the portfolio of different design points when maximizing a set of multiple acquisition functions (\Cref{Sec53}).
We finally discuss optimization algorithms used to maximize the acquisition functions (\Cref{Sec54}) and recommend open-source software for BO implementation (\Cref{Sec55}).

\subsection{Acquisition functions for Bayesian optimization}\label{Sec51}

Two classes of acquisition functions of the generic BO include one-step look-ahead (myopic) and multi-step look-ahead (or non-myopic).
\textit{One-step look-ahead acquisition functions}, which are predominantly used in BO literature, select a new design point using a utility measure without forecasting the potential impact of all future selections beyond the immediate next selection.
They can be further classified into improvement-based (\Cref{Sec511}), optimistic (\Cref{Sec512}), and  information-based (\Cref{Sec513}) acquisition functions.
In contrast, \textit{multi-step look-ahead acquisition functions} (\Cref{Sec514}) consider the impact of future selections on the decision of next design points by adding a new term to one-step look-ahead acquisition functions.
This aims to mitigate a limitation of one-step look-ahead acquisition functions that they are deemed to prefer exploitation over exploration~\cite{Hennig2022}.

Assume that we are at iteration $k$ of BO and wish to select a new design point ${\bf x}^{k}$ by maximizing an acquisition function $\alpha({\bf x})$.
What we currently know to formulate $\alpha({\bf x})$, for example, in Line 9 of \Cref{Algo1} are the GP posterior $\hat{f}_\text{H}^k({\bf x})$ constructed from the available (HF) data $\mathcal{D}^{k-1}$ and, under noise-free observations, the best solution $\{{\bf x}_{\min}, f_{\min}\}$ we found in $\mathcal{D}^{k-1}$.

\subsubsection{Improvement-based acquisition functions}\label{Sec511}

Improvement-based acquisition functions arise from a thought experiment in which we expect the objective function at the new design point to be better (i.e., smaller) than the best-observed objective function value $f_\text{min}$.
To describe this, we define the following solution improvement measure at iteration $k$ of BO:
\begin{equation}\label{eqn37}
	I({\bf x}) = \max\{f_\text{min}-f({\bf x}),0\}.
\end{equation}

\textit{Probability of improvement} (PI)~\cite{Kushner1964}, one of the earliest improvement-based acquisition functions, measures the probability that $I({\bf x})$ is greater than a non-negative target value $\tau$.
This is equivalent to the chance of having a solution improvement.
Conditioning it on the current GP posterior $\hat{f}_\text{H}^k({\bf x})$, PI can be written in the following analytical form:
\begin{equation}\label{eqn38}
	\alpha({\bf x}) = \mathbb{P}\left[I({\bf x})>\tau|\hat{f}_\text{H}^k({\bf x})\right]=\Phi\left(\frac{f_\text{min}-\mu_\text{f}^k({\bf x})-\tau}{\sigma_\text{f}^k({\bf x})}\right),
\end{equation}
where $\tau \geq 0$ is the improvement target, $\Phi(\cdot)$ is the standard normal cumulative distribution function (CDF), and $\mu_\text{f}^k({\bf x})$ and $\sigma_\text{f}^k({\bf x})$ are the predictive mean and standard deviation of $\hat{f}_\text{H}^k({\bf x})$ given in Eqs.~(\ref{eqnA6}) and (\ref{eqnA7}), respectively.

While maximizing PI tends to reduce the solution over time, it does not necessarily result in a substantial improvement in the objective function.
This is attributed to the fact that PI is not a direct quantitative measure of an improvement in the objective function value~\cite{Kochenderfer2019}.
Additionally, we should select a value of the improvement target carefully to obtain a desired solution improvement because we favor exploitation for a small $\tau$ or exploration for a large $\tau$.
A data-driven approach to choosing $\tau$ values can be found in~\citet{Jones2001}.

\textit{Expected improvement} (EI)~\cite{Mockus1975,Jones1998} is an acquisition function that measures the solution improvement quantitatively.
As its name suggests, EI calculates the expected value of $I({\bf x})$, given the current GP posterior.
Mathematically, EI reads
\begin{equation}\label{eqn39}
		\alpha({\bf x})  = \mathbb{E}\left[I({\bf x})|\hat{f}_\text{H}^k({\bf x})\right]
		=\left( f_\text{min}-\mu_\text{f}^k({\bf x})\right)\Phi\left(\frac{f_\text{min}-\mu_\text{f}^k({\bf x})}{\sigma_\text{f}^k({\bf x})}\right)
	+\sigma_\text{f}^k({\bf x})\phi\left(\frac{f_\text{min}-\mu_\text{f}^k({\bf x})}{\sigma_\text{f}^k({\bf x})}\right),
\end{equation}
where $\phi(\cdot)$ denotes the standard normal PDF.
This analytical form is derived using integration by parts~\cite{Jones1998,Kochenderfer2019}.

EI provides a way to conceptualize the balance between exploitation and exploration in optimization in the face of uncertainty.
The first term of EI embodies exploitation, guiding the search toward a new design point with a high probability of improvement.
The second term focuses on exploration, directing the search to regions where there is considerable uncertainty in the prediction of the objective function.
However, EI does not allow direct control over the exploitation-exploration balance when necessary.
For example, it is preferable to bias exploitation if the objective function tends to be unimodal.
Conversely, exploration can work well if the objective function is extremely multimodal~\cite{Sobester2005}.

To gain control over the exploitation-exploration balance, \citet{Sobester2005} introduced the \textit{weighted expected improvement} (WEI) as a weighted sum of the two terms of EI, such that
\begin{equation}\label{eqn40}
    \alpha({\bf x})  = w\left( f_\text{min}-\mu_\text{f}^k({\bf x})\right)\Phi\left(\frac{f_\text{min}-\mu_\text{f}^k({\bf x})}{\sigma_\text{f}^k({\bf x})}\right)
    +(1-w)\sigma_\text{f}^k({\bf x})\phi\left(\frac{f_\text{min}-\mu_\text{f}^k({\bf x})}{\sigma_\text{f}^k({\bf x})}\right),
\end{equation}
where the weighting factor $w \in [0,1]$.

If the new design point is selected by maximizing WEI in some set of candidate solutions, then this only results in selecting ${\bf x}^k$ in the maximal non-dominated set that simultaneously optimizes the predictive mean and standard deviation for a relatively small range of $w$, regardless the fact that $w$ can vary over $[0,1]$~\cite{Ath2021}.
By carefully examining this phenomenon, \citet{Ath2021} recommended restricting the values of $w$ to the interval $[0.185,0.5]$, which guarantees the selection of a new design point on the Pareto frontier that trades off exploration and exploitation.

\textit{Knowledge gradient} (KG)~\cite{Frazier2008} is another improvement-based acquisition function that is closely related to EI.
By introducing a departure from one of the key assumptions underlying EI that the observations are noise-free, KG operates independently of the best-observed objective function value.
This is justified by considering that the best solution might possess some level of uncertainty.

Let $\mu_\text{f}^{k+1}({\cdot})$ be the unknown predictive posterior mean obtained from the data of iteration $k$, i.e., $\mathcal{D}^{k} = \mathcal{D}^{k-1} \cup \{{\bf x}^{k},f_\text{H}({\bf x}^{k})\}$.
Let ${\bf x}_{\star}^k$ denote the minimizer of the predictive posterior mean $\mu_\text{f}^k({\cdot})$, which allows searching the current best solution over the design variable space (i.e., global search) rather than the available data points (i.e., local search).
The KG acquisition function for finding ${\bf x}^k$ is defined as~\cite{Frazier2008}
\begin{equation}\label{eqn41}
	\alpha({\bf x}) = \mathbb{E}\left[\mu_\text{f}^k({\bf x}_{\star}^k)-\mu_\text{f}^{k+1}({\bf x})\right],
\end{equation}
which requires integration over all possible points $\{{\bf x}^k,f_\text{H}({\bf x}^k)\}$ for a given ${\bf x}^k$ under the posterior predictive PDF of $\hat{f}_\text{H}^k({\bf x})$.
The simplest way to estimate KG at a specific value ${\bf x}$ is via the Monte-Carlo integration.

Rather than balancing the exploitation and exploration, KG balances the so-called influence of alternative ${\bf x}$ and variance, where the benefit of variance is the same as that of exploration~\cite{Frazier2008}.

\subsubsection{Optimistic acquisition function}\label{Sec512}
In multi-armed bandit problems where we have to select between multiple choices (arms), each with an unknown PDF of selection rewards, to maximize the total reward over a series of trials,
the solution to the exploitation-exploration trade-off is almost intractable.
Here exploitation, based on the arms selected so far, chooses the arm that currently seems to offer the highest reward, while exploration focuses on selecting different arms to gather information about the reward distribution.
An efficient, simple solution approach is to use confidence bounds to handle the exploitation-exploration trade-off~\cite{Lai1985,Auer2002}.
This approach is underpinned by the principle of optimism in the face of uncertainty,
which takes greedy actions based on optimistic estimates of their rewards~\cite{Shahriari2016}.

In the context of GP-based optimization, \citet{Srinivas2010} proposed the \textit{GP upper confidence bound} (GP-UCB) to maximize the sum of rewards while handling the exploitation-exploration trade-off in optimization.
Based on GP-UCB, we can define the following \textit{GP negative lower confidence bound} (GP-LCB) as an optimistic acquisition function for minimization problems:
\begin{equation}\label{eqn42}
	\alpha({\bf x}) = -\left[\mu_\text{f}^k({\bf x})-\sqrt{\beta^k} \sigma_\text{f}^k({\bf x})\right],
\end{equation}
where $\sqrt{\beta^k} \geq 0$ is the tuning parameter to control the exploitation-exploration trade-off.

The convergence proof by~\citet{Srinivas2010} is based on the scheduled values of $\sqrt{\beta^k}$.
In particular, $\sqrt{\beta^k}$ increases as logarithm of the number of past evaluations of the objective function.
This means, the search biases toward exploration after each iteration of BO.

\subsubsection{Information-based acquisition functions}\label{Sec513}

Information-based acquisition functions focus on gaining the solution information from the posterior PDF of unknown minimizer ${\bf x}^\star$ or the posterior PDF of unknown minimum $f({\bf x}^\star)$.
Two popular policies in this class include Thompson sampling and entropy search.

\textit{Thompson sampling} (TS), also known as posterior sampling or randomized probability matching, selects the next action in a multi-armed bandit problem by maximizing the reward with respect to a reward function randomly drawn from the associated posterior~\cite[see e.g.,][]{Chapelle2011,Agrawal2012,Bijl2016}.
More specifically, TS randomly selects each action in proportion to the posterior PDF of the optimal action~\cite{Scott2010}.

In the context of BO, instead of maximizing an explicit acquisition function, TS selects the new design point using a randomized strategy that samples a function $q_\text{H}^k({\bf x})$ from the current GP posterior $\hat{f}_\text{H}^k({\bf x})$ and then finds the minimizer of this function.
This two-stage implementation is indeed the random generation of the next design point from the posterior PDF of unknown minimizer ${\bf x}^\star$. What leads to the posterior PDF of ${\bf x}^\star$ is our lack of knowledge about the objective function, which is modeled by a GP.

Sampling $q_\text{H}^k({\bf x})$ from GP posteriors for use of optimization is nontrivial. To sample from a stationary GP posterior, an approach is via GP spectral sampling that approximates the GP prior using a Bayesian generalized linear model of the weighted sum of random features~\cite{Rahimi2007,Hensman2018}. 
Here the random features are determined based on the Fourier duality between the stationary covariance function and a spectral measure.
Then, a sample from the conditional Bayesian generalized linear model corresponding to a sample of its weight vector is considered a sample of the GP posterior $\hat{f}_\text{H}^k({\bf x})$.
In another approach to breaking the curse of dimensionality of the random Fourier features, a GP posterior sample is updated from the corresponding GP prior sample using the information from the data~\cite{Wilson2020}. Here the GP prior is approximated by a Bayesian generalized linear model of standard Gaussian weights and random Fourier features.
For a non-stationary GP, its posterior samples can be generated in a similar way in which the features can be determined from the spectral representation of the covariance function per Mercer's theorem \cite{Rasmussen2006}. 

\begin{algorithm}[t]
	\caption{Sequential Thompson sampling.}\label{Algo2}
	\begin{algorithmic}[1]
		\State \textbf{Input:} $f_\text{H}(\cdot)$, $\mathcal{X}$, $K$, $N$;
		
		\State Generate $N$ samples of ${\bf x}$;
		
		\For {$i=1:N$} 
		\State $f^i \gets f_\text{H}({\bf x}^i)$; \textcolor{black}{\Comment{Costly step}}
		\EndFor
		
		\State $\mathcal{D}^0 \gets \{{\bf x}^i,f^i\}_{i=1}^N$;
		
		\State $\{{\bf x}_{\min},f_{\min}\} \gets \min\{f^i,\, i=1,\dots, N\}$;
		
		\For {$k=1:K$} 
		\State Construct $\hat{f}_\text{H}^k({\bf x})$ from $\mathcal{D}^{k-1}$;
		\State Sample $q_\text{H}^k({\bf x})$ from $\hat{f}_\text{H}^k({\bf x})$ \label{Algo2:10}; 
		\State ${\bf x}^{k} \gets \underset{{\bf x}}{\mathrm{argmin}} \ \ q_\text{H}^k({\bf x})$ s.t. ${\bf x} \in \mathcal{X}$;
		\State $f^{k} \gets f_\text{H}({\bf x}^{k})$;
		\textcolor{black}{\Comment{Costly step}}
		\State $\mathcal{D}^k\gets\mathcal{D}^{k-1} \cup \{{\bf x}^{k},f^{k}\}$;
		\State $\{{\bf x}_{\min},f_{\min}\} \gets \min\{f_{\min},f^{k}\}$;
		\EndFor
		
		\State \Return $\{{\bf x}_{\min},f_{\min}\}$.
	\end{algorithmic}
\end{algorithm}

A pseudo-code for TS is given in~\Cref{Algo2}. We see that if the posterior $\hat{f}_\text{H}^k({\bf x})$ is highly uncertain, TS tends to sample many different posterior functions $q_\text{H}^k({\bf x})$ in Line 10, which turns out that TS favors exploration.
As the uncertainty reduces, the algorithm starts exploiting the knowledge about the true objective function.
Thus, TS handles the exploitation-exploration trade-off in a natural way. 
In fact, the exploitation ability of TS is inferior because of its inherent randomness~\cite{Scott2010}.

Theoretical results allow for establishing the connection between TS and GP-UCB via the translation from regret bounds developed for GP-UCB into Bayesian regret bounds developed for TS~\cite{Russo2014}.
It is also possible to sample multiple functions $q_\text{H}^k({\bf x})$ (Line 10 of \Cref{Algo2}) in parallel to select a batch of new design points~\cite{Kandasamy2018}.
Nevertheless, TS may address the following four problem features inadequately: problems that do not require exploration, problems that do not require exploitation, problems that are time-sensitive, and problems that require careful assessment of information gain~\cite{Russo2018}.

Information-based acquisition functions with entropy search (ES) policy select the next design points to gain information in the true minimizer ${\bf x}^\star$ or in the true minimum $f_\text{H}({\bf x}^\star)$.
The former includes information approach optimization algorithm, entropy search, and predictive entropy search.
The latter includes maximum-value entropy search.

\textit{Information approach optimization algorithm} (IAGO)~\cite{Villemonteix2009} and \textit{entropy search} (ES)~\cite{Hennig2012} define the acquisition function based on the following location-information loss:
\begin{equation}\label{eqn43}
	\lambda(\mathcal{D}^{k}) = \mathbb{H}({\bf x}^\star|\mathcal{D}^{k}).
\end{equation}
Here $\mathcal{D}^{k}$ denotes the training data obtained after adding $\{{\bf x}^k,f_\text{H}({\bf x}^k)\}$ to $\mathcal{D}^{k-1}$.
$\mathbb{H}(\cdot)$ is the entropy of the random variable inside the parentheses, measuring the spread (uncertainty) of its PDF.
The lower the entropy, the higher the information gain.
Thus, if a new design point ${\bf x}^k$ is selected by minimizing $\lambda(\mathcal{D}^{k})$, then it minimizes the entropy of ${\bf x}^\star$ given the data in the next iteration. 

Let ${\bf U}$ denote a vector of continuous random variables distributed according to $p({\bf u})$, and ${\bf V}$ denote a vector of discrete random variables, which take values in a discrete set $\mathcal{V}$.
The mathematical descriptions of the information entropy values of ${\bf U}$ and ${\bf V}$ are 
\begin{subequations}\label{eqn44}
	\begin{align}
		\mathbb{H}({\bf U})& =-\int p({\bf u}) \log{p({\bf u})} \text{d}{\bf u}
		\label{eqn44-1},\\
		\mathbb{H}({\bf V}) & = -\sum_{{\bf v} \in \mathcal{V}}p({\bf V}={\bf v}) \log{p({\bf V}={\bf v})}
		\label{eqn44-2},
	\end{align}
\end{subequations}
where ${\bf u}$ and ${\bf v}$ are realizations of ${\bf U}$ and ${\bf V}$, respectively.

Since $\lambda(\mathcal{D}^{k})$ depends on the unknown design point ${\bf x}^k$, we define IAGO and ES using the expected value of $\lambda(\mathcal{D}^{k})$, such that
\begin{equation}\label{eqn45}
		\alpha({\bf x}) = -\mathbb{E}_{f_\text{H}}\left[\lambda(\mathcal{D}^{k})\right]
		 = -\int \mathbb{H}({\bf x}^\star|\mathcal{D}^{k})  p(f_\text{H}|{\bf x},\mathcal{D}^{k-1}) \text{d}f_\text{H}
		 = -\mathbb{E}_{f_\text{H}}\left[\mathbb{H}({\bf x}^\star|{\bf x},f_\text{H},\mathcal{D}^{k-1})\right],
\end{equation}
where $``-"$ reformulates the minimization of entropy to the maximization of the corresponding acquisition function, $p(f_\text{H}|{\bf x},\mathcal{D}^{k-1})$ denotes the PDF of $f_\text{H}$ conditioned on $\bf x$ and $\mathcal{D}^{k-1}$, and $\mathbb{E}_{f_\text{H}}\left[\mathbb{H}({\bf x}^\star|{\bf x},f_\text{H},\mathcal{D}^{k-1})\right]$ is the expected entropy of ${\bf x}^\star$ given that the new design point and the corresponding value of the objective function has been added to $\mathcal{D}^{k-1}$.

Since computing the entropy for a continuous ${\bf x}^\star$ is generally analytically intractable, discretizing \cref{eqn45} makes the computation tractable.
While IAGO and ES share the same acquisition function form, they differ in their ways of discretizing \cref{eqn45}.

\textit{Predictive entropy search} (PES)~\cite{HernandezLobato2014} is another information-based acquisition function derived from $\lambda(\mathcal{D}^{k})$.
PES arises from an important observation that the maximizer of $-\mathbb{E}_{f_\text{H}}\left[\mathbb{H}({\bf x}^\star|{\bf x},f_\text{H},\mathcal{D}^{k-1})\right]$ is identical to that of $\mathbb{H}({\bf x}^\star|\mathcal{D}^{k-1})-\mathbb{E}_{f_\text{H}}\left[\mathbb{H}({\bf x}^\star|{\bf x},f_\text{H},\mathcal{D}^{k-1})\right]$.
This is due to the fact that $\mathbb{H}({\bf x}^\star|\mathcal{D}^{k-1})$ does not depend on ${\bf x}^k$.
By further leveraging the mutual information $\mathcal{I}({\bf x}^\star;f_\text{H})$ between ${\bf x}^\star$ and $f_\text{H}$, we obtain the following relation~\cite{Houlsby2012}:
\begin{equation}\label{eqn46}
		 \mathcal{I}({\bf x}^\star;f_\text{H})
		= \mathbb{H}({\bf x}^\star|\mathcal{D}^{k-1})-\mathbb{E}_{f_\text{H}}\left[\mathbb{H}({\bf x}^\star|{\bf x},f_\text{H},\mathcal{D}^{k-1})\right]
		 = \mathbb{H}(f_\text{H}|{\bf x},\mathcal{D}^{k-1})-\mathbb{E}_{{\bf x}^\star}\left[\mathbb{H}(f_\text{H}|{\bf x}^\star,{\bf x},\mathcal{D}^{k-1})\right],
\end{equation}
where $\mathbb{E}_{{\bf x}^\star}\left[\mathbb{H}(f_\text{H}|{\bf x}^\star,{\bf x},\mathcal{D}^{k-1})\right]$ is the expected entropy of $f_\text{H}$ given ${\bf x}$, ${\bf x}^\star$, and $\mathcal{D}^{k-1}$.
Based on \cref{eqn46}, PES makes use of the following acquisition function:
\begin{equation}\label{eqn47}
		\alpha({\bf x}) 
		= \mathbb{H}(f_\text{H}|{\bf x},\mathcal{D}^{k-1})-\mathbb{E}_{{\bf x}^\star}\left[\mathbb{H}(f_\text{H}|{\bf x}^\star,{\bf x},\mathcal{D}^{k-1})\right].
\end{equation}

The calculation of PES is much easier than that of ES because the first and second terms of $\alpha({\bf x})$ in \cref{eqn47} only require evaluating the entropy of univariate Gaussians.
Nevertheless, the calculation of the second term is still intricate as it necessitates marginalization over the multivariate posterior $p({\bf x}^\star|\mathcal{D}^{k-1})$ which can be approximated by a finite set of its samples generated by TS.

Rather than using the entropy of unknown minimizer ${\bf x}^\star$, \textit{maximum-value entropy search} (MES)~\cite{Hoffman2015,WangZ2017} formulates its acquisition function using the entropy of unknown minimum $f_\text{H}({\bf x}^\star)$.
In fact, MES modifies PES in \cref{eqn47} by replacing the minimizer ${\bf x}^\star$ of the second term with the minimum $f^\star = f_\text{H}({\bf x}^\star)$.
Accordingly, MES reads
\begin{equation}\label{eqn48}
	\alpha({\bf x})
	= \mathbb{H}(f_\text{H}|{\bf x},\mathcal{D}^{k-1})-\mathbb{E}_{f^\star }\left[\mathbb{H}(f_\text{H}|f^\star ,{\bf x},\mathcal{D}^{k-1})\right].
\end{equation}
MES has advantages
in implementation over PES as its second term only requires marginalization over the univariate posterior $p(f^\star|\mathcal{D}^{k-1})$.

\subsubsection{Multi-step look-ahead acquisition functions}\label{Sec514}

Although they are simple and computationally efficient, one-step look-ahead acquisition functions tend to favor exploitation~\cite{Hennig2022}.
To address this, multi-step look-ahead acquisition functions consider the influence of future selections on the decision of next design points~\cite[see e.g.,][]{Streltsov1999,Ginsbourger2010}.
However, it still remains challenging to handle an exact multi-step look-ahead problem because it requires marginalizing uncertain objective function value and design point location in each step~\cite{Gonzalez2016}.
What we can expect is to use either
two-step look-ahead acquisition functions via the Monte-Carlo integration~\cite{WuJ2019} or multi-step look-ahead acquisition functions via approximation techniques, such as rollout~\cite{Lam2016,Lee2020}, GLASSES~\cite{Gonzalez2016}, and multi-step~\cite{Jiang2020}.
In the following, we briefly describe some multi-step look-ahead acquisition functions.
For the detailed implementation, the reader is encouraged to refer to the corresponding references.

\begin{figure*}[t]
	\centering
	\includegraphics[width=0.75\textwidth]{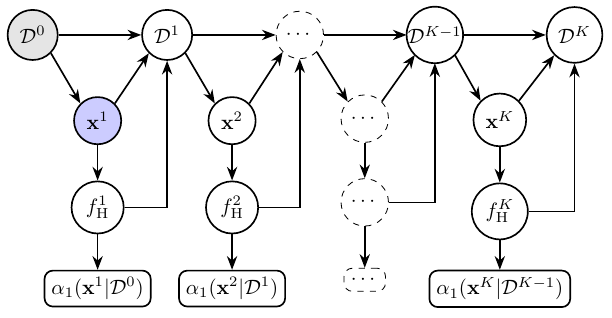}
	\caption{Illustration of a $K$-step look-ahead problem. Node $\mathcal{D}^0$ represents the data we observed so far, and node ${\bf x}^1$ is the design point we wish to find. Finding ${\bf x}^1$ is influenced by future selections of ${\bf x}^2,\dots,{\bf x}^K$.}
	\label{Fig-8}
\end{figure*}

\Cref{Fig-8} describes BO as a multi-stage decision problem for finding ${\bf x}^1,{\bf x}^2,\dots,{\bf x}^{K}$.
Our objective is to select ${\bf x}^1$ based on the information available in the observed data $\mathcal{D}^0$.
This selection is influenced by future selections of ${\bf x}^2,\dots,{\bf x}^{K}$ as we consider the overall outcome of the whole BO process.

For any data $\mathcal{D}^k$, $k=0,\dots,K$, we define
\begin{equation}\label{eqn49}
	u(\mathcal{D}^k) = \underset{({\bf x},f_{\text{H}}) \in \mathcal{D}^k}{\min}\,f_{\text{H}}({\bf x}),
\end{equation} 
which returns the best solution found among the data points of $\mathcal{D}^k$.
Based on $u(\cdot)$, we can define the one-step look-ahead acquisition function $\alpha_1({\bf x}^1|\mathcal{D}^{0})$ for selecting ${\bf x}^1$, which reads
\begin{equation}\label{eqn50}
	\alpha_1({\bf x}^1|\mathcal{D}^0)
	=\mathbb{E}_{f^1_\text{H}}\left[\max\left(u(\mathcal{D}^0) - u(\mathcal{D}^1),0\right)
	|{\bf x}^1,\mathcal{D}^0\right],
\end{equation} 
where $f^1_\text{H} = f_\text{H}({\bf x}^1)$ and $\alpha_1({\bf x}^1|\mathcal{D}^{0})$ is EI given in \cref{eqn39}.

The Bellman's principle of optimality allows the computation of a \textit{$K$-step look-ahead acquisition function} recursively, such that~\cite{Bellman1952}
\begin{equation}\label{eqn51}
	\alpha_K({\bf x}^1|\mathcal{D}^0)
	=\alpha_1({\bf x}^1|\mathcal{D}^0) + \mathbb{E}_{f^1_\text{H}}\left[\underset{{\bf x}^2}{\max}\,\alpha_{K-1}({\bf x}^2|\mathcal{D}^1)\right],
\end{equation}
where $\alpha_K(\cdot)$ and $\alpha_{K-1}(\cdot)$ represent the $K$- and $(K-1)$-step look-ahead acquisition functions, respectively.
The Bellman equation in \cref{eqn51} indicates that selecting a series of the new design points optimally is to first select ${\bf x}^1$ optimally, then select the remaining design points optimally given the objective value from the first selection.
Moreover, since $\alpha_1({\bf x}^1|\mathcal{D}^0)$ tends to favor exploitation, we can view the first and second terms of \cref{eqn51} as exploitation and exploration terms, respectively.

\citet{Jiang2020} used the following $K$-step look-ahead acquisition function for finding ${\bf x}^1$:
\begin{equation}\label{eqn52}
    \alpha_K({\bf x}^1|\mathcal{D}^0) = \alpha_1({\bf x}^1|\mathcal{D}^0) +\mathbb{E}_{f_\text{H}^1}\Bigg[\underset{{\bf x}^2}{\max}\,\Big(\alpha_1({{\bf x}}^2|\mathcal{D}^1) + 
    \mathbb{E}_{f_\text{H}^2}\Big[\underset{{\bf x}^3}{\max}\,\alpha_1({\bf x}^3|\mathcal{D}^2)+\dots  \Bigg],
\end{equation}
where $f^k_\text{H} = f_\text{H}({\bf x}^k)$, $k=1,\dots,K$.

\citet{WuJ2019} proposed \textit{two-step look-ahead acquisition function} $\alpha_2({\bf x}^1|\mathcal{D}^0)$ that, via the Monte-Carlo integration, can be computationally efficient.
Accordingly, $\alpha_2({\bf x}^1|\mathcal{D}^0)$ reads
\begin{equation}\label{eqn53}
		\alpha_2({\bf x}^1|\mathcal{D}^0) =\alpha_1({\bf x}^1|\mathcal{D}^0) + \mathbb{E}_{f^1_\text{H}}\left[\underset{{\bf x}^2}{\max}\,\alpha_1({\bf x}^2|\mathcal{D}^1)\right].
\end{equation}

\citet{Gonzalez2016} introduced \textit{GLASSES} acquisition function by assuming a joint PDF of the future selections from which a batch of design points can be generated in each iteration.
GLASSES reads
\begin{equation}\label{eqn54}
		\alpha_K({\bf x}^1|\mathcal{D}^0) \approx \alpha_1({\bf x}^1|\mathcal{D}^0) + \mathbb{E}_{f^1_\text{H}}\left[\Lambda_{K-1}(\textbf{X}_\text{x}|\mathcal{D}^1)\right],
\end{equation}
where 
$\textbf{X}_\text{x} \in \mathbb{R}^{(K-1) \times d}$ is a matrix whose rows represent the future design points ${\bf x}^2,\dots,{\bf x}^K$. $\Lambda_{K-1}(\textbf{X}_\text{x}|\mathcal{D}^1)$ is a batch value function for the matrix  $\textbf{X}_\text{x}$ conditioned on the unknown data $\mathcal{D}^1$, such that 
\begin{equation}\label{eqn55}
    \Lambda_{K-1}(\textbf{X}_\text{x}|\mathcal{D}^1)
    =\mathbb{E}_{f_\text{H}^2,\dots,f_\text{H}^K}\Bigg[\max\Big(u(\mathcal{D}^1)-u(\mathcal{D}^K),0  \Big)|\textbf{X}_\text{x},\mathcal{D}^1\Bigg].
\end{equation}

Alternatively, \textit{rollout strategies}~\cite{Lam2016,Lee2020} formulate the selection of new design points as a Markov
decision process.
Thus, the multi-step look-ahead acquisition function is the expected total reward of this Markov
decision process.
Then, the maximum of such reward can be found by approximate dynamic programming~\cite{Powell2011}.

\subsection{Acquisition functions considering fidelities}\label{Sec52}

In MF BO, we replace $\hat{f}_\text{H}^k({\bf x})$ in Line 9 of \Cref{Algo1} with one of the MF surrogates described in~\Cref{Sec4}.
This raises a further question of how to incorporate the information about model fidelities, i.e., $t \in \{1,\dots,T\}$, into the acquisition function so that we can select a new design point and an appropriate computational model for estimating the corresponding objective function value. 
 
The use of different MF surrogates and/or different acquisition functions of the generic BO results in a variety of MF acquisition functions. Nevertheless, we can categorize the MF acquisition functions into the following three groups of approaches: 
\begin{itemize}
	\item No-fidelity consideration (\Cref{Sec521}).
	
	\item Heuristic approach (\Cref{Sec522}).
	
	\item Sequential selection (\Cref{Sec523}).
	 
\end{itemize}

\subsubsection{No-fidelity consideration}\label{Sec521}

The acquisition functions for this approach only depend on design variables $\bf x$.
Once the new design point ${\bf x}^k$ has been found, it is fed to $T$ computational models associated with $T$ fidelities for predictions of the corresponding objective function values.
These predictions are used for updating the current solution and MF surrogate.
\Cref{Algo3} shows a pseudo-code for MF BO without considering fidelities.
$u(\mathcal{D}^k)$, $k=0,\dots,K$, in Lines~3 and 10 is defined in \cref{eqn49}.
The MF acquisition function in Line 6 can be any acquisition function of the generic BO.
For example, \citet{Forrester2007} used EI with $f_{\min}$ selected from the highest fidelity, i.e., $\mathcal{D}^k$ in Lines~\ref{Algo3:3} and \ref{Algo3:10} was fixed at $\mathcal{D}^K$.
\citet{Perdikaris2016} formulated EI, but with $f_{\min}$ defined as the best objective value among those from all fidelities.
 
\begin{algorithm}[t]
	\caption{MF BO, no-fidelity consideration.}\label{Algo3}
	\begin{algorithmic}[1]
		\State \textbf{Input:} $f_t(\cdot)$, $\mathcal{X}$, $K$, $\mathcal{D}_t$ $(t=1,\dots,T)$;
		\State $\mathcal{D}^0 \gets \mathcal{D}_1 \cup \dots \cup \mathcal{D}_T$;
		\State $\{{\bf x}_{\min},f_{\min}\} \gets u(\mathcal{D}^0)$; \label{Algo3:3}
		
		\For {$k=1:K$} 
		\State Construct MF surrogate $\hat{f}_\text{H}^k({\bf x})$ from $\mathcal{D}^{k-1}$;
		\State Formulate MF acquisition function $\alpha({\bf x})$; \label{Algo3:6}
		\State ${\bf x}^{k} \gets \underset{{\bf x}}{\mathrm{argmax}} \ \ \alpha({\bf x})$ s.t. ${\bf x} \in \mathcal{X}$;
		\State $\mathcal{D}_t \gets \mathcal{D}_t \cup \{{\bf x}^{k},f_t({\bf x}^{k})\}, \ \ t=1,\dots,T$;
		\State $\mathcal{D}^k \gets \mathcal{D}_1 \cup \dots \cup \mathcal{D}_T$;
		\State $\{{\bf x}_{\min},f_{\min}\} \gets u(\mathcal{D}^k)$; \label{Algo3:10}
		\EndFor
		
		\State \Return $\{{\bf x}_{\min},f_{\min}\}$.
	\end{algorithmic}
\end{algorithm}

\subsubsection{Heuristic approach}\label{Sec522}

The acquisition functions for this approach depend on both $\bf x$ and $t$.
They are often derived by using auxiliary functions to modify one of the one-step look-ahead acquisition functions of the generic BO.
These auxiliary functions consider the computational cost of each fidelity and/or how the selection of each fidelity for handling the new design point affects the accuracy improvement of MF surrogate.
\Cref{Algo4} shows a pseudo-code for MF BO using the heuristic approach.
In Lines~\ref{Algo4:7}--\ref{Algo4:10}, a new design point ${\bf x}^k_t$ is found for each enumerated value of $t$, then the pair $\{{\bf x}^k_t,t\}$ that provides the best acquisition function value ${\alpha}^k_t=\alpha({\bf x}^k_t,t)$ is selected for the next iteration.
Due to its simplicity, the heuristic approach has been widely used in engineering design optimization~\cite[see e.g.,][]{Huang2006smo,Poloczek2017,Ghoreishi2019,Fiore2021,Grassi2023}.

\begin{algorithm}[t]
	\caption{MF BO using heuristic approach.}\label{Algo4}
	\begin{algorithmic}[1]
		\State \textbf{Input:} $f_t(\cdot)$, $\mathcal{X}$, $K$, $\mathcal{D}_t$ $(t=1,\dots,T)$;
		\State $\mathcal{D}^0 \gets \mathcal{D}_1 \cup \dots \cup \mathcal{D}_T$;
		\State $\{{\bf x}_{\min},f_{\min}\} \gets u(\mathcal{D}^0)$;
		
		\For {$k=1:K$} 
		\State Construct MF surrogate $\hat{f}_\text{H}^k({\bf x})$ from $\mathcal{D}^{k-1}$;
		\State Formulate MF acquisition function $\alpha({\bf x},t)$;
		
		\For {$t=1:T$} \label{Algo4:7}
		\State $\{{\bf x}_t^{k},t,\alpha_t^k\} \gets {\max}\, \alpha({\bf x},t)$ s.t. ${\bf x} \in \mathcal{X}$;
		\EndFor
		
		\State $\{{\bf x}^{k},t\} \gets\max\{\alpha_t^k,\, t=1,\dots, T\}$ \label{Algo4:10}
		\State $\mathcal{D}_t \gets \mathcal{D}_t \cup \{{\bf x}^{k},f_t({\bf x}^{k})\}$;
		\State $\mathcal{D}^k \gets \mathcal{D}_1 \cup \dots \cup \mathcal{D}_T$;
		\State $\{{\bf x}_{\min},f_{\min}\} \gets u(\mathcal{D}^k)$;
		\EndFor
		
		\State \Return $\{{\bf x}_{\min},f_{\min}\}$.
	\end{algorithmic}
\end{algorithm}

Let $\mu^k_{f_\text{H}}(\bf x)$ and $\sigma_{f_\text{H}}^{2,k}(\bf x)$ denote the predictive mean and predictive variance of the MF surrogate $\hat{f}_\text{H}^k(\bf x)$, respectively.
Let $c(t)$, $t=1,\dots,T$, denote the computational cost associated with the model of fidelity $t$.

One of the first heuristic MF acquisition functions by \citet{Huang2006smo} was defined as the product of the so-called augmented EI (AEI)~\cite{Huang2006jgo}, developed for the generic BO with noisy objective functions, and two auxiliary functions $\alpha_1({\bf x},t)$ and $\alpha_2(t)$ for considering the influence of fidelities.
This heuristic MF acquisition function reads
\begin{equation}\label{eqn56}
	\alpha({\bf x},t) = \text{AEI}({\bf x}) \alpha_1({\bf x},t) \alpha_2(t).
\end{equation}
Here $\text{AEI}({\bf x})$ becomes $\text{EI}({\bf x})$ when the objective function is noise-free, and $\alpha_1({\bf x},t)$ is the correlation between the posterior PDF of MF surrogate $t$ and that of MF surrogate $T$.
Since $\alpha_1 = 1$ when $t=T$, $\alpha_1$ tends to promote high fidelities for maximizing $\alpha({\bf x},t)$. 
Meanwhile, $\alpha_2(t) = c(T)/c(t)$ is the ratio between the computational cost per model on the highest fidelity $T$ and that on the fidelity $t$.
In contrast to $\alpha_1$, $\alpha_2$ favors low fidelities for maximizing $\alpha({\bf x},t)$.
It is worth noting that the computational cost $c(t)$ associated with fidelity $t$ is assumed to be independent of the design variables ${\bf x}$.
This may be not realistic, especially for engineering design optimization problems where computational costs due to numerical solvers, which require initial guesses of solutions, often vary across the design variable space. 

Attempts to modify~\cref{eqn56} are problem-dependent.
One of them is to replace the standard deviation constituting $\text{AEI}({\bf x})$, which depends on ${\bf x}$ only, with a new standard deviation that is a function of both ${\bf x}$ and $t$, given that only two fidelities are considered~\cite{ZhangY2018,Ruan2020}.
Several attempts introduce different forms of $\alpha_1({\bf x},t)$.
For example, \citet{Sacher2021} used $\alpha_1({\bf x},t) = \max\left(0,1-\sigma_{f_\text{H}}^{2,k+1}({\bf x}|t)/\sigma_{f_\text{H}}^{2,k}({\bf x})\right)$, where $\sigma_{f_\text{H}}^{2,k+1}({\bf x}|t)$ denotes the predictive variance of the updated MF surrogate, which is formed by adding the new design point generated by fidelity $t$ to the current training dataset.  
Other attempts define the MF acquisition function as the ratio between an acquisition function of the generic BO and the computational cost per model on the fidelity~\cite{Winter2023,Foumani2023}.
Beyond the constraint of most one-step look-ahead heuristic MF acquisition functions, the empirical performance of two-step look-ahead EIs has recently been assessed~\cite{Ghoreishi2019,Fiore2023}.

\subsubsection{Sequential selection}\label{Sec523}

The sequential selection approach consists of two steps: (1) select the new design point ${\bf x}^k$ and (2) select the fidelity $t$ with ${\bf x}^k$ found in step (1).
The main difference between this approach and the heuristic approach is that the selection of ${\bf x}^k$ is independent of that of $t$.
This means, the solution improvement is separated from the consideration of computational cost and the accuracy improvement of the MF surrogate.
\Cref{Algo5} shows the pseudo-code for MF BO using the sequential selection approach.

\begin{algorithm}[t]
	\caption{MF BO using sequential selection.}\label{Algo5}
	\begin{algorithmic}[1]
		\State \textbf{Input:} $f_t(\cdot)$, $\mathcal{X}$, $K$, $\mathcal{D}_t$ $(t=1,\dots,T)$;
		\State $\mathcal{D}^0 \gets \mathcal{D}_1 \cup \dots \cup \mathcal{D}_T$;
		\State $\{{\bf x}_{\min},f_{\min}\} \gets u(\mathcal{D}^0)$;
		
		\For {$k=1:K$} 
		\State Construct MF surrogate $\hat{f}_\text{H}^k({\bf x})$ from $\mathcal{D}^{k-1}$;
		\State Formulate MF acquisition function $\alpha({\bf x})$;
		\State ${\bf x}^{k} \gets \underset{{\bf x}}{\mathrm{argmax}} \ \ \alpha({\bf x})$ s.t. ${\bf x} \in \mathcal{X}$; \label{Algo5:7}
		\State Formulate fidelity-query function $\gamma({\bf x}^{k},t)$; \label{Algo5:8}
		\State $t \gets \underset{t}{\mathrm{argmax}} \ \ \gamma({\bf x}^{k},t)$ s.t. $t \in \{1,\dots,T\}$; \label{Algo5:9}
		\State $\mathcal{D}_t \gets \mathcal{D}_t \cup \{{\bf x}^{k},f_t({\bf x}^{k})\}$; \label{Algo5:10}
		\State $\mathcal{D}^k \gets \mathcal{D}_1 \cup \dots \cup \mathcal{D}_T$;
		\State $\{{\bf x}_{\min},f_{\min}\} \gets u(\mathcal{D}^k)$;
		\EndFor
		
		\State \Return $\{{\bf x}_{\min},f_{\min}\}$.
	\end{algorithmic}
\end{algorithm}

One of the first methods of the sequential approach by \citet{Chen2016} used the so-called pre-posterior analysis to formulate a fidelity-query function $\gamma({\bf x}^k,t)$ (Line 8 of \Cref{Algo5}) after maximizing EI for ${\bf x}^k$ in Line 7.
The pre-posterior analysis was to examine how the standard deviation of the MF surrogate at ${\bf x}^k$ reduces when fictitious simulation data for each fidelity are added to the currently real training data.
Because they used the auto-regressive approach, \citet{Chen2016} could generate the samples constituting the fictitious simulation data for fidelity $t$ at ${\bf x}^k$ from a Gaussian characterized by the predictive mean and predictive variance at ${\bf x}^k$.
In this way, a reduction of the standard deviation associated with fidelity $t$ was defined as $\sigma_{f_\text{H}}^k({\bf x}^k)-\bar{\sigma}({\bf x}^k,t)$, where $\bar{\sigma}({\bf x}^k,t)$ denotes the expected standard deviation from the updated GP posterior after the fictitious simulation data for fidelity $t$ are added to the currently real training data several times.
By further considering the computational cost per model on the fidelity, the fidelity-query function was defined as
\begin{equation}\label{eqn57}
    \gamma({\bf x}^k,t) = \frac{\sigma_{f_\text{H}}^k({\bf x}^k)-\bar{\sigma}({\bf x}^k,t)}{\sigma_{f_\text{H}}^k({\bf x}^k)-\bar{\sigma}({\bf x}^k,T)} \frac{c(T)}{c(t)}.
\end{equation}
We see that, a large value of $\gamma({\bf x}^k,t)$ indicates a large reduction in the prediction uncertainty per unit of computational cost at $({\bf x}^k,t)$.
The computational cost associated with each fidelity is also deterministic and independent of the design variables. 
A similar fidelity-query function can be found in~\citet{Tran2020jcp,Tran2020cise}.

Alternatively, \citet{Kandasamy2017} defined $\gamma({\bf x}^k,t)$ as the negative value of the computational cost per model on the fidelity.
They also imposed two constraints on the maximization of $\gamma({\bf x}^k,t)$.
The first constraint, aiming at promoting exploration, required the posterior variance associated with $({\bf x}^k,t)$ should be larger than a pre-specified threshold value.
The second constraint placed a condition that the maximizer of $\gamma({\bf x}^k,t)$ should be found in a neighborhood of $T$, which can shrink over time.
As a result, the constraints in Line 9 of~\Cref{Algo5} should consist of the aforementioned two constraints and a bound constraint on $t$ because $t$ is considered a continuous variable~\cite{Kandasamy2017}; see~\Cref{Sec461}.

In another attempt, \citet{Meliani2019} formulated $\gamma({\bf x}^k,t)$ based on the idea that the use of LF data favors exploration, and that of HF data favors exploitation.
Thus, multiple fidelities can be used for updating MF surrogates simultaneously.
More specifically, $\gamma({\bf x}^k,t)$ can be defined as the ratio between the total uncertainty reduction when adding the new data generated from models whose fidelities are not greater than $t$ and the total computational cost of these models.
As a result, the models whose fidelities are not greater than $t$ are invoked in Line 10 of~\Cref{Algo5}, which differs from the above-mentioned methods of the sequential approach.
This method is a type of the no-fidelity consideration approach when $t=T$.

\subsection{Portfolio of acquisition functions}\label{Sec53}

No acquisition function works well on all classes of problems because the preferred search strategy may vary during different phases of a sequential optimization process~\cite{Shahriari2016}.
To address this, a promising solution involves employing a \textit{portfolio of multiple acquisition functions}~\cite{Hoffman2014,Shahriari2014}.
The idea is to leverage the interaction between different acquisition functions to safeguard against the potential failure of any single search strategy.
This collaborative interaction is quantified via a unified portfolio metric, which can take the form of a meta-criterion~\cite{Hoffman2014} or an entropy search metric~\cite{Shahriari2014}.
The approach then requires two steps.
First, it finds a collection of new design points by maximizing individual acquisition functions within the portfolio.
Second, from the collection of design points, it selects the actual design point that maximizes the portfolio metric.

\subsection{Maximization of acquisition functions}\label{Sec54}

Because it is difficult to examine the convexity of most acquisition functions, we wish to maximize $\alpha({\bf x})$ or $\alpha({\bf x},t)$ using a global optimization algorithm~\cite{Neumaier2004}.
Nevertheless, robust derivative-based algorithms can be used if an acquisition function has an analytical form.
Acquisition functions via Monte-Carlo integration can also be optimized via derivative-based optimization \cite{Wilson2018}.
When maximizing an information-based acquisition function, it may be useful to discretize the design variable space or to use a sampling method.
\Cref{Table5} lists several optimization algorithms and techniques that are found in the literature to maximize the acquisition functions (and their variants) described in Sections~\ref{Sec51} and \ref{Sec52}.
Other important aspects of maximizing acquisition functions such as optimizing in latent spaces and on combinatorial domains are discussed in Section 9.2 of \citet{Garnett2023}.

\begin{table*}[t]
  \centering
  \begin{threeparttable}
	\caption{Summary of optimization algorithms for maximizing acquisition functions.}
	\begin{tabular}{lll}
      \toprule
      Acquisition & Optimization algorithm & Reference\\
      \midrule
      EI\tnote{*} & Branch-and-bound algorithm & \cite{Jones1998}\\
                  & Nelder-Mead simplex method (NM) & \cite{Huang2006smo,Huang2006jgo}\\
                  & Broyden-Fletcher-Goldfarb-Shanno & \cite{Sobester2005}\\
                  & Limited-memory BFGS (L-BFGS) & \cite{Frazier2018,Bonfiglio2018b}\\
                  & Discretizing design variable space & \cite{Ghoreishi2019,Grassi2023}\\
                  & Genetic algorithm (GA) & \cite{Forrester2007,Chen2016,ZhangY2018,Bailly2019,Do2022}\\
                  & Particle swarm optimization & \cite{Kontogiannis2020b,Ribeiro2023}\\
                  & Evolution w/ covariance matrix adaptation & \cite{Tran2020jcp,Tran2020cise}\\
      GP-UCB\tnote{*} & Direct optimization algorithm & \cite{Kandasamy2016,Kandasamy2017}\\
      PI\tnote{*} & GA & \cite{Ruan2020}\\
      KG & Multi-start stochastic gradient& \cite{WuJ2016}\\
      IAGO & Discretizing design variable space & \cite{Villemonteix2009}\\
      ES & Sampling + L-BFGS & \cite{Hennig2012}\\
      PES & Sampling + (local search)& \cite{HernandezLobato2014}\\
      MES & Sampling + (local search or NM)& \cite{WangZ2017}\\
      \bottomrule
	\end{tabular}
    \begin{tablenotes}
    \item[*] \small{and its variants.}
    \end{tablenotes}
  \label{Table5}
  \end{threeparttable}
\end{table*}

\subsection{Software implementations}\label{Sec55}

There exist dozens of open-source BO libraries and most of them implement one-step look-ahead acquisition functions.
We recommend the Emukit package that provides a fully-featured sublibrary for BO and supports PI, EI, GP-LCB, ES, and MES~\cite{Paleyes2019}.
Other notable packages include BoTorch~\cite{Balandat2020} and SMAC3~\cite{Lindauer2022}.

A few BO libraries implement multi-step look-ahead acquisition functions. 
We recommend a repository for two-step look-ahead EI~\cite{WuJ2019} and a repository for rollout dynamic programming~\cite{Lee2020}.

\section{Illustrative examples}\label{Sec6:examples}
To demonstrate how MF BO can achieve efficient and accurate optimization, we optimize two test functions and the shapes of two standard airfoils using the KOH auto-regressive model combined with EI, PI, or GP-LCB acquisition function (\cref{Sec51}), following the no-fidelity consideration approach (\cref{Sec521}).
Two fidelity levels are considered for the objective function of each of these problems.
The optimization results obtained from MF BO are compared against those from the corresponding generic BO applied to the HF objective function.
Note that comparing the optimization results from all the MF BO methods reviewed in this work is impractical
because the combination of the MF surrogates and acquisition functions described above
results in numerous MF BO algorithms, and there is currently no full-featured MF BO software available for this purpose.
Specific comparisons between subsets of MF BO algorithms in various engineering applications are available in the works listed in \cref{Table4}.

\begin{figure*}[t!]
	\centering
	\includegraphics[width=0.9\textwidth]{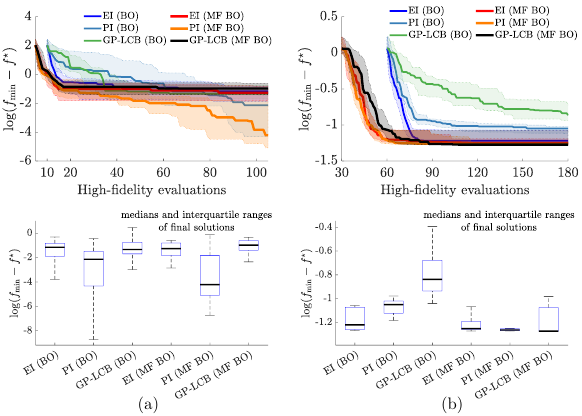}
	\caption{Medians and interquartile solution ranges from 20 trials of each BO and MFBO method for (a) 2d Levy and (b) 6d Hartmann functions.}
	\label{Fig-9}
\end{figure*}

\subsection{Test functions}\label{Sec61:testfunctions}
We first minimize the 2d Levy and 6d Hartmann functions \cite{LiS2020,Surjanovic2013}. Analytical expressions for the HF and LF models and the global minimum of each function are given in Appendix C.

For each problem, we randomly generate 20 initial datasets to initialize each method of MF BO and BO.
Thus, we have a total of 20 trials for each of the considered methods to enable a fair comparison of their performance.
For the 2d Levy function, each initial dataset fed to MF BO and BO has five HF + five LF observations and ten HF observations, respectively.
For the 6d Hartmann function, each initial dataset has 30 HF + 30 LF observations for MF BO, and 60 HF observations for BO.
In each optimization iteration, we record the best-found observation value of the error $\log(f_{\min} - f^\star)$ and the corresponding input variable value, where $f_{\min}$ and $f^\star$ are the best observation of the HF objective function found in each iteration and its true minimum value, respectively.

We use the ooDACE toolbox ~\cite{Couckuyt2014jmlr} to construct GP and KOH auto-regressive models, which are characterized by a zero-mean function and the squared exponential covariance function. To maximize the acquisition function in each iteration (for both MF BO and BO), we implement a multistart derivative-based optimization algorithm with 1000 random starting points. This algorithm is configured with a tolerance of $10^{-12}$ for both the first-order optimality measure and the constraint satisfaction.

\Cref{Fig-9} shows the medians and interquartile solution ranges from 20 trials of each MF BO and each BO method for the 2d Levy and 6d Hartmann functions.
On the 2d Levy function, MF BO with PI achieves the best solutions among the considered methods.
The performances of EI and GP-LCB are comparable for both MF BO and BO.
On the 6d Hartmann function, all optimization trials from MF BO using EI, PI, and GP-LCB quickly converge at the same solution that is better than those from BO. BO with PI and GP-LCB cannot reach the true optimal solution as they terminate.

\begin{figure*}[t]
	\centering
	\includegraphics[width=0.9\textwidth]{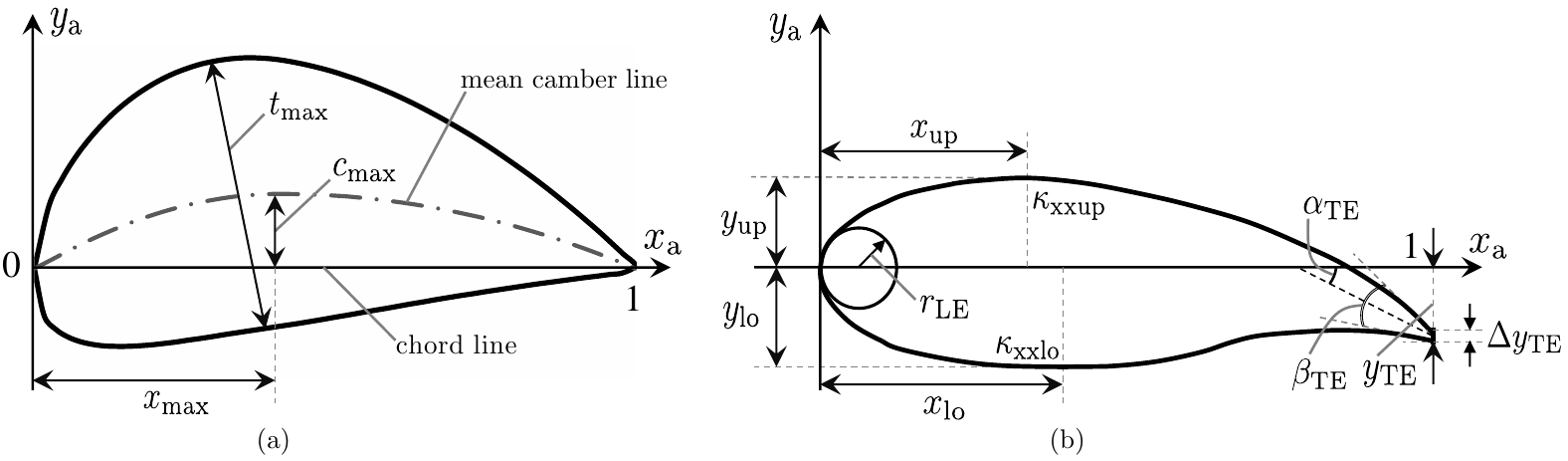}
	\caption{Parameterization of (a) NACA 4-digit and (b) PARSEC airfoils.}
	\label{Fig-10}
\end{figure*}

\begin{table}[t]
  \centering 
    \caption{Descriptions of the design variables for PARSEC airfoils, and upper and lower bounds for each variable.}
    \begin{tabular}{cllcc}
      \toprule
      No. & Parameter & Description & Lower bound & Upper bound \\
      \midrule
        $1$ & $r_\text{LE}$ & Leading-edge radius &  0.005 & 0.06 \\
        $2$ & $x_\text{up}$ & Upper crest position in horizontal coordinates &  0.25 & 0.5 \\
        $3$ & $y_\text{up}$ &  Upper crest position in vertical coordinates &  0.05 & 0.15 \\
        $4$ & $\kappa_\text{xxup}$ & Upper crest curvature  &  -1 & -0.4 \\
        $5$ & $x_\text{lo}$ & Lower crest position in horizontal coordinates &  0.35 & 0.5 \\
        $6$ & $y_\text{lo}$ & Lower crest position in vertical coordinates &  -0.12 & -0.04 \\
        $7$ & $\kappa_\text{xxlo}$ & Lower crest curvature  &  0.3 & 1 \\
        $8$ & $y_\text{TE}$ & Trailing-edge vertical coordinate &  -0.02 & 0.02 \\
        $9$ & $\Delta y_\text{TE}$ & Trailing-edge thickness &  0 & 0 \\
        $10$ & $\alpha_\text{TE}$ & Trailing-edge direction &  -3 & -8 \\
        $11$ & $\beta_\text{TE}$ & Trailing-edge wedge angle  &  4 & 8 \\
      \bottomrule
	\end{tabular}
  \label{Table6}
\end{table}

\begin{table}[t]
  \centering
    \caption{Operating conditions and weightings for calculating the multi-point lift and drag coefficients.}
    \begin{tabular}{ccccc}
      \toprule
      Condition & Mach no. & Angle of attack ($^{\circ}$) & Reynolds no. & Weight value $w_i$\\
      \midrule
        $1$ & $0.5$ & $2$ & $6.3 \times 10^6$ &  0.4 \\
        $2$ & $0.55$ & $2.2$ & $6.3 \times 10^6$ &  0.2 \\
        $3$ & $0.5$ & $2.5$ & $6.3 \times 10^6$ &  0.2 \\
        $4$ & $0.55$ & $2$ & $6.3 \times 10^6$ &  0.2 \\
      \bottomrule
	\end{tabular}
  \label{Table7}
\end{table}

\subsection{Airfoil shapes}\label{Sec62:airfoils}
We now maximize the lift-to-drag ratio for finding optimal shapes of two standard airfoils: NACA 4-digit \cite{Jacobs1933} and PARSEC \cite{Sobieczky1999}. Alternative objective functions for the shape optimization of airfoils can be found in Table~4 of \citet{LiJ2022}.

\Cref{Fig-10} shows the parameterization of NACA 4-digit and PARSEC airfoils. The detailed equations are given in Appendix D. The shape of NACA 4-digit airfoils is parameterized by three parameters, namely $c_{\max}$--the height measured from the chord line to the point where the mean camber line has the largest curvature (in absolute value), $x_{\max}$--the position of $c_{\max}$ in horizontal coordinates, and $t_{\max}$--the maximum thickness of the airfoil.
Additionally, the cosine spacing is used to generate the coordinates distribution of NACA 4-digit airfoil shapes with open trailing edges.
For PARSEC airfoils, the shape is defined by a set of 11 parameters as detailed in \cref{Fig-10}(b) and \cref{Table6} \cite{Jeong2005,Pierluigi2014}.
Specific to this work, we fix the trailing-edge thickness at $\Delta y_\text{TE} = 0$.
Thus, there are ten design variables for the shape optimization of PARSEC airfoils.
Note that all the length-type parameters are expressed in terms of their ratios to the chord length.

We adopt a multi-point design for the airfoil shapes \cite{Toal2023}. 
This design philosophy aims to achieve airfoil shapes that perform efficiently across various operating conditions, rather than at a single design condition. This design philosophy, therefore, can be considered a type of robust optimization (\Cref{Sec731}) under uncertainty in operating conditions.
Specifically, we consider four distinct operating conditions, as detailed in \cref{Table7}, which require the airfoils to operate efficiently at different speeds and angles of attack for a fixed Reynolds number.
As a result, we define the HF objective function for the shape optimization problem as the weighted sum of negative lift-to-drag ratios across the four operating conditions.
The optimization problem for NACA 4-digit and PARSEC airfoils is formulated as follows: 
\begin{equation}\label{eqn58}
	\begin{aligned}
		\min \ \ & f_\text{H}({\bf x})\\
		\textrm{s.t.} \ \ 
		& {\bf x} \in \left[ \underline{{\bf x}},\overline{{\bf x}}\right]. 
	\end{aligned}
\end{equation}
Here ${\bf x} = \left[ c_{\max}, x_{\max}, t_{\max} \right]^\intercal$ and ${\bf x} = \left[ r_\text{LE}, x_\text{up}, y_\text{up}, \kappa_\text{xxup}, x_\text{lo}, y_\text{lo}, \kappa_\text{xxlo},y_\text{TE},\alpha_\text{TE},\beta_\text{TE} \right]^\intercal$ are the vectors of design variables for NACA 4-digit and PARSEC airfoils, respectively. $\underline{{\bf x}}$ and $\overline{{\bf x}}$ are the lower and upper bounds of ${\bf x}$, where $\underline{{\bf x}} = [0,0,0.1]^\intercal$ and $\overline{{\bf x}} = [0.08,0.8,0.25]^\intercal$ for NACA 4-digit airfoils, while those for PARSEC airfoils are given in \cref{Table6}. The HF objective function reads $f_\text{H}({\bf x}) = -\sum_{i=1}^{4} w_i c_{\text{L},i}({\bf x})/c_{\text{D},i}({\bf x})$, where $c_{\text{L},i}({\bf x})$ and $c_{\text{D},i}({\bf x})$ represent the lift and drag coefficients at the $i$-th operating condition, respectively, $w_i$ listed in \cref{Table7} is the weight for the $i$-th operating condition, and the negative sign is to transform a maximization problem of the lift-to-drag ratio into a minimization problem of $f_\text{H}({\bf x})$.

To compute the lift and drag coefficients at each operating condition, we use the 2D panel code XFOIL \cite{Drela1989}.
For inviscid analyses, XFOIL employs potential flow to model the external flow around the airfoil, which simplifies the governing equation for the flow field as the Laplace of the velocity potential that also satisfies the velocity boundary condition at the airfoil surface.
Since we are interested in evaluating the surface velocity and the surface pressure, we only need a flow that satisfies the velocity boundary condition.
It follows that the airfoil surface is discretized into a series of straight panels, each with a linear distribution of vorticity defined by vortex strengths at endpoints. 
By discretizing the velocity boundary condition using the vortex strengths of all panels and imposing the Kutta condition at the trailing edge to ensure smooth flow separation, the vortex strengths at the nodes can be obtained. 
For viscous analyses, XFOIL adds a boundary layer model represented by integral momentum and kinetic energy shape parameter equations based on dissipation closure for both laminar and turbulent flows \cite{Drela1987}.
The $e^9$-type amplification formulation is used for modeling the laminar-to-turbulent flow transition \cite{Drela1987}.
For inviscid/viscous coupling, XFOIL uses a full Newton method to simultaneously solve the boundary layer model equations, transition equations, and the inviscid flow field equations.
In XFOIL, the compressibility effects on the pressure coefficient follow the Karman-Tsien rule, which works well for subsonic conditions \cite{Drela1989}.

We define the LF objective function as the negative lift-to-drag ratio for the first operating condition, i.e., $f_\text{L}({\bf x}) = -c_{\text{L},1}({\bf x})/c_{\text{D},1}({\bf x})$.
Compared to the HF objective function in problem~(\ref{eqn58}), which requires solving fluid flow models for all four operating conditions, the LF objective function is more computationally efficient as it only needs a single fluid flow model to be solved for the first operating condition. It, however, cannot fully capture the expected operating conditions required by the multi-point design.

For optimizing NACA 4-digit airfoils, we generate ten initial datasets for ten different trials of each optimization method. Each dataset has 15 initial HF $+$ 15 initial LF observations for MF BO, and 20 HF observations for BO.
We also generate ten initial datasets for optimizing PARSEC airfoils, each for MF BO having 20 initial HF $+$ 20 initial LF observations, and 30 HF observations for BO.
Note that there is no strict guideline for determining the number of initial HF observations to start BO and MF BO. We can start each method with an empty dataset \cite{Garnett2023}, a dataset of $10d$ HF observations \cite{Jones1998,Loeppky2009}, or a dataset of any size between these two extremes, where $d$ is the number of design variables. Here we initiate MF BO with fewer HF observations than BO as we hypothesize that once both methods reach a sufficiently large number of HF observations, MF BO, leveraging useful information from LF observations at a minor extra cost, would yield a solution that is comparable to (when both methods converge) or better than that of BO, despite requiring fewer initial HF observations. This is justified because we adopt the no-fidelity consideration MF BO approach that calls both HF and LF models after each optimization iteration.
In each optimization iteration, we record the best-found
observation value $f_{\min}$ of the HF objective function and the corresponding design variables.

\begin{figure*}[t!]
	\centering
	\includegraphics[width=0.9\textwidth]{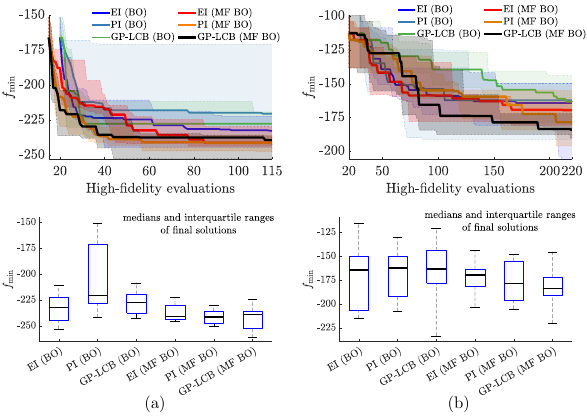}
	\caption{Medians and interquartile solution ranges from ten trials of each BO and MFBO method for (a) NACA 4-digit and (b) PARSEC airfoils.}
	\label{Fig-11}
\end{figure*}

\begin{figure*}[t]
	\centering
	\includegraphics[width=\textwidth]{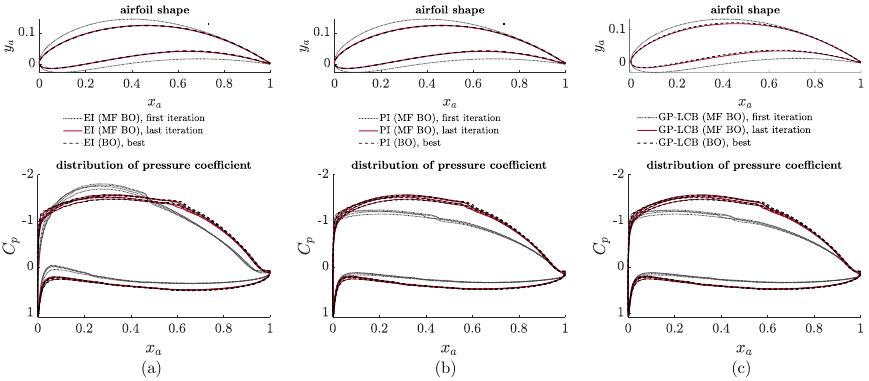}
	\caption{NACA 4-digit: Airfoil shapes and pressure coefficient distributions for four operating conditions from the first and last iterations of the MF BO trial that provides the best solution using (a) EI, (b) PI, and (c) GP-LCB.}
	\label{Fig-12}
\end{figure*}

\begin{figure*}[t!]
	\centering
	\includegraphics[width=\textwidth]{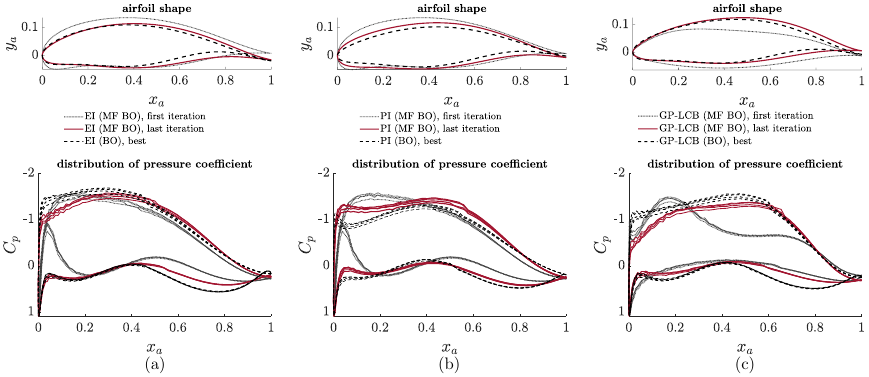}
	\caption{PARSEC: Airfoil shapes and pressure coefficient distributions for four operating conditions from the first and last iterations of the MF BO trial that provides the best solution using (a) EI, (b) PI, and (c) GP-LCB.}
	\label{Fig-13}
\end{figure*}

\Cref{Fig-11} shows the medians and interquartile ranges of solutions from ten trials of each MF BO and each BO method for NACA 4-digit and PARSEC airfoils.
As expected, MF BO outperforms BO in both solution quality and the consistency of the final solutions. 
For NACA 4-digit airfoils, the three considered MF BO methods are comparable. 
For PARSEC 4-digit airfoils, MF BO with GP-LCB yields the best solution.
This variation in the performance of the MF BO methods across the two optimization problems highlights that no acquisition function works well on all classes of problems; see \Cref{Sec53}.

At this point, an important question is how varying the number of initial HF and LF observations affects the optimization performance of the considered MF BO methods.
If we increase the number of HF observations (and thus computational cost) without adjusting the number of LF observations, we have a higher chance to accelerate the convergence of MF BO as a larger portion of the HF objective function space is sampled.
However, the effect of increasing the number of LF observations while keeping the number of HF observations unaltered on the optimization results is likely problem-dependent. Additional LF observations could slow the convergence if there is weak correlation between the HF and LF models near the solution, which misdirects the search. Meanwhile, a strong HF-LF correlation across the design variable space could facilitate faster convergence with useful information from the additional LF observations.

\Cref{Fig-12,Fig-13} illustrate the airfoil shapes and corresponding pressure coefficient distributions, $C_\text{p}$, for the best designs of NACA 4-digit and PARSEC airfoils obtained from each optimization method.
We see that the trial of each MF BO method that provides the best design considerably changes the airfoil shape from an inefficient design to an optimized design with an improved lift characteristic under four operating conditions.
While the best NACA 4-digit airfoil designs by MF BO and BO with the same acquisition function are almost identical, the best PARSEC airfoil designs by these methods exhibit slight differences in their geometries.

\section{Further research topics}\label{Sec7}

Recent advances in MFO and especially in BO have focused on solving intricate yet important design optimization problems, which stem from the nature of problems or from the inherent limitations of the generic BO.  
In this section, we briefly describe sophisticated BO and several MFO approaches to addressing some of these problems, including constrained optimization~(\Cref{Sec71}), high-dimensional optimization~(\Cref{Sec72}), optimization under uncertainty~(\Cref{Sec73}), and multi-objective optimization~(\Cref{Sec74}).
By doing so, we expect to shed light on potential opportunities for future research in MF BO.

\subsection{Constrained optimization}\label{Sec71}

Constrained design optimization is formulated as
\begin{equation}\label{eqn59}
	\begin{aligned}
		\underset{{\bf x} \in \mathcal{X}}{\min} \ \ & f_\text{H}(\bf x)\\
		\textrm{s.t.} \ \ 
		& g_{\text{H},i}({\bf x}) \leq 0,\ \ i=1,\dots,I, 
	\end{aligned}
\end{equation}
where the constraint functions are assumed to be conditionally independent given a vector of design variables and costly to compute, therefore can be estimated via HF evaluations $g_{\text{H},i}({\bf x})$.

There exist three common BO-based approaches to solving problem~(\ref{eqn59}).
The first approach formulates constrained acquisition functions $\alpha_\text{c}({\bf x})$ that consider the influence of constraints on our selection of new design points.
The second approach reformulates problem~(\ref{eqn59}) as an unconstrained problem that can be addressed by the generic BO.
The third approach employs TS, as described in~\Cref{Algo2}, via realizations of GP posteriors (i.e., sample paths) for the objective and constraint functions.    

An attempt of the first approach formulated a \textit{constrained expected improvement} (CEI) acquisition function~~\cite{Schonlau1998} as the product of EI and the so-called probability of feasibility, which is the probability that all inequality constraints are satisfied.
CEI reads
\begin{equation}\label{eqn60}
	\alpha_\text{c}({\bf x}) = \text{EI}({\bf x}) \mathbb{P}[g_{\text{H},1}({\bf x}) \leq 0,\dots,g_{\text{H},I}({\bf x}) \leq 0].
\end{equation}
Here the probability of feasibility $\mathbb{P}[g_{\text{H},1}({\bf x}) \leq 0,\dots,g_{\text{H},I}({\bf x}) \leq 0]$ tightens the search space for selecting the new design point by penalizing unfeasible regions of an approximate problem of problem~(\ref{eqn59}).
This approach, however, may put more weight on the feasible regions that are away from the boundary of feasible space, thereby overlooking the boundary where the optimal solution has a high chance of being found.

When conditioning $g_{\text{H},i}({\bf x})$ on the GP posteriors $\hat{g}_{\text{H},i}({\bf x})$, CEI becomes
\begin{equation}\label{eqn61}
	\alpha_\text{c}({\bf x}) = \text{EI}({\bf x}) \prod_{i=1}^I\Phi\left(\frac{-\mu_{\text{g},i}({\bf x})}{\sigma_{\text{g},i}({\bf x})}\right),
\end{equation}
where $\mu_{\text{g},i}({\bf x})$ and $\sigma_{\text{g},i}({\bf x})$ are the predictive mean and standard deviation of $\hat{g}_{\text{H},i}({\bf x})$, respectively.
The performance of CEI has been verified by~\citet{Gardner2014}, \citet{Sobester2014}, and \citet{Kontogiannis2020b}.

Note that it is not necessary to formulate $\alpha_\text{c}({\bf x})$ if the constraint functions are easy to compute.
In this case, we can simply maximize any acquisition function of the generic BO under a set of the easy-to-check constraints for finding a new design point.

MF BO has recently incorporated the probability of feasibility or its variants into heuristic MF acquisition functions to address several design optimization problems~\cite[see e.g.,][]{Ghoreishi2019,Ruan2020,Ribeiro2023}.
However, decoupling the selection of a new design point and the selection of fidelity for constrained optimization is still an open issue.

\citet{Gramacy2011} introduced the concept of \textit{integrated expected conditional improvement} (IECI) for solving problem~(\ref{eqn59}) when it involves inexpensive constraint functions.
This concept was then extended to solving problems with expensive constraint functions.
IECI reads
\begin{equation}\label{eqn62}
	\alpha_\text{c}({\bf x}) = \int_{\mathcal{X}} \left[\text{EI}({\bf x'}) - \text{EI}({\bf x'}|{\bf x})\right] p({\bf x'}) \text{d}{\bf x'},
\end{equation}
where $\text{EI}({\bf x'})$ is EI at a reference point $\bf x'$ that is distributed according to $p({\bf x'})$, and $\text{EI}({\bf x'}|{\bf x})$ is EI at $\bf x'$ given that $\{{\bf x},f_\text{H}({\bf x})\}$ is added to the current data.

We see that, IECI handles the constraints via $p({\bf x'})$, which can be uniformly distributed over a bounded feasible region ${\bf x'} \in \mathcal{X}$ and zero otherwise.
When the constraint functions are costly, their surrogates are used for the Monte-Carlo simulation to approximate $p({\bf x'})$~\cite{Gramacy2011}.
An interesting property of IECI is that it, via flexible choices of $p({\bf x'})$, allows the selection of infeasible design points at the cost of gaining useful information about the objective function for improving future rewards.

Two considerations should be taken into account if one wishes to use IECI for MF BO: (1) how to select $p({\bf x'})$ when considering the fidelities and (2) how to reason about the selection of an appropriate fidelity if an infeasible design point is found after maximizing $\alpha_\text{c}({\bf x})$.  

\textit{Expected volume minimization} (EVM) by ~\citet{Picheny2014} is another constrained acquisition function for solving problem~(\ref{eqn59}) via BO.
By integrating the product of the probability of improvement and the probability of feasibility, EVM is defined as
\begin{equation}\label{eqn63}          
	\begin{aligned}
		\alpha_\text{c}({\bf x}) & = \int_{\mathcal{X}} \mathbb{P}\left[f_\text{H}({\bf x}) \leq \min\{f_\text{H}({\bf x}),f_{\min}\}\right] 
		\mathbb{P}[g_{\text{H},1}({\bf x}) \leq 0,\dots,g_{\text{H},I}({\bf x}) \leq 0] \text{d}{\bf x}  \\
		& + \int_{\mathcal{X}} \mathbb{P}\left[f_\text{H}({\bf x}) \leq f_{\min}\right]
		\left(1-\mathbb{P}[g_{\text{H},1}({\bf x}) \leq 0,\dots,g_{\text{H},I}({\bf x}) \leq 0]\right) \text{d}{\bf x}.
	\end{aligned} 
\end{equation}
The first and second integrals correspond to the probability of improvement considering the feasibility and unfeasibility of the new design point, respectively.
If the new design point is feasible, the best value of the objective found so far is $\min\{f_\text{H}({\bf x}),f_{\min}\}$.
Otherwise, the best value of the objective found so far is still $f_{\min}$.
Although it is an attractive concept, maximizing EVM is costly because it requires numerical integration over $\mathcal{X}$.

\begin{algorithm}[t]
	\caption{Augmented Lagrangian method.}\label{Algo6}
	\begin{algorithmic}[1]
		\State \textbf{Require:} $f_\text{H}(\cdot)$, $g_{\text{H},i}(\cdot)$ ($i=1,\dots,I$), $\rho^0_0$, $\boldsymbol{\lambda}^0$ , bounded region $\mathcal{X}$, $\eta_1>0$, $\eta_2>0$; \label{Algo6:1}
		
		\For {$k=1,2,\dots$} 
		\State Formulate $\mathcal{L}({\bf x}|\boldsymbol{\lambda}^{k-1},\rho^{k-1}_0)$; \label{Algo6:3}
		\State ${\bf x}^{k} \gets \underset{{\bf x}}{\mathrm{argmin}} \ \ \mathcal{L}({\bf x}|\boldsymbol{\lambda}^{k-1},\rho^{k-1}_0)$ s.t. ${\bf x} \in \mathcal{X}$; \label{Algo6:4}
		\State ${\lambda}^k_i \gets \max \{0,g_{\text{H},i}({\bf x}^k)/\rho^{k-1}_0\}$;
		
		\If{$\Vert \max\{0,g_{\text{H},1}({\bf x}^{k}),\dots,g_{\text{H},I}({\bf x}^{k})\} \Vert  \leq \eta_1$ or $\Vert \nabla f_{\text{H}}({\bf x}^{k})+ \sum_{i=1}^I {\lambda}^k_i \nabla g_{\text{H},i}({\bf x}^{k}) \Vert  \leq \eta_2$} \label{Algo6:6}
			\State ${\bf x}_{\min} \gets {\bf x}^{k}$
		\Else
			\If{${\bf g}_\text{H}({\bf x}^{k}) \leq {\bf 0}$}
				\State $\rho^k_0 \gets \rho^{k-1}_0$;
			\Else
				\State $\rho^k_0 \gets 0.5\rho^{k-1}_0$;
			\EndIf  	
		\EndIf 
		\EndFor
		
		\State \Return ${\bf x}_{\min}$.
	\end{algorithmic}
\end{algorithm}

In an attempt of the reformulation approach, \citet{Gramacy2016} adopted BO to minimize the \textit{augmented Lagrangian} of problem~(\ref{eqn59}), which reads
\begin{equation}\label{eqn64}
		\mathcal{L}({\bf x}|\boldsymbol{\lambda},\rho_0)  = f_\text{H}({\bf x}) + \sum_{i=1}^{I} \lambda_i g_{\text{H},i}({\bf x}) 
		+ \frac{1}{2 \rho_0} \sum_{i=1}^{I} \max\{0,g_{\text{H},i}({\bf x})\}^2, 
\end{equation}
where  $\rho_0>0$ is a penalty parameter and $\boldsymbol{\lambda}=[\lambda_1,\dots,\lambda_I]^\intercal$ a vector of non-negative Lagrange multipliers, with $\lambda_1 \geq 0,\dots,\lambda_I \geq 0$.

The augmented Lagrangian method starts with initial values of $\rho_0$ and $\boldsymbol{\lambda}$.
A sufficiently large initial value of $\rho_0$ is required to enforce feasibility.
It then sequentially finds the minimizer of the augmented Lagrangian corresponding to the given values of $\boldsymbol{\lambda}$ and $\rho_0$, and updates them for the next iteration with the resulting minimizer.
\Cref{Algo6} implements the augmented Lagrangian method (not in the BO setting), where
$\eta_1$ and $\eta_2$ in Line 1 are two small thresholds for checking the termination conditions in Line 6.    

In the context of BO, the idea is to construct in Line 3 of~\Cref{Algo6} a GP for the augmented Lagrangian $\mathcal{L}({\bf x}|\boldsymbol{\lambda}^{k-1},\rho^{k-1}_0)$ using observations of several design variable vectors and the corresponding values of the objective and constraint functions.
Based on this GP, the next design point ${\bf x}^{k}$ in Line 4 is found by maximizing EI, and the new parameters $\rho_0$ and $\boldsymbol{\lambda}$ are updated accordingly.
The advantage of this approach is that it, via BO, can increase the chance of finding a global solution as compared with the traditional augmented Lagrangian method, which is a local optimization algorithm.

The reformulation approach enables a straightforward application of MF BO in solving constrained design optimization problems.
Once the augmented Lagrangian has been formulated, we can use any established unconstrained MF BO algorithms, which are more mature than their constrained counterparts.

The third approach to solving problem~(\ref{eqn59}) is via TS~\cite{Eriksson2021}.
Let $q^k_{\text{H},i}({\bf x})$ and $h^k_{\text{H},i}({\bf x})$ denote the realizations (i.e., sample paths) of $\hat{f}_{\text{H}}({\bf x})$ and $\hat{g}_{\text{H},i}({\bf x})$ at iteration $k$ of TS, respectively.
Assume that we perform Line 10 of \Cref{Algo2} in a batch setting of TS and obtain a total of $q$ new candidates, denoted by ${\bf x}^k_1,\dots, {\bf x}^k_q$.
Let $\mathcal{F}^k = \{{\bf x}^k_l|\,h^k_{\text{H},i}({\bf x}) \leq 0,\forall l \in \{1,\dots q\}, \forall i \in \{1,\dots I\}\}$ be a set of points whose realizations are feasible.
If $\mathcal{F}^k \neq \emptyset$, we select one of its elements that minimizes $q^k_{\text{H},i}({\bf x})$.
Otherwise, we select a candidate with a minimum value of the total violation
$\sum_{i=1}^{I} \max\{0,h_{\text{H},i}({\bf x})\}$.

\subsection{High-dimensional optimization}\label{Sec72}

Based on the size of optimization problems, we can classify them into the following three classes: 
\begin{itemize}
	\item Small-dimensional problems, which have five or fewer design variables and constraints.
	
	\item Moderate-dimensional problems, which have from five to a thousand design variables and constraints.
	
	\item High-dimensional problems, which have thousands or even millions of variables and constraints.
\end{itemize}
Although this classification is not rigid, it can reflect the fundamental differences in solution approach associated with varying problem sizes~\cite{Luenberger2008}.

Extending the BO framework to high-dimensional optimization problems faces two computational challenges.
First, a conventional GP model scales cubically with the number of training data points~\cite{Rasmussen2006}, which often increases as the number of design variables increases.
Second, maximizing high-dimensional acquisition functions is nontrivial because they are mostly flat at high dimension~\cite{Rana2017}.
These two challenges make high-dimensional BO related to, but distinct from scalable GP modeling approaches, which only focus on the first challenge~\cite{LiuH2020}. 

Recent advances in high-dimensional BO have followed three main directions~\cite{Daulton2022MOBO}: 
\begin{itemize}
	\item Exploiting low, active/effective dimensional subspace of the design variables (\Cref{Sec721}).
	
	\item Exploiting the additive structures of the objective and/or constraint functions (\Cref{Sec722}).
	
	\item Trust-region BO (\Cref{Sec723}).
\end{itemize}

\subsubsection{Subspace-based approach}\label{Sec721}

This approach constructs surrogates for the objective/constraint functions in an unknown low-dimensional subspace of the design variables, followed by the implementation of BO in this subspace.
Methods from this approach, e.g., sensitivity analysis, active subspace construction, random linear embedding, and nonlinear embedding, differ in their ways of defining the low-dimensional subspace.

Sensitivity analysis~\cite{Spagnol2019} is a classical method of the subspace-based approach that selects the most influential design variables and ignores the less influential ones.
A drawback of this method is that it cannot capture the models varying most prominently along the directions that are not aligned with the original coordinate system of the design variable space~\cite{Constantine2014}, leading to the quest for active/effective subspace.

An active subspace for GP applications can be determined using the eigenvalue decomposition of a covariance matrix associated with the gradient of the function of interest~\cite{Constantine2014}.
Specifically, the first $d_\text{e}$ eigenvectors are selected to form a reduced-order basis.
This is justified because an active subspace represents the directions of the largest variability of a function.
However, a limitation of this method is that calculating the exact gradient covariance matrix is impossible. This requires the use of the Monte-Carlo integration, which in turn demands a large number of HF data points.  

Random linear embedding~\cite{WangZ2016} is one of the important methods of the subspace-based approach.
It states that for any design variable vector ${\bf x} \in \mathbb{R}^d$ with an unknown active subspace dimension of $d_\text{e}$, there exists, with probability 1, a vector ${\bf y} \in \mathbb{R}^{d_\text{s}}$ ($d_\text{s} \geq d_\text{e}$) such that $f_\text{H}({\bf x}) = f_\text{H}({\bf Ay})$, where ${\bf A} \in \mathbb{R}^{d \times d_\text{s}}$ is a random projection matrix whose independent elements are sampled from $\mathcal{N}(0,1)$.
This enables performing BO in the low-dimensional space of ${\bf y}$ once ${\bf A}$ and the domain of ${\bf y}$ have been determined.
The choice of covariance function to construct GPs in the space of ${\bf y}$ is another important aspect that affects the performance of the random linear embedding. 
Theoretically, any GP-based BO algorithm runs on the embedded low-dimensional space as it would if it was run on an unknown active subspace~\cite{Nayebi2019}.
\citet{Letham2020} further listed several crucial issues and misconceptions about the use of random linear embedding for high-dimensional BO.

Apart from random linear embedding methods, nonlinear embedding techniques have been used to explore more effective low-dimensional spaces.
For example, \citet{Snoek2015} showed excellent performance on tasks of hyperparameter tuning by defining the output layer of a deep neural network as a set of nonlinear basis functions for Bayesian linear regression.
\citet{GomezBombarelli2018} learned a low-dimensional space using variational autoencoders.
\citet{Moriconi2020} performed high-dimensional BO via a nonlinear feature mapping for dimensionality reduction and a reconstruction mapping for evaluating the objective function. 
A drawback of the nonlinear embedding techniques is that they require a large number of HF data points for embedding learning.

To adopt the subspace-based approach for MF BO, in addition to formulating an MF surrogate and an MF acquisition function in an approximate low-dimensional subspace, addressing the following questions is of interest: 
\begin{itemize}
	\item Is it possible to facilitate learning the approximate low-dimensional subspace using the information from multiple fidelities?
	
	\item How can we select a new design point and a fidelity level to gain the information used for updating the approximate subspace as much as possible after each iteration?
\end{itemize}

\subsubsection{Additive structure approach}\label{Sec722}

This approach makes a strong assumption that the objective function can be written in an additive form, such that~\cite{Kandasamy2015}
\begin{equation}\label{eqn65}
	f_\text{H}({\bf x}) = \sum_{m=1}^{M}f_m({\bf x}_m), 
\end{equation}
where ${\bf x}_m \in \mathcal{X}_m$ are disjoint lower dimensional components of a high-dimensional vector of design variables.

Using the assumption in \cref{eqn65}, we can construct component GP models for $f_m({\bf x}_m)$ and maximize the acquisition functions formulated from these models to progress BO, regardless of our knowledge on the structure of $f_m({\bf x}_m)$.
If ${\bf x}^k_m$ is the new point from maximizing the acquisition function $m$, then the new design point is ${\bf x}^k = \bigcup_{m=1}^M {\bf x}^k_m$.
Alternatively, ${\bf x}^k$ can be found by maximizing an acquisition function derived from an overall GP for $f_\text{H}({\bf x})$ whose mean (or covariance) function is assigned as the sum of mean (or covariance) functions of the component GPs. 

Unfortunately, it is difficult to reason about a good, unknown decomposition structure, especially when the objective function is non-additive. 
Recent works have attempted to find possible model decompositions via Markov chain Monte-Carlo (MCMC) algorithms.
For example, a large number of possible additive structures that well explain the data in each BO iteration can be sampled from a model posterior via Metropolis–Hastings algorithm~\cite{Gardner2017}. 
Gibbs sampling can be performed to construct local GPs for use of the so-called ensemble BO after the input space is partitioned via a Mondrian process~\cite{WangZ2018}.

Different combinations of HF and LF data may completely change the additive form in \cref{eqn65} because learning possible additive structures is data-driven.
Thus, examining the sensitivity of decomposition structures with respect to the data used may be the first step if one wishes to use the additive structure approach for MF BO.

\subsubsection{Trust-region BO (TuRBO) approach}\label{Sec723}

This approach simultaneously uses independent local BO runs for global optimization~\cite{Eriksson2019}.
Each local BO relies on a local GP constructed from a trust region that is defined as a hyperrectangle centered at the best solution found from a set of data points.
The size of such a trust region can be expanded when better solutions are found after several consecutive iterations, or be contracted when no better solution is found after several consecutive iterations.

Assume that TuRBO maintains $m$ trust regions, i.e., $\text{TR}_1,\dots,\text{TR}_m$, at its iteration $k$.
The local GPs constructed from these trust regions are denoted by $\mathcal{GP}_1^k,\dots,\mathcal{GP}_m^k$.
In each iteration, TuRBO selects a batch of $q$ new design points, i.e., $\{{\bf x}_1^k,\dots,{\bf x}_q^k\}$, which are drawn from the union of the trust regions, and then updates the local GPs associated with the trust regions from which the new design points are drawn~\cite{Eriksson2019}.
This can be done via performing TS described in~\Cref{Algo2}.

To select a new design point $i$ in iteration $k$ of TuRBO, i.e., ${\bf x}_i^k$, TS randomly draws $m$ functions $q_{\text{H},1}^{i,k}({\bf x}), \dots,q_{\text{H},m}^{i,k}({\bf x})$ from  $\mathcal{GP}_1^k,\dots,\mathcal{GP}_m^k$, respectively.
Then, it selects ${\bf x}_i^k$, $i=1,\dots,q$, as the point that minimizes the  function value across all $m$ sampled functions~\cite{Eriksson2019}, such that
\begin{equation}\label{eqn66}
	{\bf x}_i^k = \underset{l \in \{1,\dots,m\}}{\mathrm{argmin}}\,\underset{{\bf x} \in \text{TR}_l }{\mathrm{argmin}} \ \ q_{\text{H},l}^{i,k}({\bf x}), \ \ i=1,\dots,q.
\end{equation}
By solving problem~(\ref{eqn66}), we obtain the information about ${\bf x}_i^k$ as well as the trust region from which ${\bf x}_i^k$ is drawn.
This information informs the updates of the size of the trust regions and the local GPs.
Note that while problem~(\ref{eqn66}) is still a high-dimensional minimization problem, solving it in an efficient manner is not the focus of TuRBO.

The following are two possible ways to develop a trust-region MF BO algorithm based on the TuRBO approach:
\begin{itemize}
	\item Use local GP-based MF surrogates for local BO runs and then adopt the no-fidelity consideration approach (\Cref{Sec521}) to update these surrogates after ${\bf x}_i^k$ is obtained by solving problem~(\ref{eqn66}).
	
	\item Use local GP-based MF surrogates for local BO runs and then adopt the sequential selection approach (\Cref{Sec523}) to select fidelity variables after ${\bf x}_i^k$ is obtained by solving problem~(\ref{eqn66}).
	In this way, the fidelity-query function $\gamma({\bf x}_i^k,t)$ described in \cref{eqn57} can be formulated for the trust region from which ${\bf x}_i^k$ is drawn.
\end{itemize}   

\subsection{Optimization under uncertainty}\label{Sec73}

Managing uncertainty is one of the important tasks in scientist computing~\cite{Oberkampf2010,Roy2011} and engineering design~\cite{Beyer2007}.
To perform this task, we can follow two key steps.
\begin{itemize}
	\item First, determine and capture the sources of uncertainty for the problem of interest. These sources may include model parameters, the mathematical model itself, measurement noise, and the lack of knowledge of the modelers and/or engineers. Each source can be described by a PDF or an interval-valued quantity if it is classified into aleatory or epistemic uncertainty, respectively.  
	
	\item Second, select a method to propagate the uncertainty through the problem for evaluating quantified uncertainty in the quantities of interest.
	The choice of the propagation method depends on the quality of information we have in the first step.
	This choice is also problem-dependent.
\end{itemize}

Optimization under uncertainty is crucial because the optimal solution is sensitive to even a small change in the problem.
This may be attributed to, for example, the fact that optimal solutions are often on the boundary of the feasible space.

Although there is a rich literature on techniques for optimization in the face of uncertainty, processing this class of optimization is still a challenging task in engineering design.
In this section, we briefly review recent applications of MF modeling methods and BO in two branches of optimization under uncertainty: robust optimization (RO) (\Cref{Sec731}) and reliability-based optimization (RBO) (\Cref{Sec732}).
We also attempt to provide possible directions for future research on MF BO to address these problems.

Let ${\bf s}$ denote a vector of random parameters that encapsulates uncertainty in our problem.
If a design variable is random, then it can be defined as a sum of a nominal variable, which is considered a design variable, and a random parameter before the optimization problem is formulated. 
We describe uncertainty in ${\bf s}$ using an interval $[{\bf s}_\text{l},{\bf s}_\text{u}]$ for set-based uncertainty and a PDF $p({\bf s})$ for probabilistic uncertainty.
The latter is the focal point of our discussion below. 

\subsubsection{Robust optimization}\label{Sec731}

An RO problem is often formulated using one of the following three concepts: absolute robustness, relative robustness, and less variance~\cite{Kanno2020}.
The absolute robustness formulates the RO problem based on the worst values of the objective and constraint functions under set-based uncertainty with a fixed set $[{\bf s}_\text{l},{\bf s}_\text{u}]$.
This problem is to minimize the worst value of the objective function under the constraints on the worst values of constraint functions, which is also called the worst-case scenario or minimax approach~\cite[see e.g.,][]{BenTal2009,Elishakoff2010}.
The relative robustness formulates the problem under set-based uncertainty to maximize the gap between the nominal value and the worst-case value of the objective function.
The less variance concept formulates the problem based on probabilistic uncertainty.

Note that there exists another robustness concept that formulates the problem using the information-gap decision theory~\cite{Hemez2004}.
In this concept, the uncertainty set, often defined as a closed Euclidean ball of radius centered at a nominal vector, can vary by adjusting the size of the uncertainty set.  
A design is considered robust if it remains feasible for large uncertainty sets.

In the less variance concept, a design that is less sensitive to uncertainty is considered more robust.
Thus, we wish to minimize the mean and variance of the objective function simultaneously, thereby formulating a
bi-objective RO problem which will be discussed in~\Cref{Sec74}.
Nevertheless, we can reformulate this bi-objective RO problem using the following weighted-sum formulation:
\begin{equation}\label{eqn67}
	\begin{aligned}
		\underset{{\bf x} \in \mathcal{X}}{\min} \ \ & w \mathbb{E}_{\bf s}\left[f_\text{H}({\bf x},{\bf s})\right]+(1-w)\sqrt{\mathbb{V}_{\bf s}\left[f_\text{H}({\bf x},{\bf s})\right]}\\
		\textrm{s.t.} \ \ 
		&\mathbb{E}_{\bf s}\left[g_{\text{H},i}({\bf x},{\bf s})\right] 
		+ \beta_i \sqrt{\mathbb{V}_{\bf s}\left[g_{\text{H},i}({\bf x},{\bf s})\right]} \leq 0, \ \ i=1,\dots,I, 
	\end{aligned}
\end{equation}
where $\mathbb{V}_{\bf s}[\cdot]$ denotes the variance of the quantity inside the brackets under uncertainty in ${\bf s}$, $w \in (0,1)$ is a weight value, and $\beta_i$ is a risk attitude factor for constraint $i$.

Difficulties in solving problem~(\ref{eqn67}) arise from two aspects: (1) uncertainty propagation that calculates statistical estimates of the objective and constraint functions for each candidate solution, and (2) search strategy that involves an inner loop of uncertainty propagation.

For uncertainty quantification, the Monte-Carlo integration~\cite{Caflisch1998} and polynomial chaos expansion~\cite{Crestaux2009} are traditional methods.
However, they come at cost of the curse of dimensionality.
The Taylor series approximation~\cite{Anderson2012} and Bayes-Hermite quadrature~\cite{OHagan1991} may serve as alternatives if the uncertain functions are differentiable and the random parameters $\bf s$ are normally distributed.
For the search strategy, derivative-free methods~\cite{Larson2019} are often used because they do not require derivative information of the statistical estimates, which are too expensive to extract.
Only a few works in the literature have solved RO problems via BO~\cite{Do2021,Daulton2022RMOBO}. 

As an early attempt to solve problem~(\ref{eqn67}) via MFO, \citet{Ng2014} estimated the mean and variance at candidate solutions via the Monte-Carlo integration while adjusting solutions using a derivative-free method.
Given two computational models with different infidelities and under a fixed computational budget, the MF mean was evaluated from the HF and LF mean values, which required $n$ calls of the HF model and $m$ calls of the LF model, where $m>n$.
The MF variance was evaluated in a similar way.

Other attempts to solve problem~(\ref{eqn67}) via MFO differ in their ways of using MF surrogates, uncertainty propagation methods, and optimization solvers~\cite[see e.g.,][]{Shah2015,Fusi2015,Chakraborty2017,Zhou2018}.
For example, \citet{Shah2015}, under the model-then-optimize approach, used a composition method via polynomial chaos for the mean and variance estimates in each iteration of the sequential least squares quadratic programming.
\citet{Zhou2018} incorporated the hierarchical Kriging model and the Monte-Carlo integration into a GA solver.

While MFO approaches can enhance the feasibility of addressing problem~(\ref{eqn67}), it does so at the expense of additional uncertainty induced by the use of LF models.
This uncertainty can misguide the optimization process or render the RO solutions infeasible if not considered meticulously. 
Hence, it is of great interest to devise efficient methods that can account for this uncertainty if one wishes to solve problem~(\ref{eqn67}) via an MFO approach.

\subsubsection{Reliability-based optimization}\label{Sec732}

Under probabilistic uncertainty in ${\bf s}$, RBO or chance-constrained optimization~\cite{Campi2011} minimizes the mean of the objective function under probabilistic constraints.
The general formulation of RBO is
\begin{equation}\label{eqn68}
	\begin{aligned}
		\underset{{\bf x} \in \mathcal{X}}{\min} \ \ & \mathbb{E}_{\bf s}\left[f_\text{H}({\bf x},{\bf s})\right]\\
		\textrm{s.t.} \ \ 
		& \mathbb{P}_{\bf s}\left[g_{\text{H},i}({\bf x},{\bf s}) \leq 0\right] \geq 1-\epsilon_i,\ \ i=1,\dots,I, 
	\end{aligned}
\end{equation}
where $\epsilon_i \in (0,1)$ is a prescribed risk level associated with probabilistic constraint $i$.
Large values of $\epsilon_i$ sacrifice the reliability of the solution, while small values lead to conservative designs.
We also assume that the constraint functions are conditionally independent given a vector of design variables under which problem~(\ref{eqn68}) is an RBO with individual probabilistic constraints, which differs from another class of RBO problems that involve joint probabilistic constraints~\cite{Xie2018}.

Although problem~(\ref{eqn68}) is very important for optimization under uncertainty, it is too difficult to solve. 
The difficulty arises from three aspects.
First, checking the feasibility of a candidate solution is nontrivial because the computation of probabilistic constraints is generally an NP-hard problem~\cite{Geng2019}.
Second, finding an exact optimal solution may be impossible because the feasible region defined by the probabilistic constraints is generally non-convex~\cite{Nemirovski2012}.
Third, estimating the mean of the objective is computationally demanding, which is discussed in~\Cref{Sec731}.

The Monte-Carlo integration is a traditional method for estimating the probabilistic constraint functions.
While this method is simple to implement and able to provide unbiased statistical estimates, its rate of convergence and convergence stability often depend on the quality of random generators~\cite{Melchers2018}, leading to the difficulty in capturing the sensitivity of constraint functions.
The Monte-Carlo integration also introduces a considerable computational burden when evaluating the feasibility of a candidate solution, which violates a reasonable expectation of achieving greater computational efficiency through an increased number of optimization iterations, rather than dedicating excessive time to each iteration.

\begin{figure*}[t]
	\centering
	\includegraphics[width=0.85\textwidth]{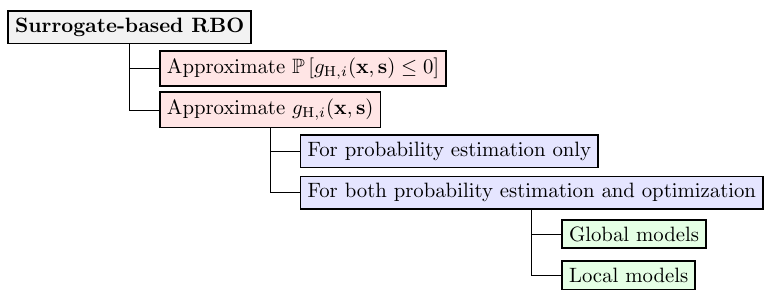}
	\caption{Different schemes of surrogate-based RBO~\cite{Moustapha2019}.}
	\label{Fig-14}
\end{figure*}

In engineering design optimization, problem~(\ref{eqn68}) is often solved approximately.
Classical RBO approximation techniques fall into three distinct categories: top-level, mono-level, and decoupled approaches.
A benchmark study and a comprehensive review of methods in each of these approaches can be found in~\citet{Aoues2010} and \citet{Valdebenito2010}, respectively.
Other approximation techniques solve problem~(\ref{eqn68}) sequentially by leveraging surrogates for $\mathbb{P}_{\bf s}\left[g_{\text{H},i}({\bf x},{\bf s}) \leq 0\right]$ or $g_{\text{H},i}({\bf x},{\bf s})$.
\Cref{Fig-14} shows different schemes of the surrogate-based RBO, which is discussed in~\citet{Moustapha2019}.  

Existing approaches in the field of mathematical optimization typically rely on several special forms of problem~(\ref{eqn68}).
For example, the scenario approach~\cite{Calafiore2006} is formulated for linear objective functions and continuous, convex constraint functions $g_{\text{H},i}({\bf x},{\bf s})$ with respect to ${\bf x}$.
The convex approximation requires that $g_{\text{H},i}({\bf x},{\bf s})$ is convex in ${\bf x}$ for every instance
of ${\bf s}$ and the set defined by the deterministic constraints is convex~\cite{Nemirovski2012}. 

In BO literature, \citet{Bichon2008} estimated failure probabilities using a modification of EGO that relies on GPs for constraint functions and an expected feasibility acquisition function.
This acquisition function aims to increase the accuracy of the GPs in the vicinity of the limit state, i.e., the boundary of the feasible region. 
Once the GPs have been refined by EGO, they can be used to estimate the probabilities via multimodal adaptive importance sampling.
\citet{Blanchard2021siamjuq} provided a class of likelihood-weighted acquisition functions to uncertainty quantification of the quantities of interest.
\citet{Huynh2023} extended the approach by~\citet{Bichon2008} and solved problem~(\ref{eqn68}) by embedding EGO in a sequential strategy of the decoupled RBO approach.
\citet{Chaudhuri2020} developed an RBO approach that couples reusing samples for a posteriori biasing density nearby designs for probability estimation with randomly exploiting in the neighborhood of the current design for optimization.

Although many efficient methods, inspired by~\citet{Bichon2008}, have adopted MF BO to facilitate reliability analyses ~\cite[see e.g.,][]{Chaudhuri2021,Patsialis2021,ZhangC2022,AshwinRenganathan2023}, the use of MF BO for addressing RBO problems is still an open issue.
This may be attributed to the conflict between the accuracy and efficiency properties of a good optimization algorithm, as mentioned in \Cref{Sec1}.
While we focus on improving the accuracy by using sophisticated MF reliability analysis methods, we unintentionally require excessive computer time or storage, thereby reducing the optimization efficiency.    

The above observations present potential opportunities for MF BO to address problem~(\ref{eqn68}).
In addition to improving reliability analyses, considerations should include other important aspects such as improving the solution quality, evaluating the feasibility, selecting fidelities, and addressing uncertainty due to the use of LF models.
We believe that the sequential selection approach (\Cref{Sec52}) would be suitable for such serious considerations because it can decouple the selections of design variables $\bf x$, random parameters $\bf s$, and fidelity variables for different purposes.   
After these considerations, MF BO may look on a far horizon to explore its possibility of solving RBO problems with joint probabilistic constraints, which are also very important for decision-making under uncertainty, but generally more difficult to solve than problem~(\ref{eqn68}).

\subsection{Multi-objective optimization}\label{Sec74}

We formulate a multi-objective (or vector-valued) optimization problem when we wish to optimize several competing objectives but do not know how to prioritize them.
A general $m$-objective optimization problem reads
\begin{equation}\label{eqn69}
	\begin{aligned}
		\underset{{\bf x} \in \mathcal{X}}{\min} \ \ & \left[f_{\text{H},1}({\bf x}),\dots,f_{\text{H},m}({\bf x})\right]\\
		\textrm{s.t.} \ \ 
		& g_{\text{H},i}({\bf x}) \leq 0,\ \ i=1,\dots,I. 
	\end{aligned}
\end{equation}

Solving problem~(\ref{eqn69}) yields a set of potentially optimal solutions lying on the so-called Pareto frontier of the objectives.
There are typically two approaches to finding the Pareto frontier: decomposition and population-based.

\textit{Decomposition approaches} convert the vector-valued objective function to a series of single-valued objective functions using constraint-based or weight-based methods.
In comparison, \textit{population-based approaches} involve the adaptation of standard population-based algorithms to incrementally improve both the accuracy and diversity of approximate Pareto frontiers. The adaptation may combine several techniques such as population partitioning, nondomination ranking, Pareto filtering, and niche methods~\cite{Kochenderfer2019}. 
One of the popular population-based multi-objective algorithms is the non-dominated sorting genetic algorithm (NSGA-II)~\cite{Deb2002}.
There also exist several algorithms, e.g., MOEA/D~\cite{ZhangQ2007}, that decompose the problem into scalar-valued optimization subproblems which are solved simultaneously by evolving a population of candidate solutions. 

Multi-objective BO (MOBO) algorithms devised for finding approximate Pareto frontiers of problem~(\ref{eqn69}) can also be classified into decomposition and population-based approaches.
The former, for example, ParEGO~\cite{Knowles2006}, converts $m$ objective functions to a single objective function via an $m$-dimensional weight vector, thereby enabling the generic BO to identify a new design point in each iteration.
By sweeping over the space of possible weight vectors and employing one value in each iteration, an approximate Pareto frontier can be incrementally constructed.
The limitation of this approach is that it only works well on small-dimensional problems and often fails in capturing nonconvex parts of the Pareto frontier when using the weighted sum of objectives~\cite{Das1997}.

The latter views the initial set of samples generated for MOBO as a population that evolves to improve approximate Pareto frontiers.
This evolution strategy is often based on a non-dominated sorting algorithm and an acquisition function.
While the sorting algorithm is to find non-dominated solutions constituting the approximate Pareto frontier in each iteration, the acquisition function is tailored to improve the quality of the approximate Pareto frontier incrementally.
Notable examples of MOBO's acquisition functions include the expected hypervolume improvement~\cite{Emmerich2006}, expected value of Euclidean distance improvement~\cite{Keane2006}, $\mathcal{S}$-metric~\cite{Ponweiser2008}, and Pareto-frontier entropy search~\cite{Suzuki2020}.

Recent works have exploited MFO and MF BO approaches to multi-objective optimization.
\citet{Singh2017}, via the model-then-optimize approach, constructed recursive MF surrogates for the objective functions to be optimized by a population-based multi-objective algorithm. 
By limiting the total computing budget allocation, \citet{Khatamsaz2021aiaa} combined a model composition and the expected hypervolume improvement.
Notably, \citet{Daulton2023} proposed a one-step look-ahead
acquisition function based on a hypervolume KG that is capable of conditioning on MF evaluations and decoupling the objective evaluations.
This allows independent evaluations of the objective functions at different fidelities, which can be efficient when the computational cost varies substantially between the objectives.

Future research on MF MOBO may consider the following questions:
\begin{itemize}
	\item Can the computational cost of multi-objective acquisition functions be effectively reduced through the application of MF approaches? This is prompted by the inherent expense associated with maximizing multi-objective acquisition functions, which often involves integration over the (feasible) space of objective functions.
	
	\item What are the possible impacts of varying fidelities within a fixed computational budget on the accuracy and diversity of approximate Pareto frontiers? Addressing this aspect could unveil insights into the development of more efficient MF multi-objective acquisition functions for MOBO.
	
	\item How to reason about multi-step look-ahead multi-objective acquisition functions when considering fidelities and do we really need them to further balance exploitation and exploration?   
\end{itemize}

\section{Conclusion}\label{Sec8}
MF BO is capable of facilitating costly design optimization problems by making efficient use of HF data while extracting all information we have about the mathematical models and/or the structures of the physical processes underlying these problems. 
As shown from the literature surveyed in this work, MF BO has achieved an important impact on engineering design optimization, specifically when it involves running HF computational models or conducting time-consuming physical experiments.
Ever since the seminal works by~\citet{Huang2006smo} and \citet{Forrester2007}, MF BO has continually evolved in three directions: (1) improving MF modeling methods to enhance the information transfer between HF and MF models, (2) devising novel acquisition functions for judicious selections of both new design points and model fidelities, and (3) applying MF BO methods to various domains of engineering design.
 
Specific to this work, we highlight two essential ingredients of MF BO that exist in the literature: GP-based MF surrogates and acquisition functions.
We first classify the existing MF modeling methods and MFO strategies to determine where MF BO is located in the rich literature on SbO and MFO algorithms.
We then survey important GP-based MF surrogates, followed by a comprehensive review of various acquisition functions of BO and how to modify them for use of MF BO.  
While describing the techniques from each ingredient, we attempt to exploit their common properties.
For example, we show that the GMGP model is more general than the KOH model while being a type of the LMC model.
We focus on the exploitation-exploration trade-off to see how it can be addressed by some popular acquisition functions, including EI, WEI, GP-LCB, and TS, and to provide a new perspective that multi-step look-ahead acquisition functions indeed aim at improving exploration by considering the impact of future selections of new design points.  
Our ultimate goal is to provide a structured understanding of MF BO.
We compare the performance of several coKriging-based MF BO methods using a minimization problem of a test function and shape optimization problems of two standard airfoils.

Additionally, we provide recent advances in MFO and BO to address intricate yet important design optimization problems, including constrained optimization, high-dimensional optimization, optimization under uncertainty, and multi-objective optimization.
For each of these problems, we attempt to reveal important aspects that require further research for applications of MF BO.
We expect to open up new avenues for MF BO, bringing it closer to being a comprehensive optimization technique.

Although they are not detailed in this work, other potential opportunities for future research on MF BO include    
\begin{itemize}
	\item MF BO for combinatorial optimization. This is a natural extension of several advanced BO algorithms that have recently been devised for solving combinatorial optimization problems. These algorithms have focused on developing covariance functions of non-continuous design variables~\cite[see e.g.,][]{GomezBombarelli2018,GarridoMerchan2020,ZhangYi2020,Vangelatos2021}. Another aspect that should be considered when using MF BO for combinatorial optimization is to find an efficient optimizer for maximizing acquisition functions in the combinatorial space of the design variables and fidelities.
	
	\item MF BO with gradients. It is shown that exploiting gradients to construct a GP for the objective function can help BO provide good solutions with fewer objective function evaluations~\cite{WuJ2017}.
 
    \item LF models as prior models. LF models can serve as priors for the surrogates we construct during the process of BO (e.g., the mean and covariance of GP models). From this perspective, we may view MF BO as the generic BO with more informative surrogate models.
        
\end{itemize}

While it is beyond the scope of this work, there is notable value in the development of open-source software that offers full-featured sublibraries for MFO, BO, and MF BO.
Such a platform would facilitate in-depth comparative investigations on the performance of various MF BO techniques reviewed in this work across a wide spectrum of engineering design.

\section*{Acknowledgments}
We thank the University of Houston for providing startup fund to support this research.

\section*{Appendix A: Univariate Gaussian processes}\label{AppA}
\setcounter{equation}{0} %
\renewcommand{\theequation}{A.\arabic{equation}} %

Consider a training dataset $\mathcal{D}=\left\{ {\bf X},{\bf Y}\right\}=\{{\bf x}^k,y^k\}_{k=1}^N$, where $\textbf{x}^k\in\mathbb{R}^d$ are $d$-dimensional vectors of design variables and $y^k\in\mathbb{R}$ the corresponding observations of the function of interest (e.g., objective or constraint function).
Based on $\mathcal{D}$, we wish to find the mapping $y({\bf x})=f({\bf x}) + \varepsilon_\text{y}: \mathbb{R}^d \mapsto \mathbb{R}$, where $f({\bf x})$ is a regression function conditioned on $\mathcal{D}$, and $\varepsilon_\text{y} \sim \mathcal{N}(0,\sigma^2_\text{y})$ is additive zero-mean Gaussian noise with variance $\sigma^2_\text{y}$.
Note that the noise is assumed to be independent and identically distributed.

GP assumes that any finite subset of an infinite set of the values of $f({\bf x})$ has a joint Gaussian distribution \cite{Rasmussen2006}.
This assumption is encoded in the following GP prior:
\begin{equation}\label{eqnA1}
	f(\cdot) \sim \mathcal{GP} \left(m(\cdot|\boldsymbol \beta_\text{m}),\kappa(\cdot,\cdot|\boldsymbol \phi_{\text{x}})\right),
\end{equation}
where  $m(\cdot|\boldsymbol \beta_{\text{m}})$: $\mathbb{R}^d \mapsto \mathbb{R}$ is a mean function characterized by the hyperparameter $\boldsymbol \beta_{\text{m}}$ and $\kappa({\bf x},{\bf x}'|\boldsymbol \phi_{\text{x}}) = \text{cov}[f({\bf x}),f({\bf x}')]: \mathbb{R}^d \times \mathbb{R}^d \mapsto \mathbb{R} $ is a positive semi-definite covariance function parameterized by the hyperparameter vector $\boldsymbol \phi_{\text{x}}$.

Thus, for the set of $N$ parameter vectors $\{{\bf x}_1,\dots,\textbf{x}_N \}$, the vector of the corresponding error function values ${\bf F}({\bf X})=[f({\bf x}^1),\dots,f({\bf x}^N)]^\intercal \in \mathbb{R}^N$ is distributed according to an $N$-variate Gaussian with parameters determined by the mean
and covariance functions, such that
\begin{equation}\label{eqnA2}
	{\bf F}({\bf X}) \sim \mathcal{N} \left({\bf M}({\bf X}),{\bf K}({\bf X},{\bf X})\right),
\end{equation}
where ${\bf M}({\bf X})=[m({\bf x}^1|\boldsymbol \beta_{\text{m}}),\dots,m({\bf x}^N|\boldsymbol \beta_{\text{m}})]^\intercal \in \mathbb{R}^N$ is the mean vector, ${\bf K}({\bf X},{\bf X}) \in \mathbb{R}^{N \times N}$ is the covariance matrix with the $(i,j)$th element $({\bf K}({\bf X},{\bf X}))_{i,j} = \kappa({\bf x}^i,{\bf x}^j|\boldsymbol \phi_{\text{x}})$, $i,j=1,\dots,N$.

Furthermore, the covariance of the observations is
\begin{equation}\label{eqnA3}
		\text{cov}[y({\bf x}^i),y({\bf x}^j)]  = \text{cov}[f({\bf x}^i),f({\bf x}^j)] + \text{cov}[\varepsilon_\text{y}^i,\varepsilon_\text{y}^j]
		 = \kappa({\bf x}^i,{\bf x}^j|\boldsymbol \phi_{\text{x}}) + \mathbbm{1}_{i,j},
\end{equation}
where $\mathbbm{1}_{i,j}$ denotes the Kronecker delta. Eqs.~(\ref{eqnA2}) and (\ref{eqnA3}) lead to
\begin{equation}\label{eqnA4}
	{\bf Y}({\bf X}) \sim \mathcal{N} \left({\bf M}({\bf X}),{\bf K}({\bf X},{\bf X})+\sigma^2_\text{y} {\bf I}_N\right),
\end{equation}
where ${\bf Y}({\bf X})=[y({\bf x}^1),\dots,y({\bf x}^N)]^\intercal \in \mathbb{R}^N$ is the vector of observations.

Let ${\boldsymbol \phi} = [\boldsymbol \beta_{\text{m}}^\intercal,\sigma^2_\text{y},\boldsymbol \phi_{\text{x}}^\intercal]^\intercal$ denote the vector of hyperparameters of the GP model.
We can find the optimal value ${\boldsymbol \phi^\star}$ of ${\boldsymbol \phi}$ by maximizing the likelihood of ${\bf Y}({\bf X})$ model derived from \cref{eqnA4}.

Once ${\boldsymbol \phi^\star}$ has been found, we wish to predict $f(\overline{\bf x})$, where $\overline{\bf x}$ is an unseen variable vector.
The posterior predictive PDF at $\overline{\bf x}$ reads
\begin{equation}\label{eqnA5}
	p\left(f(\overline{\bf x})|\overline{\bf x},\mathcal{D},\boldsymbol \phi\right) = \mathcal{N}\left(\mu_\text{f}(\overline{\bf x}),\sigma_\text{f}^2(\overline{\bf x})\right).
\end{equation}
The mean and variance of this PDF are
\begin{equation}\label{eqnA6}
	\mu_\text{f}(\overline{\bf x})=m(\overline{\bf x}|\boldsymbol \beta_{\text{m}}) 
	+ {\bf K}(\overline{\bf x},{\bf X})^\intercal \left[ {\bf K}({\bf X},{\bf X})+\sigma^2_\text{y} {\bf I}\right]^{-1} \left({\bf Y}({\bf X})-{\bf M} ({\bf X})\right),
\end{equation}
\begin{equation}\label{eqnA7}
	\sigma_\text{f}^2(\overline{\bf x})=\kappa(\overline{\bf x},\overline{\bf x}|\boldsymbol \phi_{\text{x}})
	-{\bf K}(\overline{\bf x},{\bf X})^\intercal \left[ {\bf K}({\bf X},{\bf X})+\sigma^2_\text{y} {\bf I}\right]^{-1} {\bf K}(\overline{\bf x},{\bf X}),
\end{equation}
where
\begin{equation}\label{eqnA8}
	{\bf K}(\overline{\bf x},{\bf X})= \left[\kappa(\overline{\bf x},{\bf x}^1|\boldsymbol \phi_{\text{x}}),\dots,\kappa(\overline{\bf x},{\bf x}^N|\boldsymbol \phi_{\text{x}})\right]^\intercal.
\end{equation}

\section*{Appendix B: Multiivariate Gaussian processes}\label{AppB}
\setcounter{equation}{0} %
\renewcommand{\theequation}{B.\arabic{equation}} %

We find the mapping ${\bf y}({\bf x})={\bf f}({\bf x}) + \boldsymbol{\varepsilon}_\text{y}: \mathbb{R}^d \mapsto \mathbb{R}^T$ to explain the training dataset $\mathcal{D}=\left\{ {\bf X},{\bf Y}\right\}=\{{\bf x}^k,{\bf y}^k\}_{k=1}^N$, where $\textbf{x}^k\in\mathbb{R}^d$, ${\bf y}^k\in\mathbb{R}^T$, ${\bf f}({\bf x})$ is a regression function, and $\boldsymbol{\varepsilon}_\text{y} \sim \mathcal{N}({\bf 0},\boldsymbol{\Sigma}_\text{y})$ is a zero-mean Gaussian noise with covariance matrix $\boldsymbol{\Sigma}_\text{y} \in \mathbb{R}^{T \times T}$.

Multivariate GP for vector-valued functions is a natural extension of the univariate GP.
The multivariate GP prior reads
\begin{equation}\label{eqnB1}
	{\bf f}(\cdot) \sim \mathcal{GP} \left({\bf m}(\cdot|\boldsymbol \beta_\text{m}),{\bf S}(\cdot,\cdot|\boldsymbol \phi_{\text{x}})\right),
\end{equation}
where ${\bf m}(\cdot|\boldsymbol \beta_\text{m})$: $\mathbb{R}^d \mapsto \mathbb{R}^T$ is the vector-valued mean function characterized by the hyperparameter vector $\boldsymbol \beta_\text{m}$ and
${\bf S}({\bf x},{\bf x}'|\boldsymbol \phi_{\text{x}}) = \text{cov}[{\bf f}({\bf x}),{\bf f}({\bf x}')|\boldsymbol \phi_{\text{x}}] \in \mathbb{R}^{T \times T}$ denotes the inter-group covariance matrix.

Let ${\bf F}({\bf X})=[{\bf f}^\intercal({\bf x}^1),\dots,{\bf f}^\intercal({\bf x}^N)]^\intercal \in \mathbb{R}^{TN}$ be a vector that concatenates the vectors ${\bf f}({\bf x}^i)$, $i=1,\dots,N$.
Let ${\bf M}({\bf X})=[{\bf m}^\intercal({\bf x}^1|\boldsymbol \beta_\text{m}),\dots,{\bf m}^\intercal({\bf x}^N|\boldsymbol \beta_\text{m})]^\intercal \in \mathbb{R}^{TN}$ be a vector that concatenates the mean vectors ${\bf m}({\bf x}^i)$.
According to the GP prior in \cref{eqnB1}, the distribution of ${\bf F}({\bf X})$ reads 
\begin{equation}\label{eqnB2}
	{\bf F}({\bf X}) \sim \mathcal{N} \left({\bf M}({\bf X}),{\bf K}({\bf X},{\bf X})\right),
\end{equation}
where the covariance matrix
${\bf K}({\bf X}, {\bf X}) \in \mathbb{R}^{TN \times TN}$ is the block partitioned matrix with
$\left({\bf K}({\bf X}, {\bf X})\right)_{i,j} = {\bf S}({\bf x}^i,{\bf x}^j|\boldsymbol \phi_{\text{x}})$, $i,j=1,\dots,N$.

Let ${\bf Y}({\bf X})=[{\bf y}^\intercal({\bf x}^1),\dots,{\bf y}^\intercal({\bf x}^N)]^\intercal \in \mathbb{R}^{TN}$ be a vector of output observations.
${\bf Y}({\bf X})$ is distributed according to
\begin{equation}\label{eqnB3}
	{\bf Y}({\bf X}) \sim \mathcal{N} \left({\bf M}({\bf X}),{\bf K}({\bf X},{\bf X})+\boldsymbol{\Sigma}_\text{y} \otimes {\bf I}_N\right),
\end{equation}
where $\otimes$ denotes the Kronecker product between matrices.

By maximizing the likelihood of ${\bf Y}({\bf X})$ model in \cref{eqnB3}, we obtain the optimal set of hyperparameters $\boldsymbol \phi=\{\boldsymbol \phi_{\text{x}},\boldsymbol{\Sigma}_\text{y},\boldsymbol \beta_{\text{m}}\}$.
Then, the posterior predictive PDF at $\overline{\bf x}$ can be derived as
\begin{equation}\label{eqnB4}
	p\left({\bf f}(\overline{\bf x})|\overline{\bf x},\mathcal{D},\boldsymbol \phi\right) = \mathcal{N}\left(\boldsymbol{\mu}_\text{f}(\overline{\bf x}),\boldsymbol{\Sigma}_\text{f}(\overline{\bf x})\right).
\end{equation}
The mean and variance of this PDF are
\begin{equation}\label{eqnB5}
		\boldsymbol{\mu}_\text{f}(\overline{\bf x})={\bf m}(\overline{\bf x}|\boldsymbol \beta_{\text{m}})
		+ {\bf K}(\overline{\bf x},{\bf X})^\intercal \left[ {\bf K}({\bf X},{\bf X})+\boldsymbol{\Sigma}_\text{y} \otimes {\bf I}\right]^{-1} \left({\bf Y}({\bf X})-{\bf M} ({\bf X})\right),
\end{equation}
\begin{equation}\label{eqnB6}
		\boldsymbol{\Sigma}_\text{f}(\overline{\bf x})={\bf S}(\overline{\bf x},\overline{\bf x}|\boldsymbol \phi_{\text{x}})
		-{\bf K}(\overline{\bf x},{\bf X})^\intercal \left[ {\bf K}({\bf X},{\bf X})+\boldsymbol{\Sigma}_\text{y} \otimes {\bf I}\right]^{-1} {\bf K}(\overline{\bf x},{\bf X}),
\end{equation}
where ${\bf K}(\overline{\bf x},{\bf X}) \in \mathbb{R}^{TN \times T}$, which concatenates the following matrices:
\begin{equation}\label{eqnB7}
	({\bf K}(\overline{\bf x},{\bf X}))_i = {\bf S}(\overline{\bf x},{\bf x}^i|\boldsymbol \phi_{\text{x}}), \ \ i=1,\dots,N.
\end{equation}

\section*{Appendix C: Test functions}\label{AppC}
\setcounter{equation}{0} 
\renewcommand{\theequation}{C.\arabic{equation}} 

\paragraph{2d Levy function:}
The HF and LF expressions for the 2d Levy function are given as follows \cite{Surjanovic2013,LiS2020}:
\begin{subequations}
    \begin{equation}
    	f_\text{H}({\bf x}) = \sin^2(3 \pi x_1) + (x_1 - 1)^2 \left[ 1 +  \sin^2(3 \pi x_2) \right] + (x_2 - 1)^2 \left[ 1 +  \sin^2(2 \pi x_2) \right],
    \end{equation}
    \begin{equation}
	f_\text{L}({\bf x}) =  \exp\left( 0.1 \sqrt{f_\text{H}({\bf x})} \right) + 0.1 \sqrt{1 + f^2_\text{H}({\bf x})},
    \end{equation}
\end{subequations}
where ${\bf x} \in [-10, 10]^2$. The function has a global minimum at ${\bf x}^\star = [1, 1]^\intercal$ with $f^\star = f({\bf x}^\star) = 0$.

\paragraph{6d Hartmann function:}
This function is defined over $\mathcal{X}=[0,1]^6$ \cite{Surjanovic2013}, and has a global minimum at ${\bf x}^\star =[0.20169,0.150011,0.476874,0.275332,0.311625,0.6573]^\intercal$ with $f^\star = f({\bf x}^\star) = -3.32237$. Its rescaled HF and LF expressions used for optimization are given as follows:
\begin{subequations}
    \begin{equation}
	f_\text{H}({\bf x}) = -\frac{1}{1.94}\left[2.58 + \sum_{i=1}^{4} a_i \exp \left( -\sum_{j=1}^{6} A_{ij} (x_j-P_{ij})^2 \right) \right],
    \end{equation}
    \begin{equation}
	f_\text{L}({\bf x}) = -\frac{1}{1.94}\left[2.58 + \sum_{i=1}^{3} a_i \exp \left( -\sum_{j=1}^{6} A_{ij} (x_j-P_{ij})^2 \right) \right],
    \end{equation}
\end{subequations}
where 
\begin{subequations}
    \begin{equation}
		{\bf a} = [1,1.2,3,3.2]^\intercal,
  \end{equation}
  \begin{equation}
		{\bf A}  = 
            \begin{bmatrix}
			10 & 3 & 17 & 3.5 & 1.7 & 8 \\
			0.05 & 10 & 17 & 0.1 & 8 & 14 \\
			3 & 3.5 & 1.7 & 10 & 17 & 8 \\
			17 & 8 & 0.05 & 10 & 0.1 & 14
		\end{bmatrix},
  \end{equation}
  \begin{equation}
		{\bf P}  = 10^{-4}\begin{bmatrix}
			1312 &1696 &5569 &124 &8283 &5886\\
			2329 &4135 &8307 &3736 &1004 &9991\\
			2348 &1451 &3522 &2883 &3047 &6650\\
			4047 &8828 &8732 &5743 &1091 &381
		\end{bmatrix}.
  \end{equation}
\end{subequations}

\section*{Appendix D: Airfoil shape parameterization}\label{AppD}
\setcounter{equation}{0} 
\renewcommand{\theequation}{D.\arabic{equation}} 

\paragraph{NACA 4-digit airfoils:} As described in \Cref{Sec62:airfoils}, the shapes of NACA 4-digit airfoils are parameterized by three parameters: $c_{\max}$--the height measured from the chord line to the point where the mean camber line has the largest curvature, $x_{\max}$--the position of $c_{\max}$ in horizontal coordinates, and $t_{\max}$--the maximum thickness of the airfoil, see \cref{Fig-10}(a)  \cite{Jacobs1933}.
The airfoil sections are created from the mean camber line and a thickness distribution drawn perpendicular to the mean camber line.

The mean camber line describes the curve between the leading and trailing edges. It is split into two regions: region (1) from 0 to the location of maximum camber with $0 \leq x_\text{a} < x_{\max}$ and region (2) with $x_{\max} \leq x_\text{a} \leq  1$, where $x_\text{a}$ is the distance along the chord.
Let  $y_\text{c}$ be the vertical coordinate of the camber line at a position $x_\text{a}$.
The equation for region (1) reads
\begin{equation}
    y_\text{c} = \frac{c_{\max}}{x_{\max}^2} \left(2 x_{\max} x_\text{a}- x_\text{a}^{2} \right),
\end{equation}
and that for region (2) is
\begin{equation}
    y_\text{c} = \frac{c_{\max}}{(1-x_{\max})^2} \left(1 - 2 x_{\max} + 2 x_{\max} x_\text{a} - x_\text{a}^{2} \right).
\end{equation}

The thickness distribution $y_\text{t}$ at any position $x_\text{a}$ along the chord is given by
\begin{equation}
    y_\text{t} = \frac{t_{\max}}{0.2} \left(a_0 x_\text{a}^{1/2} + a_1 x_\text{a} + a_2 x_\text{a}^{2} +a_3 x_\text{a}^{3} + a_4 x_\text{a}^{4} \right),
\end{equation}
where $a_0 = 0.2969$, $a_1 = 0.126$, $a_2 = 0.3516$, $a_3 = 0.2843$, and $a_4 = -0.1015$ for an open trailing edge or $-0.1035$ for a closed trailing edge.

As a result, the coordinates for the upper surface $(x_{\text{a,u}},y_{\text{a,u}})$ are
\begin{equation}
    x_{\text{a,u}} = x_\text{a} - y_\text{t} \sin(\theta), \quad y_{\text{a,u}} = y_\text{c} + y_\text{t} \cos(\theta),
\end{equation}
and those of the lower surface $(x_{\text{a,l}},y_{\text{a,l}})$ are
\begin{equation}
    x_{\text{a,l}} = x_\text{a} + y_\text{t} \sin(\theta), \quad y_{\text{a,l}} = y_\text{c} - y_\text{t} \cos(\theta),
\end{equation}
where $\theta = \arctan\left( \frac{d y_\text{c}}{d x_\text{a}}\right)$.

\paragraph{PARSEC airfoils:}
Let ${\bf x} = [x_1,\dots,x_{11}]^\intercal =  \left[ r_\text{LE}, x_\text{up}, y_\text{up}, \kappa_\text{xxup}, x_\text{lo}, y_\text{lo}, \kappa_\text{xxlo},y_\text{TE},\Delta y_\text{TE},\alpha_\text{TE},\beta_\text{TE} \right]^\intercal$ represent a vector containing the geometrical features of a PARSEC airfoil, see \cref{Fig-10}(b) and \cref{Table6}.
The vertical coordinates of the upper and lower surfaces of the PARSEC airfoil section are obtained from the following polynomial equations \cite{Pierluigi2014}:
\begin{equation}
   y_{\text{a,u}} = \sum_{i=1}^6 a_{\text{u},i} x_\text{a}^{i-1/2}, \quad y_{\text{a,l}} = \sum_{i=1}^6 a_{\text{l},i} x_\text{a}^{i-1/2}.
\end{equation}
Here the coefficient vectors ${\bf a}_\text{u} = \left[a_{\text{u},1},\dots, a_{\text{u},6}\right]^\intercal$ and ${\bf a}_\text{l} = \left[a_{\text{l},1},\dots, a_{\text{l},6} \right]^\intercal$ are obtained by solving the following systems of linear equations: 
\begin{equation}
  {\bf C}_\text{u} {\bf a}_\text{u} = {\bf b}_\text{u}, \quad {\bf C}_\text{l} {\bf a}_\text{l} = {\bf b}_\text{l},
\end{equation}
where
\begin{equation}
  {\bf C}_\text{u} = \left[  \begin{matrix}
1 &  1&  1&  1&  1&  1\\
 x_2^{1/2}&  x_2^{3/2}&  x_2^{5/2}&  x_2^{7/2}&  x_2^{9/2}&  x_2^{11/2}\\
 1/2&   3/2&   5/2&   7/2&   9/2&   11/2\\
 \frac{1}{2}x_2^{-1/2}&  \frac{3}{2}x_2^{1/2}&  \frac{5}{2}x_2^{3/2}&  \frac{7}{2}x_2^{5/2}&  \frac{9}{2}x_2^{7/2}&  \frac{11}{2}x_2^{9/2}\\
 -\frac{1}{4}x_2^{-3/2}&  \frac{3}{4}x_2^{-1/2}&  \frac{15}{4}x_2^{1/2}&  \frac{35}{4}x_2^{3/2}& \frac{63}{4}x_2^{5/2} &  \frac{99}{4}x_2^{7/2}\\
 1&  0&  0&  0&  0& 0
\end{matrix}\right], \quad {\bf b}_\text{u} = \left[ \begin{matrix}
 x_8 + x_9/2\\
 x_3\\
 \tan\left( x_{10}-x_{11}/2\right)\\
 0\\
 x_4\\
\sqrt{2 x_1}
\end{matrix} \right],
\end{equation}
and 
\begin{equation}
  {\bf C}_\text{l} = \left[  \begin{matrix}
1 &  1&  1&  1&  1&  1\\
 x_5^{1/2}&  x_5^{3/2}&  x_5^{5/2}&  x_5^{7/2}&  x_5^{9/2}&  x_5^{11/2}\\
 1/2&   3/2&   5/2&   7/2&   9/2&   11/2\\
 \frac{1}{2}x_5^{-1/2}&  \frac{3}{2}x_5^{1/2}&  \frac{5}{2}x_5^{3/2}&  \frac{7}{2}x_5^{5/2}&  \frac{9}{2}x_5^{7/2}&  \frac{11}{2}x_5^{9/2}\\
 -\frac{1}{4}x_5^{-3/2}&  \frac{3}{4}x_5^{-1/2}&  \frac{15}{4}x_5^{1/2}&  \frac{35}{4}x_5^{3/2}& \frac{63}{4}x_5^{5/2} &  \frac{99}{4}x_5^{7/2}\\
 1&  0&  0&  0&  0& 0
\end{matrix}\right], \quad {\bf b}_\text{l} = \left[ \begin{matrix}
 x_8 - x_9/2\\
 x_6\\
 \tan\left( x_{10}+x_{11}/2\right)\\
 0\\
 x_7\\
\sqrt{2 x_1}
\end{matrix} \right].
\end{equation}
\clearpage
\bibliographystyle{elsarticle-num-names}
\bibliography{MFBO-clean} 

\end{document}